%% file: main.tex
\documentclass[sigconf, nonacm]{acmart}

\input{sections/00_preamble}

\begin{document}

\input{sections/01_title_authors}

\input{sections/02_abstract}

\maketitle

\input{sections/03_introduction}
\input{sections/04_related_work}
\input{sections/05_methodology}
\input{sections/06_experiments}
\input{sections/07_conclusion}

\balance
\bibliographystyle{ACM-Reference-Format}
\bibliography{sections/references}

\clearpage
\appendix
\input{sections/09_01_appendix_data}
\input{sections/09_02_appendix_judge}
\input{sections/09_03_appendix_metrics}
\input{sections/09_04_appendix_supplementary_figures}

\input{sections/09_05_appendix_case_study}

\input{sections/09_06_appendix_method_details}
\input{sections/09_07_appendix_experimental_protocol}
\input{sections/09_08_appendix_statistical_robustness}
\input{sections/09_09_appendix_exposure_capacity}
\input{sections/09_10_appendix_semantic_vs_lexical}

\end{document}

%% file: sections/00_preamble.tex
\usepackage{balance}
\usepackage{booktabs}
\usepackage{multirow}
\usepackage{array}
\usepackage{subcaption}
\usepackage{xspace}
\usepackage{threeparttable}
\usepackage{placeins}
\usepackage{graphicx}
\usepackage{tabularx}
\usepackage{pifont}
\newcommand{\sigmark}{\ding{51}}

\AtBeginDocument{%
  \providecommand\BibTeX{{%
    \normalfont B\kern-0.5em{\scshape i\kern-0.25em b}\kern-0.8em\TeX}}}

\graphicspath{{./images/}}

\newcommand{\framename}{Alpha-R1\xspace}
\newcommand{\modelname}{Alpha-R1\xspace}

\newif\ifcameraready
\ifx\paperversion\submissionmode
  \camerareadyfalse
\else
  \camerareadytrue
\fi

\providecommand{\Description}[1]{}

\newcommand{\tablefontsize}{\footnotesize}
\newcommand{\tablecolsep}{4pt}
\newcommand{\fitcolumnwidth}[1]{%
  \resizebox{\ifdim\width>\columnwidth\columnwidth\else\width\fi}{!}{#1}%
}
\newcommand{\fittextwidth}[1]{%
  \resizebox{\ifdim\width>\textwidth\textwidth\else\width\fi}{!}{#1}%
}
\newcolumntype{L}[1]{>{\raggedright\arraybackslash}p{#1}}
\newcolumntype{Y}{>{\raggedright\arraybackslash}X}

%% file: sections/01_title_authors.tex
\title{{\modelname}: Alpha Screening with LLM Reasoning via Reinforcement Learning}

\author{Zuoyou Jiang}
\authornote{Both authors contributed equally to this research.}
\authornote{Work done during an internship at FinStep and StepFun.}
\affiliation{%
  \institution{Shanghai Jiao Tong University}
  \city{Shanghai}
  \country{China}
}
\email{zuoyou.jiang@sjtu.edu.cn}

\author{Li Zhao}
\authornotemark[1]
\affiliation{%
  \institution{StepFun}
  \city{Shanghai}
  \country{China}
}
\email{zhaoli@stepfun.com}

\author{Rui Sun}
\affiliation{%
  \institution{StepFun}
  \city{Shanghai}
  \country{China}
}
\email{sunrui@stepfun.com}

\author{Ruohan Sun}
\authornotemark[2]
\affiliation{%
  \institution{Shanghai Jiao Tong University}
  \city{Shanghai}
  \country{China}
}
\email{ruohan.sun@sjtu.edu.cn}

\author{Zhongjian Li}
\authornotemark[2]
\affiliation{%
  \institution{Shanghai Jiao Tong University}
  \city{Shanghai}
  \country{China}
}
\email{zhongjian.li@sjtu.edu.cn}

\author{Jing Li}
\affiliation{%
  \institution{StepFun}
  \city{Shanghai}
  \country{China}
}
\email{futureli@stepfun.com}

\author{Daxin Jiang}
\affiliation{%
  \institution{StepFun}
  \city{Shanghai}
  \country{China}
}
\email{djiang@stepfun.com}

\author{Zuo Bai}
\authornote{Corresponding authors.}
\affiliation{%
  \institution{FinStep, StepFun}
  \city{Shanghai}
  \country{China}
}
\email{baizuo@{finstep.cn, stepfun.com}}

\author{Cheng Hua}
\authornotemark[3]
\affiliation{%
  \institution{Shanghai Jiao Tong University}
  \city{Shanghai}
  \country{China}
}
\email{cheng.hua@sjtu.edu.cn}

\renewcommand{\shortauthors}{Jiang et al.}

%% file: sections/02_abstract.tex
\begin{abstract}
    Signal decay and regime shifts pose recurring challenges for data-driven investment strategies in non-stationary markets, where conventional time-series and machine learning approaches often struggle to generalize beyond historical correlations. While large language models (LLMs) offer strong capabilities for processing unstructured information, their potential to support quantitative factor screening through explicit economic reasoning remains underexplored. Existing factor-based methods typically reduce alphas to numerical time series, overlooking the semantic rationale that determines when a factor is economically relevant.

    We present \modelname, an RL-aligned LLM framework for context-aware alpha screening. Its core mechanism, semantic gating, evaluates each candidate factor's semantic profile against a dynamically constructed market state description, selecting a sparse subset of factors whose economic rationale aligns with current market conditions. The selection model is trained via group relative policy optimization (GRPO), using realized portfolio returns as the primary reward signal. Under a 12-month out-of-sample evaluation, \modelname achieves annualized returns of 47.87\% on S\&P~500 and 40.57\% on CSI~300 with Sharpe ratios of 1.62 and 2.23. These results, obtained under a bounded candidate-pool evaluation protocol, provide evidence for second-stage semantic factor reranking in non-stationary markets.
    The full implementation and resources are available at \url{https://github.com/FinStep-AI/Alpha-R1}.
\end{abstract}


%% file: sections/03_introduction.tex
\section{Introduction}
\label{sec:introduction}

Factor investing has remained a cornerstone of asset management since the seminal work of Fama and French~\cite{FAMA19933}, evolving from foundational CAPM~\cite{sharpe1964capitalassetprices} and multi-factor models~\cite{carhart1997persistence} to the modern high-dimensional ``Factor Zoo''~\cite{feng2020taming}. As the number of candidate factors has grown, so has the challenge of selecting factors that remain effective across changing market regimes---a problem that purely statistical methods struggle to solve under non-stationarity.

Despite progress in domain-specific LLMs for finance~\cite{wu2023bloomberggptlargelanguagemodel, liu2023fingpt}, numerical factor signals and textual market information are still largely treated as separate modalities connected by static rules, rather than within a unified framework that captures their semantic interactions in dynamic markets. The financial domain's inherent non-stationarity and noise~\cite{feng2020taming, he2025reinforcement} demand adaptive reasoning. Neither sparsity-based methods~\cite{freyberger2020dissecting}, which exhibit instability during regime changes, nor unaligned general-purpose LLMs~\cite{tatsat2025blackboxinterpretabilityllms} currently meet this requirement.

Existing LLM research in finance spans factor mining~\cite{wang2024llmfactorextractingprofitablefactors, shi2026alphajungle}, agent-based iterative refinement~\cite{Tang2025AlphaAgentLA, Shi2024AlphaForgeAF}, and financial reasoning enhancement~\cite{xiao2025tradingr1financialtradingllm, liu2025finr1largelanguagemodel, deng2026alphaquanter}. Yet a gap remains between broad factor generation and the path-dependent, context-conditioned reasoning required for systematic factor screening under non-stationarity.

To address this gap, we present \modelname, an RL-aligned LLM framework for dynamic second-stage factor reranking after numerical pre-screening~\cite{shao2024deepseekmathpushinglimitsmathematical, guo_deepseek-r1_2025}. Rather than pursuing end-to-end autonomous trading, \modelname focuses specifically on second-stage factor reranking through a mechanism we term \emph{semantic gating}: reasoning over heterogeneous market information and timestamped pre-decision news to assess the economic relevance of a bounded candidate set under the current market state, while producing human-readable rationales.

Our primary contributions are as follows. First, we formulate dynamic second-stage factor reranking as a context-conditioned selection problem and instantiate it with an RL-LLM framework in which semantic gating matches factor rationales to the prevailing market state. Second, we align this selection mechanism with market feedback through a GRPO-based reinforcement learning pipeline that uses realized portfolio returns and structural penalties as training signals. Third, out-of-sample backtests on primary asset pools show that under our fixed evaluation protocol \modelname consistently outperforms both traditional quantitative strategies and zero-shot LLM baselines, with the gains corroborated by block-bootstrap inference, risk-exposure attribution, and lexical-matching falsification tests.

%% file: sections/04_related_work.tex
\section{Related Work}
\label{sec:related_work}

\subsection{LLMs in Quantitative Trading}

\paragraph{Foundation Models and Adaptation.} Initial adaptation focused on domain-specific pre-training, established by BloombergGPT~\cite{wu2023bloomberggptlargelanguagemodel}, and instruction tuning via efficient methods like FinGPT~\cite{liu2023fingpt} and Instruct-FinGPT~\cite{zhang2023instructfingptfinancialsentimentanalysis}, with applications in sentiment extraction~\cite{lopezlira2023chatgpt} and financial benchmarks~\cite{lu2023bbtfincomprehensiveconstructionchinese, zhang2023xuanyuan20largechinese, xie2023pixiulargelanguagemodel}. These models provide the linguistic foundation for complex quantitative tasks.

\paragraph{Alpha Factor Generation.} LLMs have shifted alpha factor mining from manual signal crafting to automated, agent-driven discovery. Early efforts extract profitable factors directly from research context and news~\cite{wang2024llmfactorextractingprofitablefactors} and couple human--AI interaction with code-level implementation~\cite{Wang2025AlphaGPT}. Subsequent frameworks organize mining as search or evolution over the formulaic factor space: iterative refinement against backtest feedback with regularized exploration to counteract alpha decay~\cite{Tang2025AlphaAgentLA}, generative mining and dynamic combination of formulaic factors~\cite{Shi2024AlphaForgeAF}, LLM-guided Monte Carlo tree search~\cite{shi2026alphajungle}, trajectory-level evolution of the end-to-end mining pipeline~\cite{han2026quantaalpha}, and multi-agent data-centric co-optimization of factors and models~\cite{li2025rdagentquant, kou-etal-2025-automate}. A complementary trend trains the miner itself, converting backtest feedback into policy updates~\cite{tang2026alphaagentevo, zhang2026quantevolver}, and equips agents with skill and experience memory to avoid redundant factors entering an already crowded zoo~\cite{wang2026factorminer}; evaluation has likewise been systematized into backtest-free multi-dimensional protocols~\cite{Ding2025AlphaEval}. These pipelines expand the factor zoo but leave zoo-level selection to correlation thresholds or heuristic screening modules~\cite{yuan2026alphacrafter}. Our work complements them at exactly this stage: rather than generating new factors, we train the second-stage screener that selects a factor subset from an existing zoo.

\paragraph{Quant Agents.} Research has expanded toward holistic agents managing the investment lifecycle, from return forecasting with unstructured news~\cite{Guo2024FineTuningLLM} to layered-memory trading agents~\cite{li2024finmem} and multi-agent orchestration of research, risk, and execution~\cite{xiao2025tradingagentsmultiagentsllmfinancial, zhao2025alphaagents}, incorporating multimodal tool use~\cite{zhang2024multimodalfoundationagentfinancial} and self-reflection to mitigate hallucination~\cite{Koa_2024}. Recent benchmarks stress-test such agents under realistic conditions: live, leakage-free fund-investment evaluation~\cite{li2025deepfund} and two-decade cross-sectional backtests~\cite{li2026finsaber} both report that LLM-driven strategies overfit to specific market phases and degrade under regime shifts, while evolutionary multi-agent frameworks explicitly maintain strategy diversity to adapt across regimes~\cite{yun2025quantevolve}. These findings reinforce the need for state-conditioned decision making; in contrast to end-to-end trading agents, our work targets factor screening specifically, using semantic reasoning to bridge static factor definitions and dynamic market regimes.

\subsection{Methods for Screening Alpha Factors}

\paragraph{Sparsity-Driven and Non-Linear Selection.} Direct machine learning approaches to the Factor Zoo include sparsity-inducing regularization like the Lasso~\cite{tibshirani1996lasso, freyberger2020dissecting}, tree-based ensembles for capturing nonlinear interactions~\cite{gu2020empiricalassetpricing}, and deep learning techniques for dimension reduction. Recent work sharpens this toolkit: panel trees grow interpretable, economically guided splits of the asset panel whose leaf portfolios act as tradable factors~\cite{cong2025ptree}, and forest-style partitioning of the characteristics space builds cross-sections on which mainstream factor models still earn significant alpha~\cite{bryzgalova2025forest}. A parallel debate questions whether the zoo should be screened at all, arguing that returns are driven by very large factor models that favor shrinkage over selection~\cite{didisheim2024apt}, while conditional asset pricing embeds structural deep networks whose factor loadings vary with macroeconomic states~\cite{fan2022structural}. Common to these approaches is that selection conditions on numerical statistics alone and cannot incorporate the economic interpretation of the current market state. This limitation motivates a semantic approach to screening, in which factor relevance is evaluated against a structured market narrative rather than derived solely from historical performance vectors.

\subsection{Reinforcement Learning for LLMs}

\paragraph{Alignment and Reasoning Optimization.} Reinforcement learning is key to aligning LLMs with domain objectives~\cite{kaufmann2024surveyreinforcementlearninghuman}. While RLHF via PPO~\cite{ouyang2022traininglanguagemodelsfollow, schulman2017proximalpolicyoptimizationalgorithms} is standard, memory costs and reward hacking~\cite{casper2023openproblemsfundamentallimitations} have spurred lighter preference-based alternatives such as DPO~\cite{rafailov2024directpreferenceoptimizationlanguage}. For reasoning tasks, Group Relative Policy Optimization (GRPO)~\cite{shao2024deepseekmathpushinglimitsmathematical} offers a robust critic-free recipe that fosters self-evolving reasoning chains~\cite{guo_deepseek-r1_2025}; subsequent work has corrected its length- and difficulty-level optimization biases~\cite{liu2025drgrpo}, stabilized large-scale training with decoupled clipping and dynamic sampling~\cite{yu2025dapo}, lifted importance ratios from the token to the sequence level~\cite{zheng2025gspo}, and identified policy-entropy collapse as a central bottleneck of RL with verifiable rewards~\cite{cui2025entropy}.

\paragraph{RL in Finance.} This paradigm is expanding into finance, with GRPO-style tuning for end-to-end trading decisions and financial QA~\cite{xiao2025tradingr1financialtradingllm, liu2025finr1largelanguagemodel}, tool-augmented agentic RL that learns when to query which market tool~\cite{deng2026alphaquanter}, and open platforms benchmarking such RL fine-tuning at scale~\cite{zhang2026finworld}. Because financial rewards are verifiable yet noisy, recent work adds process-level reasoning verification and consistency-grounded rewards to keep aligned reasoning faithful~\cite{sun2026trader1, chen2026stockr1}. A separate line of RL-for-finance systems---FinRL~\cite{liu2020finrl}, MAPS~\cite{lee2020maps}, and SARL~\cite{ye2020sarl}---manages per-asset portfolio weights from numerical states, without semantic representations of either factors or market states. Our work differs in both task and reward: the model performs second-stage factor screening rather than trade execution, and the alignment signal is the realized portfolio return of the selected factor subset. Table~\ref{tab:positioning} (Appendix~\ref{app:positioning}) summarizes this positioning.

%% file: sections/05_methodology.tex
\section{Methodology}
\label{sec:methodology}

\begin{figure*}[htbp]
    \centering
    \includegraphics[width=\textwidth]{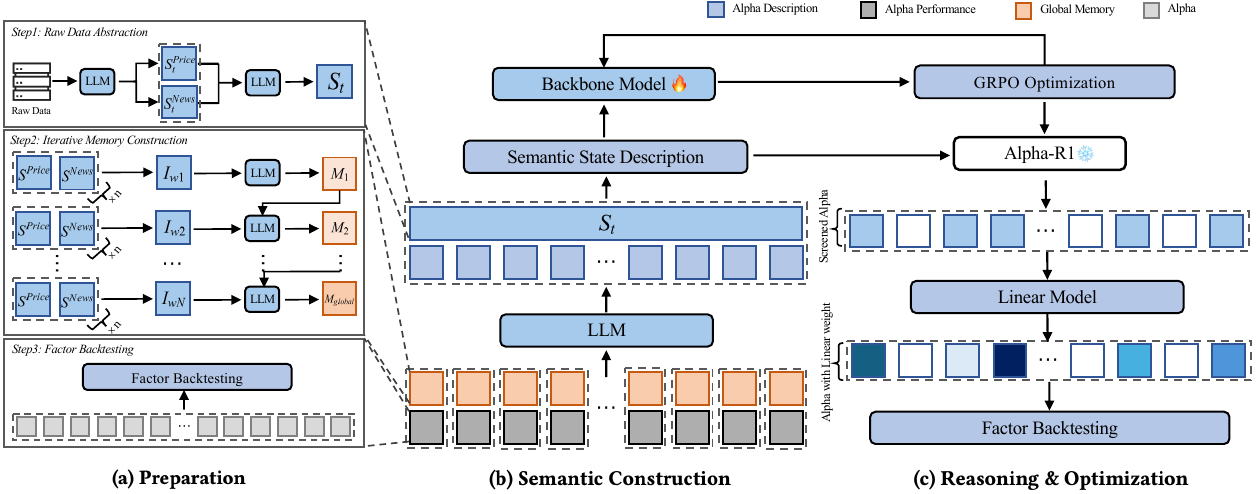}

    \caption{\framename Framework Overview. The pipeline follows three sequential stages:
    (a) Preparation: constructing a global historical memory ($M_{global}$) from raw data and running systematic factor backtesting;
    (b) Semantic Construction: mapping quantitative metrics into semantic factor profiles ($\alpha_{\text{des}}$) and dynamic market state descriptions ($S_t$);
    (c) Reasoning \& Optimization: performing semantic gating that evaluates $\alpha_{\text{des}}$ against $S_t$, with the policy refined through GRPO.}
    \Description{Framework diagram showing the preparation, semantic construction, and reasoning and optimization stages of \framename.}

    \label{fig:framework}
\end{figure*}

We present \framename, an RL-aligned framework that integrates semantic reasoning with reinforcement learning to dynamically select alpha factors in non-stationary markets. As illustrated in Figure~\ref{fig:framework}, the pipeline consists of three sequential stages: (a) Preparation, (b) Semantic Construction, and (c) Reasoning \& Optimization.

\subsection{Data, Memory, and Factor Baselines}
\label{sec:phase1_prep}

\subsubsection{Raw Data Abstraction}
We first transform heterogeneous raw data into structured textual atomic units. At each time step $t$, LLM summarization produces two complementary descriptors: $S^{\text{price}}_t$, which captures technical indicators, trading volume, and sector rotation, and $S^{\text{news}}_t$, which captures financial news and macroeconomic announcements. These atomic units are constructed under a decision-time information set: price and news inputs must be available before the portfolio action associated with day $t$, and no future labels or post-decision annotations are included. Representative price and news samples for Chinese and U.S. markets are provided in Appendix~\ref{sec:appendix_data_flow}.
\subsubsection{Iterative Memory Construction}
\label{sec:memory_construction}
We employ an iterative memory construction pipeline to capture long-term market context. This process aggregates the atomic textual units into a coherent historical narrative.
Let $I_w = \{S^{\text{price}}_t, S^{\text{news}}_t\}_{t \in w}$ denote the set of market descriptions within week $w$. The weekly market summary $M_w$ is updated recursively using a large language model:
\begin{equation}
    M_w = F_{\text{LLM}}^{\text{mem}}(I_w \oplus M_{w-1}),
    \label{eq:iterative_refinement}
\end{equation}
where $\oplus$ denotes textual concatenation and $w \in \{1, \dots, W\}$ represents the week index. After iterating through the historical observation period of $W$ weeks, we obtain the comprehensive historical market description as $M_{\text{global}} = M_W$. This encapsulates the structural evolution and regime shifts of the market, forming the long-term historical memory required for the subsequent semantic profiling.
For the 2025 test period, this historical memory is frozen before 2025.01.01 and is not updated with any observations from the evaluation window.

\subsubsection{Factor Backtesting}
\label{sec:factor_backtesting}
To obtain historical performance estimates for factor behavior, we perform a backtest on the entire factor pool $\mathcal{U}$ over the historical window.
For each factor $i$, we obtain a quantitative performance vector $P_i$, which includes key metrics such as returns, volatility, and decay characteristics.
These historical estimates serve as the empirical basis for linking market memory with factor effectiveness.

\subsection{Profiling and State Description}
\label{sec:phase2_construction}

\subsubsection{Factor Semantic Descriptions}
This stage maps quantitative signals into structured semantic representations. We combine the global market description from the historical observation period, $M_{\text{global}}$, with the factor-specific performance results $P_i$. An LLM (Claude 3.7 Sonnet; Appendix~\ref{app:experimental_protocol}) then generates a semantic profile $\alpha_{\text{des}, i}$ for each factor $i$:
\begin{equation}
    \alpha_{\text{des}, i} = F_{\text{LLM}}^{\text{profile}}(M_{\text{global}}, P_i).
\end{equation}
Here, $P_i$ is computed only from the historical factor-backtesting window 2020.01.01--2023.12.31. The resulting semantic factor profile $\alpha_{\text{des}, i}$ is generated from $(M_{\text{global}}, P_i)$ prior to the 2025 evaluation period and then kept fixed throughout test-time decision-making. Each profile articulates the factor's underlying mechanism, its suitability across market regimes (e.g., high-volatility environments), and its limitations or failure conditions. These semantic descriptions serve as an instruction manual for the reasoning core in subsequent decision-making.
\subsubsection{Asset Pool State Description}
To characterize the current investment environment, we construct a market state description.
In contrast to the long-term market memory, this representation is generated dynamically for each decision day $t$. Using the daily atomic units, ($S^{\text{price}}_t, S^{\text{news}}_t$), we synthesize the market state up to time $t$:
\begin{equation}
    S_t = F_{\text{LLM}}^{\text{state}}(S^{\text{price}}_t, S^{\text{news}}_t).
    \label{eq:state_s_t}
\end{equation}
At test time, $S_t$ is generated only from price data through day $t{-}1$'s close and news published before 30 minutes prior to market open on day $t$ (see the decision-day timeline in Appendix~\ref{sec:appendix_data_flow}), with no access to same-day intraday prices, future observations, future labels, or post-hoc annotations. The resulting state captures the prevailing index dynamics, dominant sector themes, and capital flow patterns. Given these semantic representations, the reasoning model can now perform context-conditioned factor selection.

\subsection{The \modelname Reasoning Model}
\label{sec:alpha_r1}

The \modelname model serves as the central reasoning agent for factor screening and selection. The reasoning process operates as a two-stage pipeline: Stage~1 performs coarse pre-screening on the broader Alpha101 universe using historical backtesting statistics (Rank IC), yielding a pre-screened top-40 candidate pool; Stage~2 applies semantic gating within this reduced set to produce the final context-conditioned selection. This design keeps the prompt context bounded and mitigates context-window overload. Accordingly, our empirical claims concern second-stage semantic reranking within this controlled candidate budget, not unrestricted LLM reasoning over the full factor zoo.

At each decision time $t$, a semantic decision context $C_t$ is constructed by combining two components:
\begin{equation}
    C_t = \{ \alpha_{\text{des}, i} \}_{i \in \mathcal{U}} \oplus S_t,
    \label{eq:decision_context}
\end{equation}
where $\oplus$ represents the structured concatenation of semantic descriptions into a unified prompt context. The components are defined as:
\begin{itemize}
    \item $\alpha_{\text{des}, i}$ denotes the semantic factor description for each candidate factor in the pool $\mathcal{U}$ (sourced from Alpha101~\cite{kakushadze2016101}, see Section~\ref{sec:dataset_and_factors} for details).
    \item $S_t$ is the contemporaneous market state description synthesized via $F_{\text{LLM}}^{\text{state}}$ as defined in Equation~\ref{eq:state_s_t}, which encapsulates price dynamics and news narratives at time $t$.
    \end{itemize}

    This context is instantiated into a structured prompt (detailed in Appendix~\ref{sec:appendix_prompt_sample}) that instructs the model to perform step-by-step analysis before outputting the final factor selection. Based on $C_t$, \modelname outputs the selected factor list $\mathcal{A}_t$. We refer to this mechanism as \emph{semantic gating}: unlike conventional feature selection that filters factors using numerical statistics alone, semantic gating evaluates whether a factor's economic rationale is compatible with the current market narrative. The inputs are factor semantic profiles and the market state description; the output is the selected factor subset $\mathcal{A}_t$; and the role is discrete gating rather than learning continuous factor weights---concretely, the gating decision is produced by the LLM's prompted reasoning over structured inputs, not by a separately parameterized gate network.
\subsection{Reinforcement Learning via GRPO with Market Feedback}
\label{sec:grpo}

We optimize the \modelname reasoning model using reinforcement learning. This design enables learning directly from objective market feedback, adapting the RLHF paradigm to replace subjective human preferences with performance-based financial signals. The reward function combines quantitative portfolio outcomes with assessments of reasoning quality, and training is carried out using Group Relative Policy Optimization (GRPO) to ensure stable and efficient policy updates. Additional judge details are provided in Appendix~\ref{app:judge_rubric}.

\subsubsection{Backbone Model}

We adopt Qwen3-8B as the backbone model for its strong reasoning capabilities. Its pre-trained knowledge accelerates convergence during reinforcement learning and may reduce the risk of the policy overfitting superficial heuristics, providing a stronger initialization for RL fine-tuning.

\subsubsection{Multi-Component Reward Function with Market Feedback}

We design a multi-component reward function that balances market performance with reasoning discipline:
\begin{equation}
r_{\text{final}} = r_{\text{adjusted}} - P_{\text{structural}},
\label{eq:final_reward}
\end{equation}
where $r_{\text{adjusted}}$ captures market-based performance feedback, and $P_{\text{structural}}$ denotes the structural penalty that enforces action validity (sparsity is separately controlled by the prompt schema's maximum factor count). The reward components are computed through the following pipeline.

\paragraph{Rule-based Performance Reward}
Market feedback is obtained via a backtesting procedure based on a linear factor model trained on four years of historical data.
We adopt a linear specification for three reasons: it provides a stable and interpretable mapping from factor exposures to expected returns, enables direct attribution of performance to selected factors, and avoids introducing non-stationarity into the reward signal during reinforcement learning by keeping the evaluation model fixed.

\begin{enumerate}
    \item Linear Model: We use fixed regression coefficients ${\beta_i}$ estimated from historical data.
    \item For each stock, predicted returns are computed using the selected factor set $\mathcal{A}_t$:
    \begin{equation}
    R_{\text{pred}} = \beta_0 + \sum_{i \in \mathcal{A}_t} (\beta_i \times V_i),
    \label{eq:predicted_return}
    \end{equation}
    where $V_i$ denotes the previous-day value of factor $i$, and unselected factors contribute zero.
    \item Portfolio Construction: Stocks are ranked by predicted returns, and the top $N$ are selected to form an equal-weighted portfolio.
    \item Base Reward Calculation: Compute the active return over the benchmark over a holding period $H$, scaled for the reward function:
    \begin{equation}
    r_{\text{base}} = (R_{\text{port}} - R_{\text{bench}}) \times 100,
    \label{eq:base_reward}
    \end{equation}
    where $H$ denotes the holding period (e.g., $H=5$ days). Both $R_{\text{port}}$ and $R_{\text{bench}}$ are computed under the same execution protocol described in Section~\ref{sec:execution_mechanism}, including slot rotation, VWAP fills, and transaction fees.
\end{enumerate}

\paragraph{Quality-Adjusted Reward with LLM-as-Judge Evaluation}
We incorporate reasoning quality into the reward through LLM-as-judge evaluation, used as a \emph{consistency regularizer} rather than a verifier of chain-of-thought faithfulness, factor profitability, or economic truth. A consistency penalty $P_{\text{consistency}}$ is computed as
\begin{equation}
P_{\text{consistency}} = F_{\text{LLM}}^{\text{judge}}(C_t, \mathcal{A}_t, \text{response}),
\label{eq:consistency_eval}
\end{equation}
where $F_{\text{LLM}}^{\text{judge}}$ denotes a judge LLM (Claude 3.5 Haiku) that evaluates the generated rationale against a structured rubric targeting five failure modes: logical contradictions, linguistic confusion, hallucinated content, repetition, and other irrationalities, yielding a penalty score in $[0, 10]$. Manual calibration with three finance reviewers showed high directional agreement (Appendix~\ref{app:judge_rubric}). The normalized score $P_{\text{norm}} = P_{\text{consistency}} / 10$ is applied as a proportional penalty on the absolute return:
\begin{equation}
r_{\text{adjusted}} = r_{\text{base}} - \lvert r_{\text{base}} \rvert \times P_{\text{norm}}.
\label{eq:adjusted_reward}
\end{equation}

\paragraph{Structural Penalties}
The structural penalty $P_{\text{structural}}$ enforces output validity by strictly penalizing the generation of unparsable or non-existent factors, ensuring that all reasoning outputs are executable within the quantitative backtesting framework. Sparsity is enforced separately through the prompt schema, which imposes a hard maximum of 10 selected factors per decision (see Appendix~\ref{sec:appendix_prompt_sample}). We define the validity penalty as:
\begin{equation}
P_{\text{structural}} = 5 \cdot \mathbb{I}(\text{response} \notin \Omega_{\text{valid}}),
\label{eq:structural_penalty}
\end{equation}
where $\mathbb{I}(\cdot)$ is the indicator function, and $\Omega_{\text{valid}}$ represents the set of responses that strictly adhere to the required XML output format (as detailed in Appendix~\ref{sec:appendix_prompt_sample}) and contain valid factor identifiers from the universe $\mathcal{U}$. This imposes a hard constraint to prevent format hallucinations.

\subsubsection{GRPO Optimization with Market-Aligned Objectives}

We employ standard Group Relative Policy Optimization (GRPO) to fine-tune \modelname. For each semantic decision prompt, a group of outputs is sampled and their rewards are normalized into relative advantages. Let $q$ denote the input context and $\{o_i\}_{i=1}^{G}$ the sampled group of responses. The normalized advantage of the $i$-th response is
\begin{equation}
\hat{A}_i = \frac{r_i - \mathrm{mean}(r)}{\mathrm{std}(r)},
\label{eq:grpo_advantage}
\end{equation}
where $r_i$ is the final reward of response $o_i$. The token-level policy ratio at generation step $t$ is
\begin{equation}
\rho_t^{(i)}(\theta) = \frac{\pi_\theta(o_{i,t} \mid q, o_{i,<t})}{\pi_{\theta_{\mathrm{old}}}(o_{i,t} \mid q, o_{i,<t})},
\label{eq:grpo_ratio}
\end{equation}
where $\pi_\theta$ and $\pi_{\theta_{\mathrm{old}}}$ denote the current and sampling policies, respectively. The GRPO objective combines a clipped surrogate term with KL regularization against the reference policy $\pi_{\mathrm{ref}}$:
\begin{equation}
\begin{aligned}
J_{\text{GRPO}}(\theta)
= \mathbb{E}_{q,\{o_i\}}\!\bigg[ &
    \frac{1}{G}\sum_{i=1}^{G}\frac{1}{|o_i|}\sum_{t=1}^{|o_i|}
    \min\!\Big(
        \rho_t^{(i)}(\theta)\hat{A}_i,
    \\ &
        \operatorname{clip}\!\left(\rho_t^{(i)}(\theta), 1-\epsilon, 1+\epsilon\right)\hat{A}_i
    \Big)
\bigg] \\
& - \beta \, \mathbb{E}_{q}\!\left[
    D_{\mathrm{KL}}\!\left(
        \pi_\theta(\cdot \mid q)
        \| \pi_{\mathrm{ref}}(\cdot \mid q)
    \right)
\right],
\end{aligned}
\label{eq:grpo_objective}
\end{equation}
where $\epsilon$ is the clipping coefficient and $\beta$ controls the KL regularization strength. The first term improves high-advantage responses while preventing excessively large policy updates, and the second term regularizes the policy against drifting too far from the reference policy, which is the frozen pre-trained backbone. The numerical hyperparameters are listed in Appendix~\ref{app:method_details}.
\subsection{Portfolio Construction and Execution}
\label{sec:execution_mechanism}

The selected factor set $\mathcal{A}_t$ is converted into tradable long-only positions through a fixed linear return model and a realistic execution protocol. We use slot rotation with holding period $H=5$, select the top $N=10$ stocks by predicted return into an equal-weighted long-only portfolio, execute with the first-30-minute VWAP (09:31--10:00), reject limit-locked orders, exclude newly listed stocks on their first trading day, and charge double-sided 10\,bps transaction fees. When fewer than $N$ stocks are available due to limit locks or suspensions, the unfilled portion is held in cash. The same execution-adjusted return definition is used in both GRPO reward computation and final backtesting; formulas and execution details are provided in Appendix~\ref{app:method_details}.

%% file: sections/06_experiments.tex
\section{Experiments}
\label{sec:experiments}

This section presents the empirical evaluation of \modelname against traditional quantitative strategies and zero-shot large language models (LLMs). Only \modelname receives domain-specific RL alignment; the generic LLM baselines are evaluated zero-shot under the same downstream protocol. This comparison evaluates the domain-specific RL-trained \modelname pipeline against zero-shot prompting, and does not by itself isolate how much of the gain comes from the RL algorithm itself versus the training data and reward design. Out-of-universe generalization results on Russell 2000 and CSI 1000 are reported in Section~\ref{sec:cross_universe_transfer}.

\subsection{Experimental Setup}
\label{sec:dataset_and_factors}

\subsubsection{Datasets and Temporal Split}
We evaluate on S\&P 500 and CSI 300, with out-of-universe generalization results on Russell 2000 and CSI 1000 reported in Section~\ref{sec:cross_universe_transfer}. The dynamic factor zoo is constructed from the Alpha101 library~\cite{kakushadze2016101}. We enforce strict temporal separation between factor coefficient estimation (2020--2023), model training (2024), and out-of-sample testing (2025). All asset pools use point-in-time constituent membership available as of each decision date. Detailed anti-leak\-age enforcement, factor-aug\-men\-ta\-tion strategy, and constituent-hand\-ling rules are provided in Appendix~\ref{app:experimental_protocol}.

\subsubsection{Training and Inference}
\modelname adopts Qwen3-8B as its backbone and is trained on 64 H800 GPUs for approximately 120 hours; full training and inference configuration details are provided in Appendix~\ref{app:experimental_protocol}.

\subsubsection{Baselines and Evaluation Protocol}
We compare against traditional quantitative strategies (PCA, XGBoost, LightGBM, A2C, PPO), FinRL-style RL agents (DDPG, TD3, SAC) trained in the same factor-selection environment, classic momentum/mean-reversion baselines (full grid in Appendix~\ref{app:exposure_capacity}), and zero-shot generic LLM baselines (Gemini 2.5 Pro, Claude 3.7 Sonnet, DeepSeek-R1, Qwen3-8B). All strategies follow the same downstream execution protocol described in Section~\ref{sec:execution_mechanism} ($H=5$, $TopN=10$, VWAP, double-sided 10\,bps fees), with risk-free rates of 4.5\% for the US market and 1.5\% for the Chinese market. Performance is evaluated using annualized return (AR), excess Sharpe ratio (SR), and maximum drawdown (MDD) (detailed definitions in Appendix~\ref{sec:appendix_metrics}); additional risk metrics and return-distribution diagnostics are reported in Appendices~\ref{sec:appendix_metrics} and~\ref{app:return_distribution_analysis}. All comparisons follow the same 12-month evaluation protocol over the 2025 test period. No hyperparameters were tuned on the 2025 test period: GRPO hyperparameters are standard published defaults (Appendix~\ref{app:method_details}), the reported checkpoint was selected by training-period reward, and the reported execution configuration is the code default fixed before evaluation---the sensitivity heatmaps (Figure~\ref{fig:heatmaps_main}) show it sits far below the test-surface peak (e.g., 176.03\% AR around $TopN=2$--$3$ on S\&P 500 versus the reported 47.87\%), which is the opposite of test-set tuning.

\subsection{Performance Evaluation}

\begin{figure*}[htbp]
    \centering
    \begin{subfigure}[b]{0.465\textwidth}
        \centering
        \includegraphics[width=\textwidth]{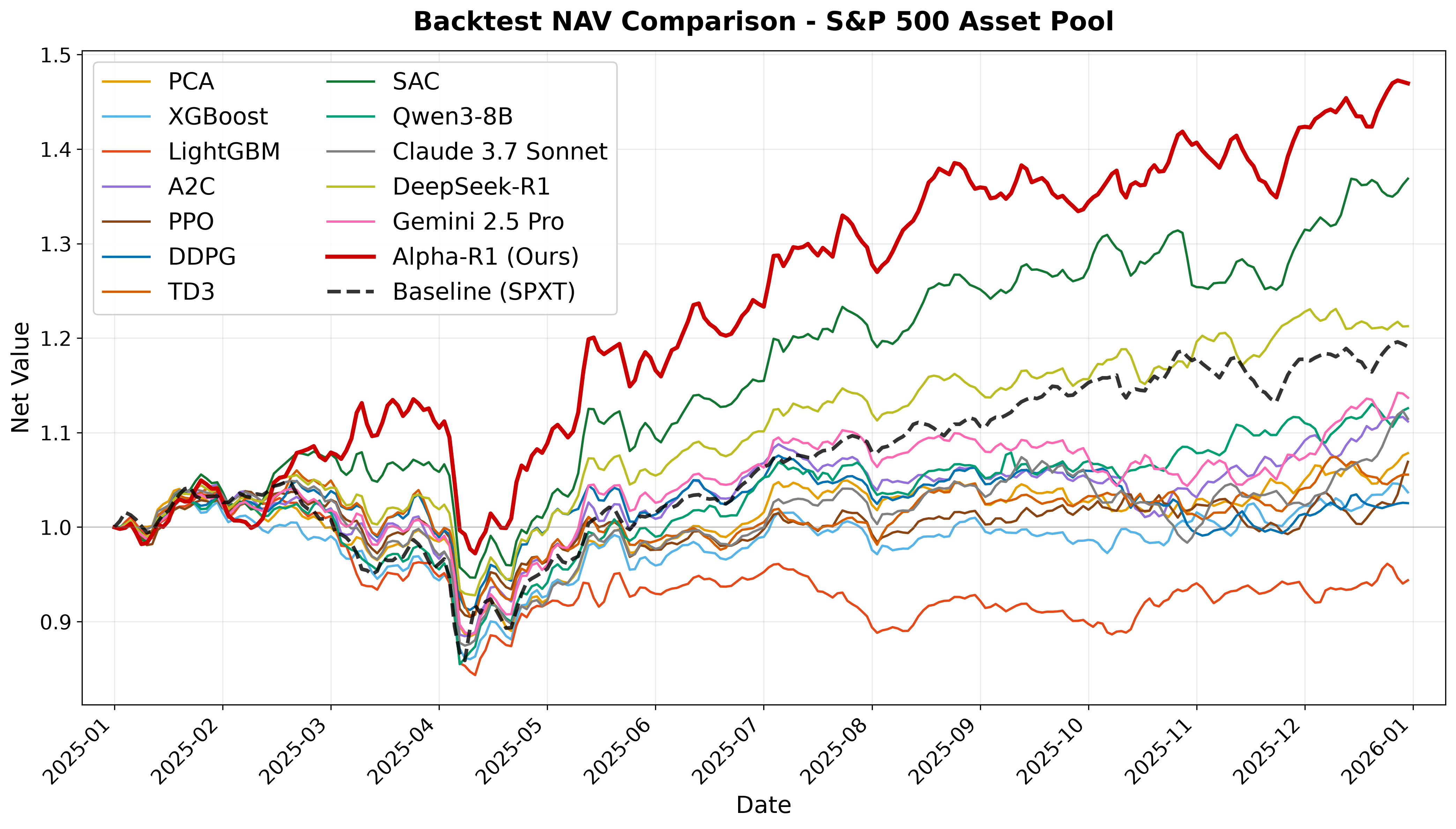}
        \caption{S\&P 500}
        \label{fig:nav_pool_a}
    \end{subfigure}
    \hfill
    \begin{subfigure}[b]{0.465\textwidth}
        \centering
        \includegraphics[width=\textwidth]{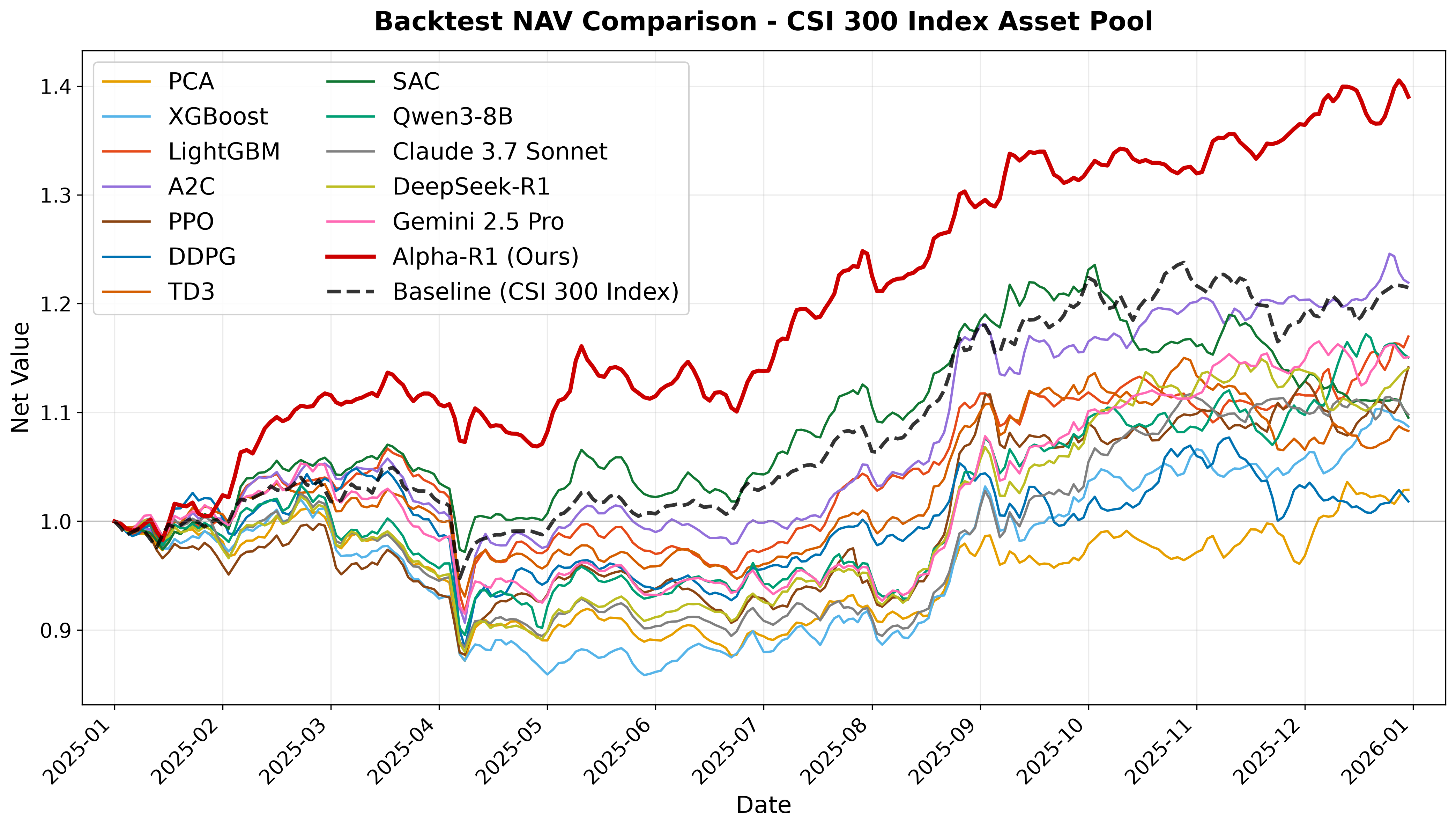}
        \caption{CSI 300}
        \label{fig:nav_pool_b}
    \end{subfigure}
    \caption{Main Experiment Cumulative Net Value Curves over the 12-month 2025 testing set. (a) S\&P 500: \modelname produces the highest cumulative return among evaluated strategies. (b) CSI 300: \modelname achieves the highest cumulative return with the smallest maximum drawdown. The DDPG, TD3, and SAC curves are the FinRL-style RL agents trained in the same factor-selection environment (full FinRL and momentum/mean-reversion grids in Appendices~\ref{app:finrl_baselines} and~\ref{app:momentum_reversal}).}
    \Description{Two cumulative net value line charts comparing \modelname with traditional quantitative strategies and zero-shot LLM baselines on S\&P 500 and CSI 300.}
    \label{fig:nav_curves_main}
\end{figure*}

We first evaluate the capacity of \modelname to generate active returns against the benchmark. Table~\ref{tab:main_results} details the quantitative metrics, while Figure~\ref{fig:nav_curves_main} visualizes the wealth accumulation trajectories.

\begin{table}[htbp]
    \centering
    \caption{Main Experiment Results. Performance comparison of \framename against two primary asset pools (S\&P 500 and CSI 300) over the 12-month testing period 2025.01.01 -- 2025.12.31. We report annualized return (AR), excess Sharpe ratio (SR), and maximum drawdown (MDD). The best results are highlighted in bold. FinRL-style RL agents and the momentum\slash mean-reversion grid follow the identical protocol (Appendices~\ref{app:finrl_baselines} and~\ref{app:momentum_reversal}).}
    \label{tab:main_results}
    \begingroup
    \tablefontsize
    \setlength{\tabcolsep}{\tablecolsep}
    \fitcolumnwidth{
    \begin{tabular}{llcccccc}
        \toprule
        \multirow{2}{*}{\textbf{Type}} & \multirow{2}{*}{\textbf{Method}} & \multicolumn{3}{c}{\textbf{Asset Pool S\&P 500}} & \multicolumn{3}{c}{\textbf{Asset Pool CSI 300}} \\
        \cmidrule(lr){3-5} \cmidrule(lr){6-8}
         & & AR (\%) & SR & MDD (\%) & AR (\%) & SR & MDD (\%) \\
        \midrule
        \multirow{9}{*}{Non-LLM} & Buy \& Hold & 19.34 & 0.80 & 18.75 & 22.16 & 1.31 & 10.49 \\
         & PCA & 7.98 & 0.27 & 17.30 & 2.93 & 0.17 & 14.46 \\
         & XGBoost & 3.49 & 0.03 & 18.45 & 8.99 & 0.50 & 16.26 \\
         & LightGBM & -5.42 & -0.43 & 20.93 & 18.44 & 1.05 & 14.92 \\
         & A2C & 10.82 & 0.40 & 17.70 & 22.96 & 1.20 & 14.86 \\
         & PPO & 7.68 & 0.25 & 14.97 & 14.96 & 0.81 & 12.95 \\
         & DDPG & 2.53 & -0.02 & 15.04 & 1.97 & 0.12 & 16.54 \\
         & TD3 & 5.54 & 0.14 & 16.58 & 8.66 & 0.52 & 10.26 \\
         & SAC & 37.60 & 1.44 & 15.18 & 9.77 & 0.56 & 11.68 \\
        \midrule
        \multirow{5}{*}{LLM} & Gemini 2.5 Pro & 14.23 & 0.55 & 17.01 & 16.29 & 0.90 & 14.01 \\
         & Claude 3.7 Sonnet & 10.92 & 0.40 & 18.88 & 10.13 & 0.57 & 14.49 \\
         & DeepSeek-R1 & 21.94 & 0.93 & \textbf{14.36} & 14.66 & 0.81 & 14.60 \\
         & Qwen3-8B & 12.85 & 0.47 & 19.52 & 15.44 & 0.79 & 14.38 \\
         & \textbf{\modelname (Ours)} & \textbf{47.87} & \textbf{1.62} & 16.91 & \textbf{40.57} & \textbf{2.23} & \textbf{6.58} \\
        \bottomrule
    \end{tabular}
    }
    \endgroup
\end{table}

\subsubsection{Main Results Analysis}
As shown in Table~\ref{tab:main_results} and Figure~\ref{fig:nav_curves_main}, \modelname achieves the strongest performance under the shared 12-month evaluation protocol. All conventional quantitative baselines trail substantially on both asset pools. Against zero-shot LLM baselines, the domain-specific RL-trained \modelname pipeline achieves annualized returns of 47.87\% on S\&P 500 and 40.57\% on CSI 300, with excess Sharpe Ratios of 1.62 and 2.23, compared to the strongest generic LLM (DeepSeek-R1) at 21.94\% and 14.66\%. These results are consistent with the value of context-conditioned factor reranking.

The two added baseline families reinforce this reading. Every FinRL-style agent trails \modelname by at least 10.3\,pp AR in every universe---the strongest, SAC on S\&P 500 at 37.60\%, still trails by 10.3\,pp---and most trail even the plain A2C baseline from the same algorithm family (full results in Appendix~\ref{app:finrl_baselines}). The best classic momentum\slash mean-reversion baseline per universe (Momentum-60 on CSI 300 at 29.60\%, Momentum-120 on S\&P 500 at 24.34\%) trails by 10.97\,pp or more, and momentum variants are deeply negative on the small-cap universes (-34.0\% to -42.7\% on CSI 1000; Momentum-20 at -56.9\% on Russell 2000; Appendix~\ref{app:momentum_reversal}), which is inconsistent with a momentum-repackaging account of the reported gains.
\subsection{Cross-Universe Transfer}
\label{sec:cross_universe_transfer}

To assess whether the gains observed on the primary asset pools extend to out-of-domain universes, we evaluate the same trained \modelname policy and baselines on Russell~2000 and CSI~1000 without retraining or hyperparameter tuning. These experiments reuse the same pre-screened factor pool and evaluation pipeline as the primary experiments.

\begin{table}[htbp]
    \centering
    \caption{Cross-Universe Transfer Results. Performance on out-of-domain asset pools (Russell 2000 and CSI 1000) over the 12-month testing period 2025.01.01 -- 2025.12.31 within the shared factor framework. We report annualized return (AR), excess Sharpe ratio (SR), and maximum drawdown (MDD). The best results are highlighted in bold. FinRL-style RL agents and the momentum\slash mean-reversion grid follow the identical protocol (Appendices~\ref{app:finrl_baselines} and~\ref{app:momentum_reversal}).}
    \label{tab:zeroshot_results}
    \begingroup
    \tablefontsize
    \setlength{\tabcolsep}{\tablecolsep}
    \fitcolumnwidth{
    \begin{tabular}{llcccccc}
        \toprule
        \multirow{2}{*}{\textbf{Type}} & \multirow{2}{*}{\textbf{Method}} & \multicolumn{3}{c}{\textbf{Asset Pool Russell 2000}} & \multicolumn{3}{c}{\textbf{Asset Pool CSI 1000}} \\
        \cmidrule(lr){3-5} \cmidrule(lr){6-8}
         & & AR (\%) & SR & MDD (\%) & AR (\%) & SR & MDD (\%) \\
        \midrule
        \multirow{9}{*}{Non-LLM} & Buy \& Hold & 13.76 & 0.48 & 23.79 & 32.49 & 1.33 & 16.87 \\
         & PCA & 14.04 & 0.47 & 28.15 & 0.63 & 0.06 & 20.99 \\
         & XGBoost & 0.52 & -0.04 & 27.00 & 0.71 & 0.05 & 22.24 \\
         & LightGBM & -8.43 & -0.37 & 36.46 & -4.82 & -0.23 & 20.88 \\
         & A2C & 55.74 & 1.91 & \textbf{16.46} & -8.03 & -0.43 & 22.90 \\
         & PPO & -25.83 & -1.12 & 42.99 & 15.28 & 0.72 & 20.39 \\
         & DDPG & -21.50 & -0.88 & 32.01 & 9.60 & 0.45 & 22.13 \\
         & TD3 & 18.95 & 0.59 & 33.99 & -0.30 & 0.04 & 22.26 \\
         & SAC & 6.72 & 0.23 & 34.50 & 8.29 & 0.39 & 22.41 \\
        \midrule
        \multirow{5}{*}{LLM} & Gemini 2.5 Pro & 19.95 & 0.64 & 29.24 & 0.49 & 0.06 & 28.37 \\
         & Claude 3.7 Sonnet & 30.14 & 0.72 & 27.92 & -1.92 & 0.08 & 23.86 \\
         & DeepSeek-R1 & 30.50 & 0.93 & 31.12 & -6.45 & -0.27 & 30.65 \\
         & Qwen3-8B & -9.41 & -0.43 & 32.97 & 14.48 & 0.61 & 23.02 \\
         & \textbf{\modelname (Ours)} & \textbf{80.54} & \textbf{2.46} & 17.67 & \textbf{73.52} & \textbf{2.80} & \textbf{13.75} \\
        \bottomrule
    \end{tabular}
    }
    \endgroup
\end{table}

\begin{figure*}[htbp]
    \centering
    \begin{subfigure}[b]{0.49\textwidth}
        \centering
        \includegraphics[width=\textwidth]{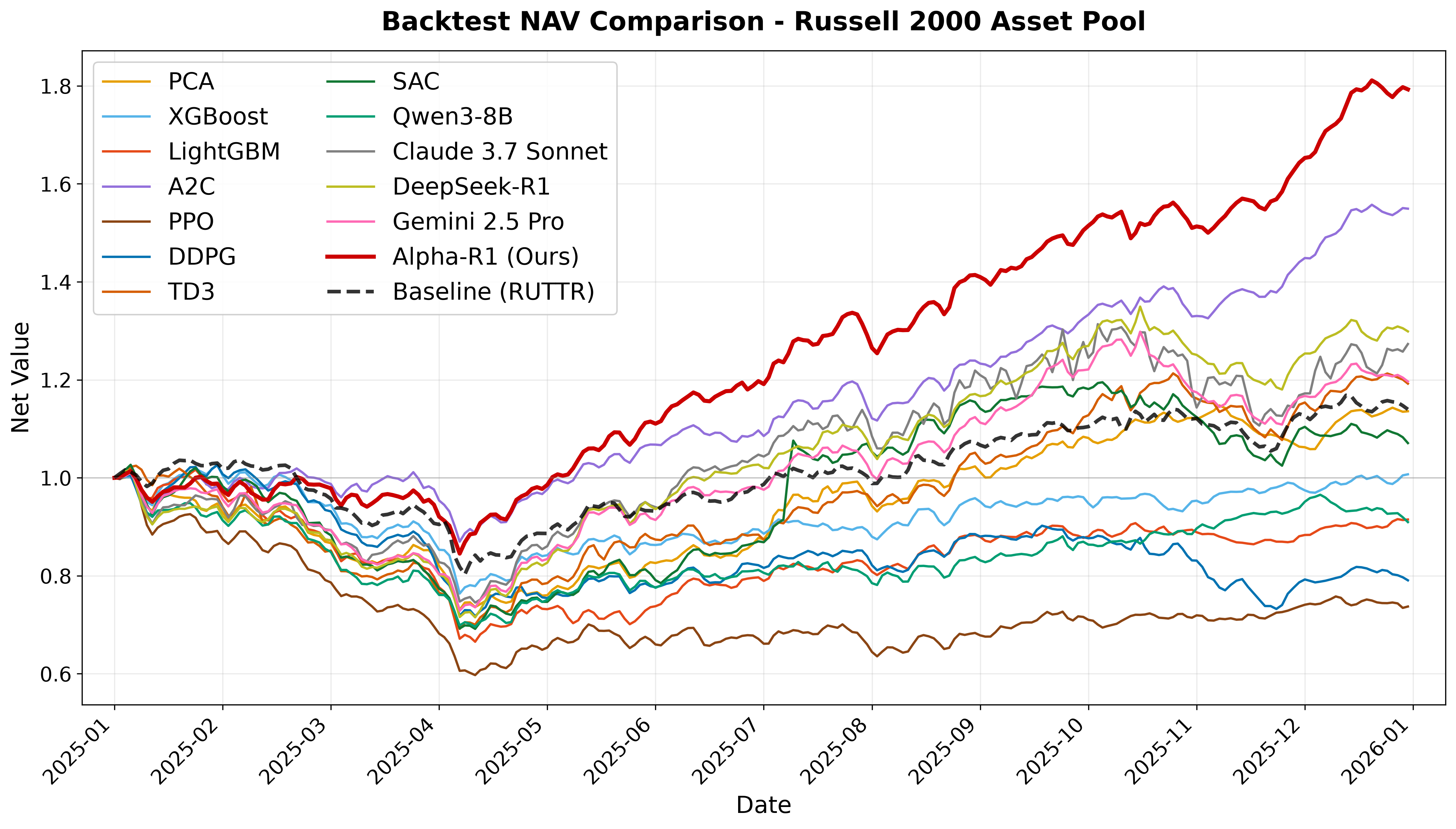}
        \caption{Russell 2000}
        \label{fig:nav_pool_c}
    \end{subfigure}
    \hfill
    \begin{subfigure}[b]{0.49\textwidth}
        \centering
        \includegraphics[width=\textwidth]{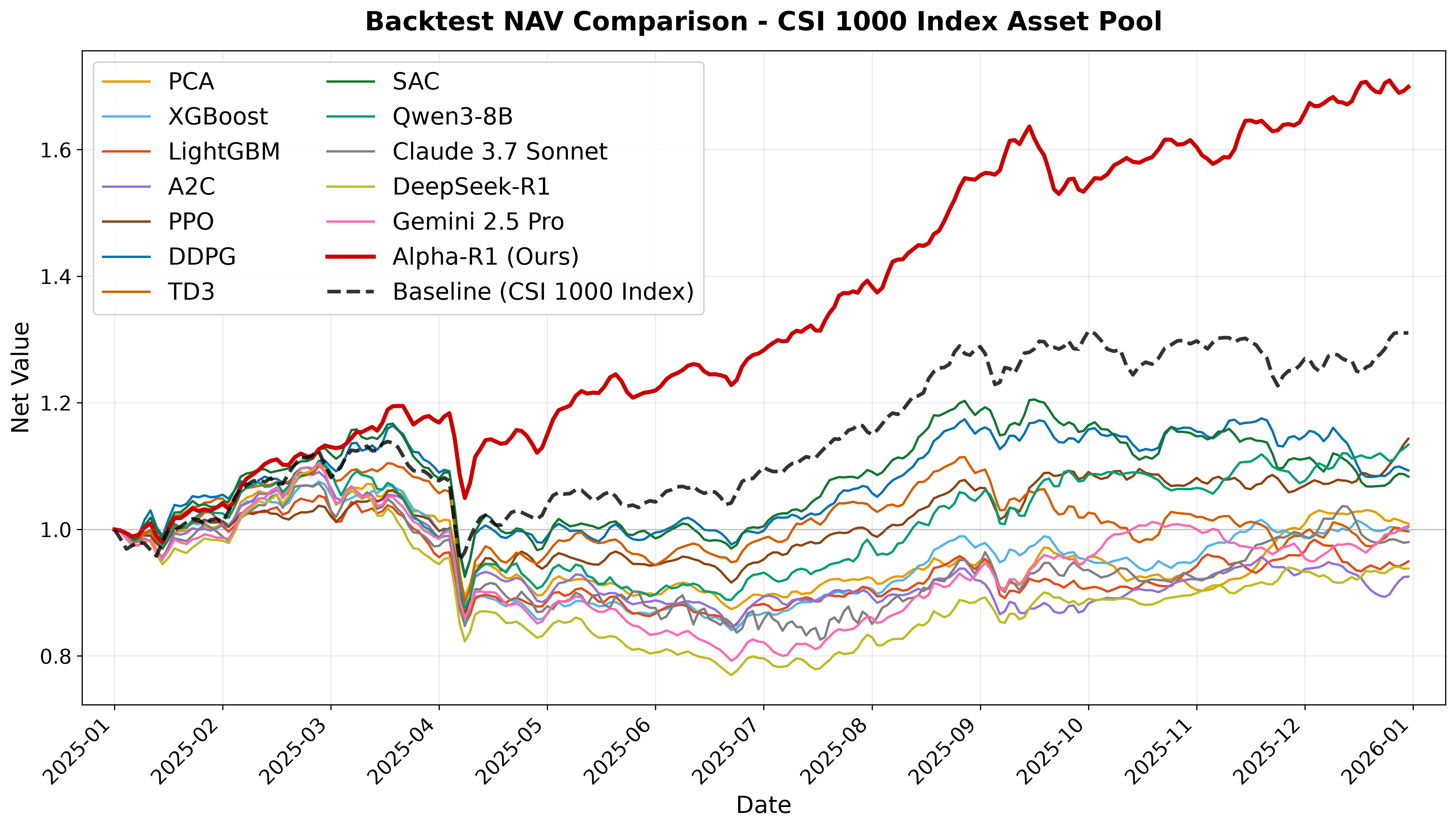}
        \caption{CSI 1000}
        \label{fig:nav_pool_d}
    \end{subfigure}
    \caption{Cross-Universe Transfer Curves over the 12-month 2025 testing set. Comparison of wealth accumulation trajectories on out-of-domain asset pools within the shared factor framework: (a) Russell 2000 and (b) CSI 1000. The DDPG, TD3, and SAC curves are the FinRL-style RL agents trained in the same factor-selection environment (full grids in Appendix~\ref{app:exposure_capacity}).}
    \Description{Two cumulative net value line charts comparing \modelname with baselines on Russell 2000 and CSI 1000.}
    \label{fig:nav_curves_zeroshot}
\end{figure*}

On Russell~2000, \modelname achieves an annualized return of 80.54\% with an excess Sharpe Ratio of 2.46 (Table~\ref{tab:zeroshot_results}). On CSI~1000, it delivers 73.52\% annualized return and a Sharpe Ratio of 2.80. Figure~\ref{fig:nav_curves_zeroshot} visualizes the corresponding wealth accumulation trajectories. The large performance gap over zero-shot LLM and traditional baselines suggests that the semantic gating policy learned on S\&P~500 and CSI~300 transfers to materially different market-cap segments.

\subsection{Ablation Study}

To dissect the contributions of individual components, we conduct a series of ablation experiments on both CSI 300 and S\&P 500 asset pools. All ablation variants are trained from scratch using the modified framework to ensure a fair comparison.

\subsubsection{Ablation Variants}
We evaluate four variants to dissect the contribution of each component: (1) w/o Market Price, which removes market price trends from $S_{t}$; (2) w/o News, which excludes market news information from the state space; (3) w/o Factor Semantics, where natural language factor descriptions are replaced with their raw mathematical formulas; and (4) w/o RL Optimization, which uses the base backbone model (Qwen3-8B) without the reinforcement learning alignment pipeline.

\begin{table}[htbp]
    \centering
    \caption{Ablation Study Results. Performance contribution of different components on S\&P 500 and CSI 300 over the 12-month testing period 2025.01.01 -- 2025.12.31. We report annualized return (AR), excess Sharpe ratio (SR), and maximum drawdown (MDD). The best results are highlighted in bold.}
    \label{tab:ablation}
    \begingroup
    \tablefontsize
    \setlength{\tabcolsep}{\tablecolsep}
    \fitcolumnwidth{
    \begin{tabular}{lcccccc}
        \toprule
        \multirow{2}{*}{\textbf{Method}} & \multicolumn{3}{c}{\textbf{S\&P 500}} & \multicolumn{3}{c}{\textbf{CSI 300}} \\
        \cmidrule(lr){2-4} \cmidrule(lr){5-7}
         & AR (\%) & SR & MDD (\%) & AR (\%) & SR & MDD (\%) \\
        \midrule
        \textbf{\modelname (Full)} & \textbf{47.87} & \textbf{1.62} & \textbf{16.91} & \textbf{40.57} & \textbf{2.23} & 6.58 \\
        \midrule
        Buy \& Hold & 19.34 & 0.80 & 18.75 & 22.16 & 1.31 & 10.49 \\
        w/o Market Price & 28.83 & 1.02 & 17.59 & 27.93 & 1.55 & 7.21 \\
        w/o News & 34.63 & 1.20 & 17.60 & 31.79 & 1.75 & \textbf{6.51} \\
        w/o Factor Semantics & 40.45 & 1.37 & 17.16 & 35.67 & 1.94 & 6.51 \\
        w/o RL Optimization & 12.85 & 0.47 & 19.52 & 15.44 & 0.79 & 14.38 \\
        \bottomrule
    \end{tabular}
    }
    \endgroup
\end{table}

\subsubsection{Results and Analysis}
Table~\ref{tab:ablation} quantifies component contributions over the same 12-month horizon. RL alignment has the largest effect: removing it (equivalent to zero-shot Qwen3-8B) reduces AR from 47.87\% to 12.85\% on S\&P 500 ($-$35.02\,pp) and from 40.57\% to 15.44\% on CSI 300 ($-$25.13\,pp), with Sharpe Ratios dropping below 0.80 on both pools. Among the remaining components, the market-price stream contributes the most: removing it reduces AR by 19.04\,pp (S\&P 500) and 12.64\,pp (CSI 300), followed by the news stream at 13.24\,pp and 8.78\,pp. Replacing natural-language factor descriptions with raw formulas (w/o Factor Semantics) costs a further 7.42\,pp and 4.90\,pp. Although this is the smallest of the component deltas, it is the arm that most directly isolates the semantic channel---the state text and all numerical inputs are left untouched---and the mechanism-level tests in Appendix~\ref{app:semantic_vs_lexical} (BM25 strawman and profile obfuscation) show that this contribution is not reducible to lexical overlap.

The corresponding cumulative net value curves are provided in Appendix~\ref{app:supplementary_figures} (Figure~\ref{fig:ablation_curves}). On both S\&P 500 and CSI 300, the full model preserves the steepest upward trend, whereas the variants without RL optimization or factor semantics show materially weaker cumulative growth. These trajectory-level comparisons provide additional support that the gains in Table~\ref{tab:ablation} are not visually concentrated in only one short segment of the 2025 test window.

\subsection{Semantic vs. Baseline Gating Strategies}

To validate semantic gating, Table~\ref{tab:gating_comparison} compares \modelname against eight factor-gating baselines under the same candidate pool and downstream execution protocol: heuristic selectors (Random, Rolling IC, Top-k), search selectors (Greedy, Beam), and learned selectors (Lasso, MLP, Transformer).

\begin{table}[htbp]
    \centering
    \caption{Gating Baseline Comparison on S\&P 500 and CSI 300 under the 12-month evaluation protocol. We report annualized return (AR), excess Sharpe ratio (SR), and maximum drawdown (MDD). The best results are highlighted in bold.}
    \label{tab:gating_comparison}
    \begingroup
    \tablefontsize
    \setlength{\tabcolsep}{\tablecolsep}
    \fitcolumnwidth{
    \begin{tabular}{llcccccc}
        \toprule
        \multirow{2}{*}{\textbf{Type}} & \multirow{2}{*}{\textbf{Method}} & \multicolumn{3}{c}{\textbf{S\&P 500}} & \multicolumn{3}{c}{\textbf{CSI 300}} \\
        \cmidrule(lr){3-5} \cmidrule(lr){6-8}
         & & AR (\%) & SR & MDD (\%) & AR (\%) & SR & MDD (\%) \\
        \midrule
        \multirow{3}{*}{Heuristic} & Random & 5.35 & 0.13 & 15.48 & 0.75 & 0.05 & 14.51 \\
         & Rolling IC & 15.30 & 0.61 & 14.78 & 11.15 & 0.63 & 13.84 \\
         & Top-k & 1.33 & -0.07 & 15.60 & -10.22 & -0.44 & 19.55 \\
        \midrule
        \multirow{2}{*}{Search} & Greedy & -12.62 & -0.82 & 22.50 & 24.92 & 1.17 & 11.26 \\
         & Beam & 16.59 & 0.61 & 16.20 & 23.01 & 1.09 & 11.02 \\
        \midrule
        \multirow{3}{*}{Learned} & Lasso & 0.50 & -0.12 & 18.88 & 6.42 & 0.46 & 8.75 \\
         & MLP & 16.53 & 0.69 & 14.67 & -6.16 & -0.30 & 16.61 \\
         & Transformer & 20.57 & 0.86 & \textbf{14.29} & 8.75 & 0.45 & 14.48 \\
        \midrule
        Semantic & \textbf{\modelname (Ours)} & \textbf{47.87} & \textbf{1.62} & 16.91 & \textbf{40.57} & \textbf{2.23} & \textbf{6.58} \\
        \bottomrule
    \end{tabular}
    }
    \endgroup
\end{table}

Across both S\&P 500 and CSI 300 under the shared evaluation protocol, \modelname delivers the strongest annualized return and excess Sharpe ratio. The best non-semantic alternatives remain substantially behind: Transformer reaches only 43\% of \modelname's AR on S\&P 500 and 22\% on CSI 300, while Beam performs competitively on CSI 300 but still relies on a static validation-selected subset. Overall, the comparison indicates that conditioning factor screening on semantic market descriptions provides a distinct adaptation mechanism beyond numerical IC signals, static search, or lightweight learned gating. A BM25 lexical-matching strawman---which scores the same profiles against the same state text with no reasoning---trails \modelname by 24.7--56.7\,pp in every universe (Appendix~\ref{app:bm25_strawman}; on Russell 2000 the paired interval includes zero, reported as-is), indicating the gating advantage is not reducible to keyword overlap. Full gating trajectories and case-level rationales are provided in Appendix~\ref{app:supplementary_figures} and Appendix~\ref{app:case_study}.

\subsection{Parameter Sensitivity Analysis}
\label{sec:parameter_sensitivity}

\begin{figure*}[htbp]
    \centering
    \begin{subfigure}[b]{0.32\textwidth}
        \centering
        \includegraphics[width=\textwidth]{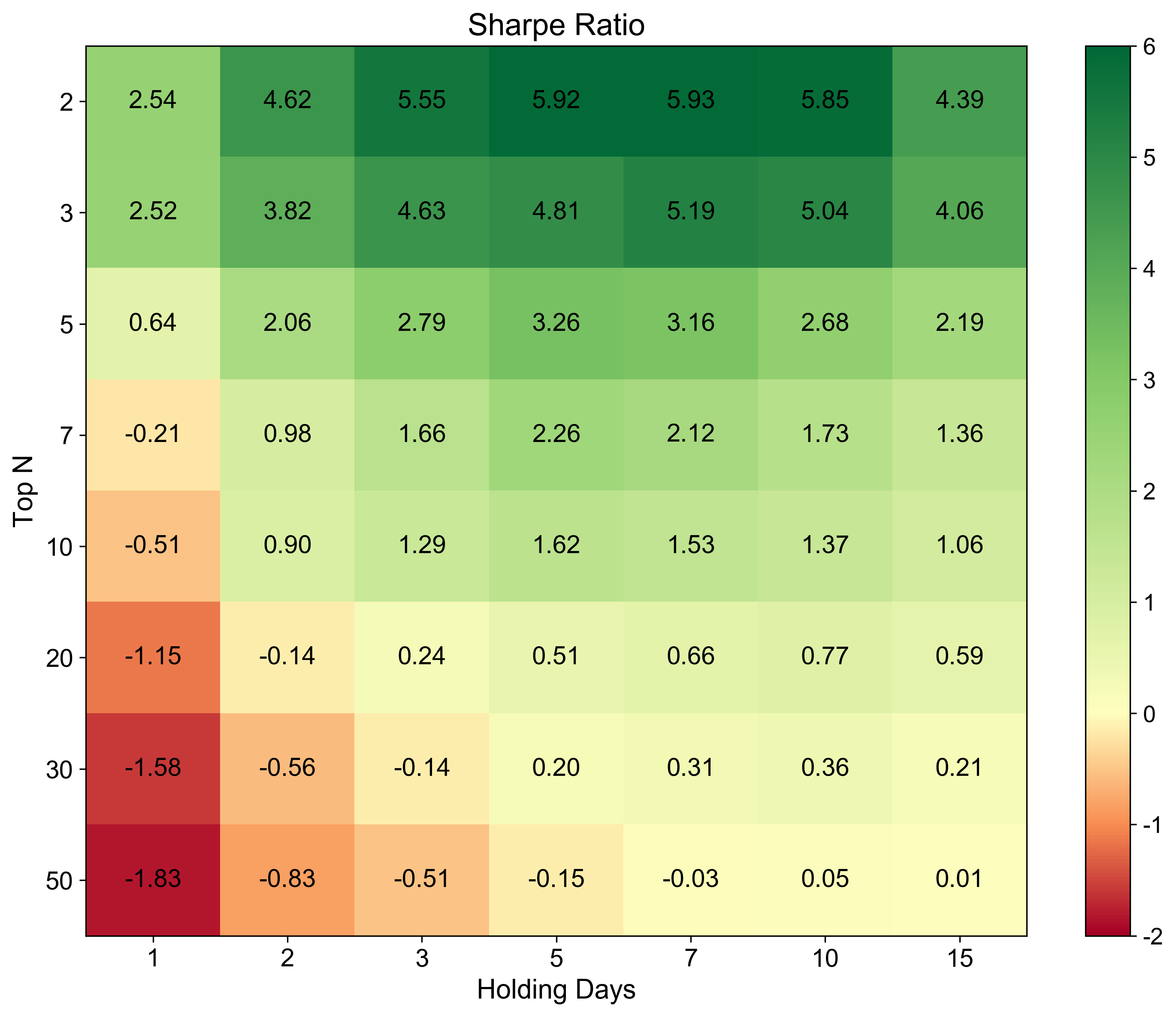}
        \caption{S\&P 500: Sharpe Ratio}
    \end{subfigure}
    \hfill
    \begin{subfigure}[b]{0.32\textwidth}
        \centering
        \includegraphics[width=\textwidth]{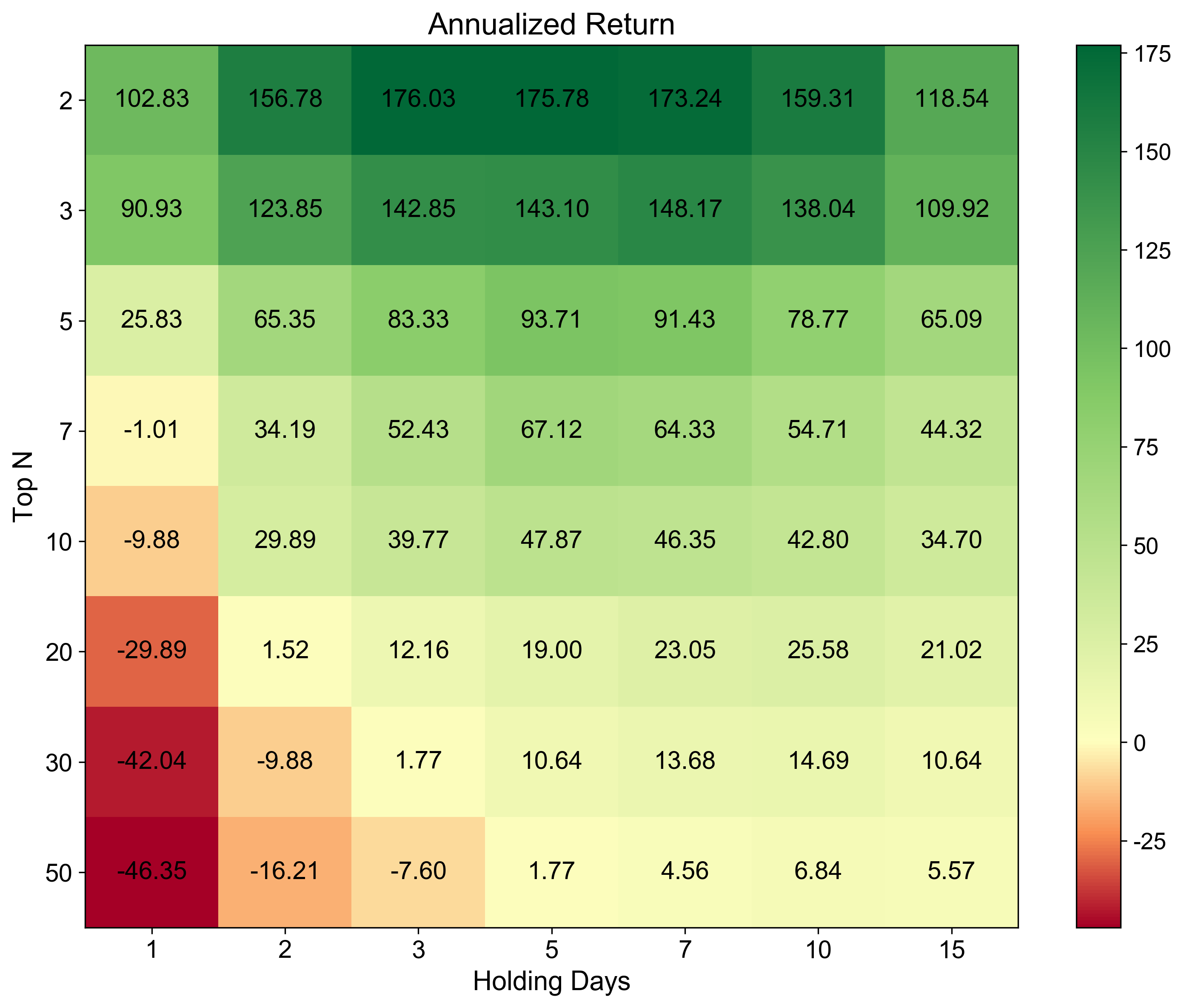}
        \caption{S\&P 500: Annualized Return}
    \end{subfigure}
    \hfill
    \begin{subfigure}[b]{0.32\textwidth}
        \centering
        \includegraphics[width=\textwidth]{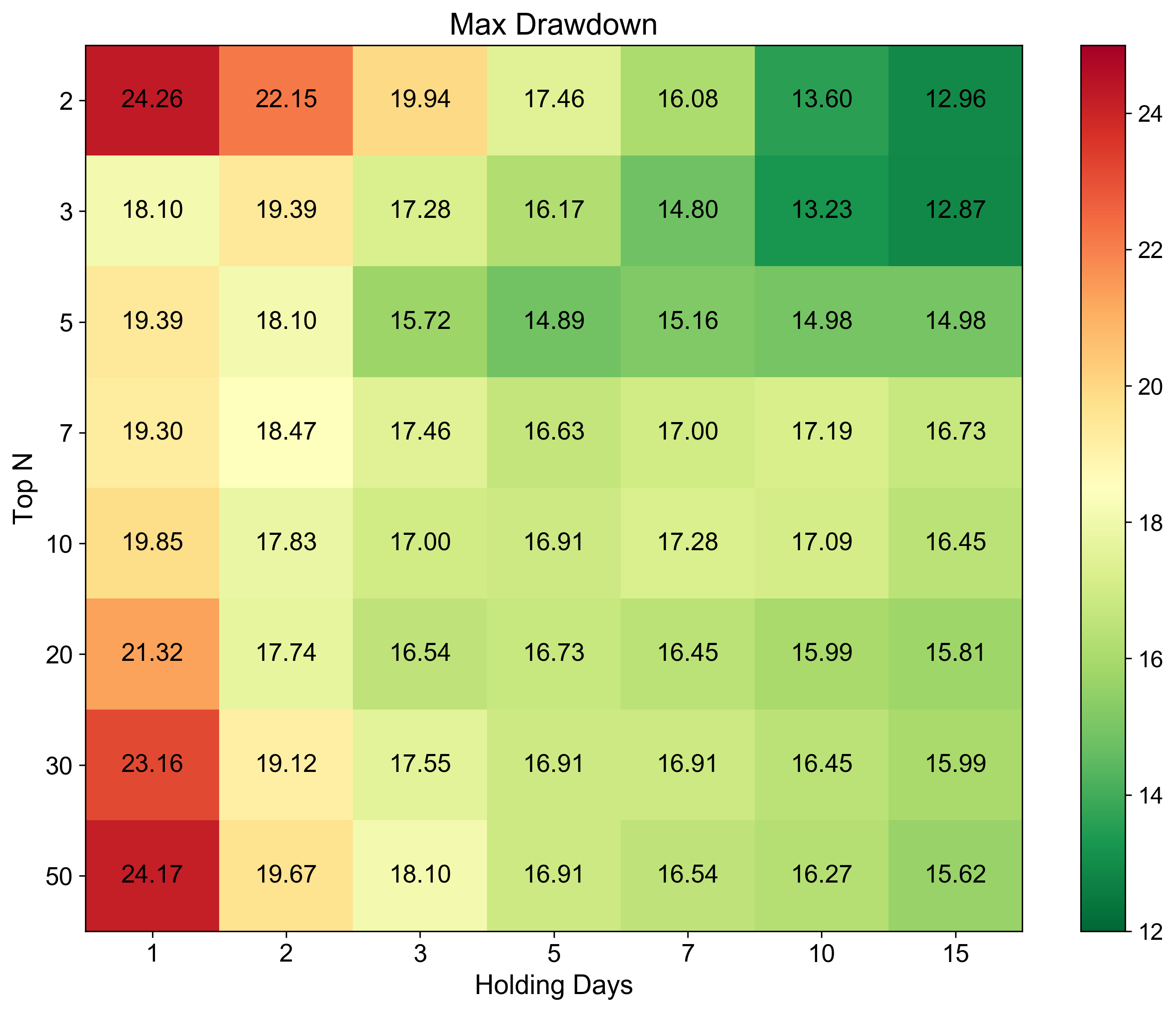}
        \caption{S\&P 500: Max Drawdown}
    \end{subfigure}

    \vspace{0.12cm}

    \begin{subfigure}[b]{0.32\textwidth}
        \centering
        \includegraphics[width=\textwidth]{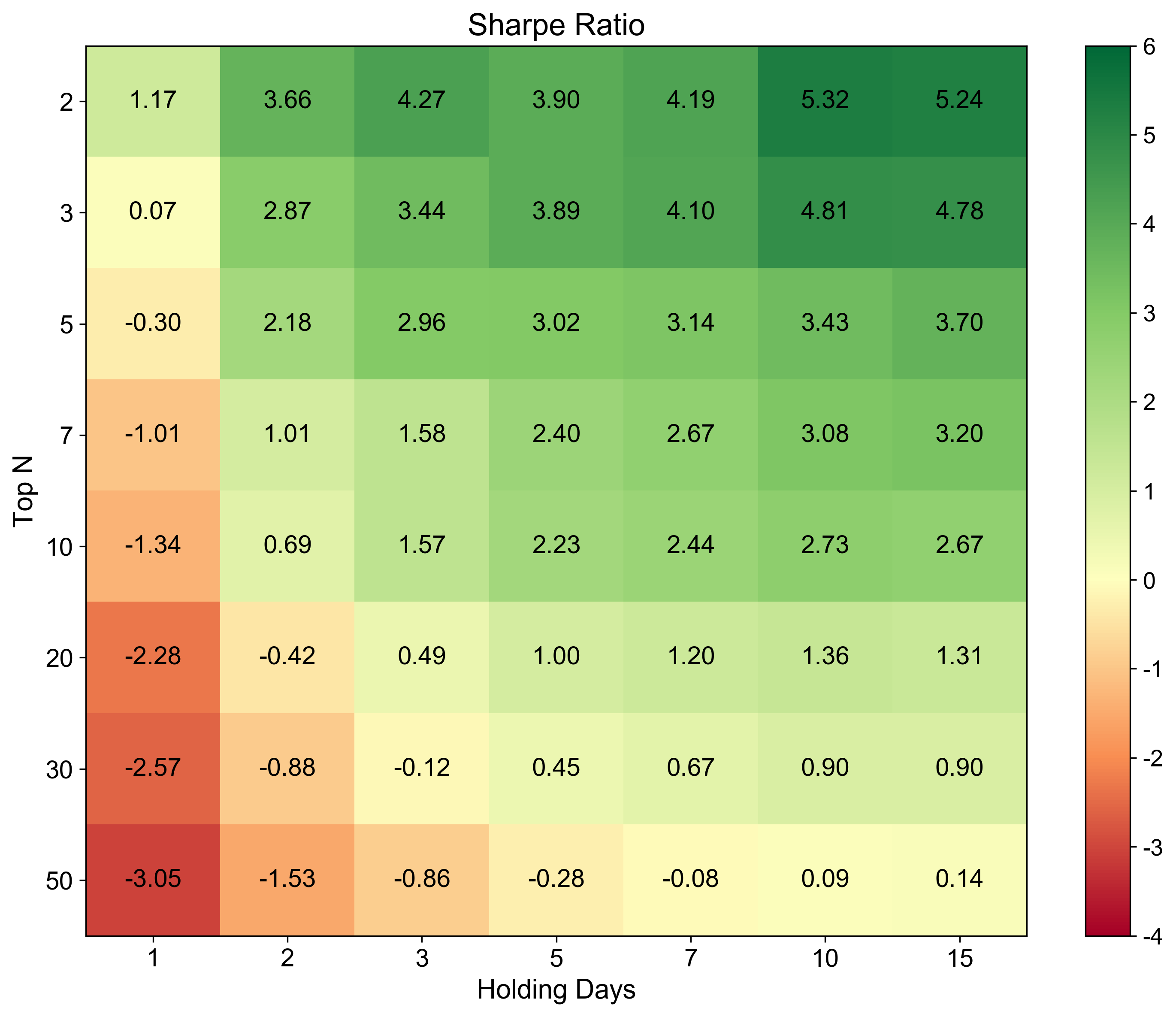}
        \caption{CSI 300: Sharpe Ratio}
    \end{subfigure}
    \hfill
    \begin{subfigure}[b]{0.32\textwidth}
        \centering
        \includegraphics[width=\textwidth]{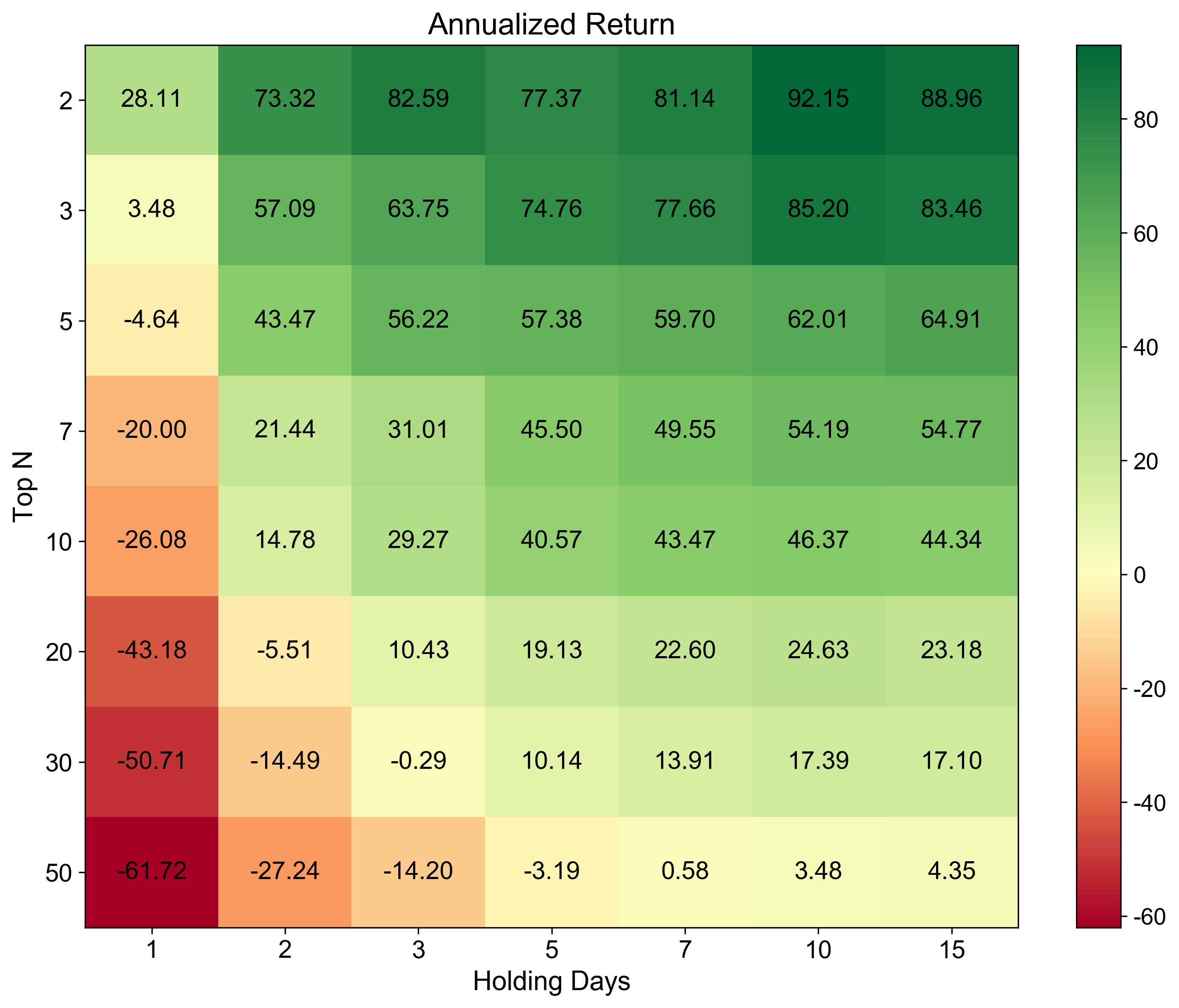}
        \caption{CSI 300: Annualized Return}
    \end{subfigure}
    \hfill
    \begin{subfigure}[b]{0.32\textwidth}
        \centering
        \includegraphics[width=\textwidth]{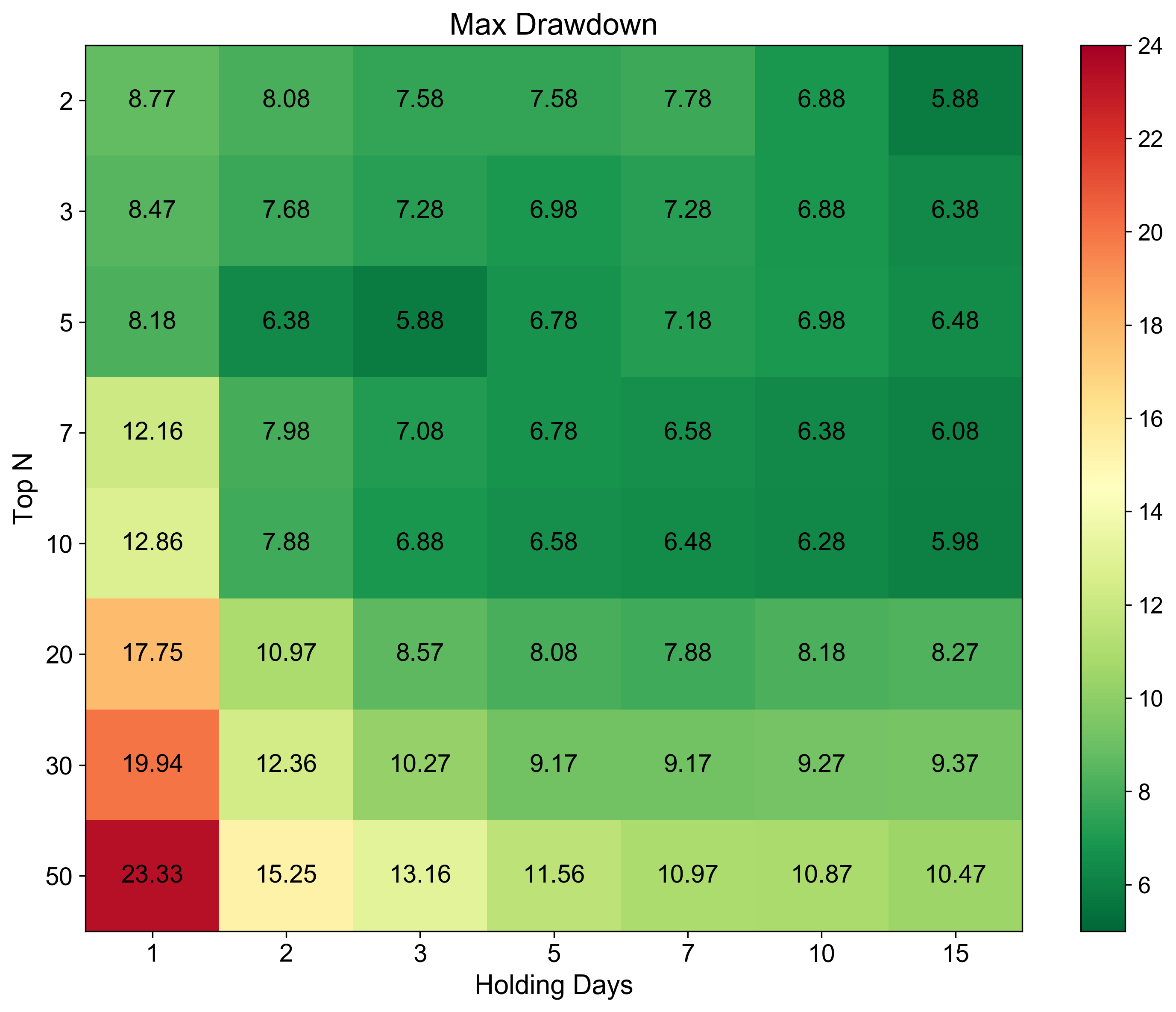}
        \caption{CSI 300: Max Drawdown}
    \end{subfigure}

    \caption{Parameter sensitivity analysis on the main asset pools. The heatmaps vary the selected-stock count $TopN$ and the holding horizon $H$ while keeping the remaining evaluation protocol fixed. Green regions indicate favorable performance (high Sharpe and return, low drawdown), while red regions indicate weaker configurations.}
    \Description{Six heatmaps showing how Sharpe ratio, annualized return, and maximum drawdown vary with holding horizon and selected-stock count for S\&P 500 and CSI 300.}
    \label{fig:heatmaps_main}
\end{figure*}

We examine robustness to the two execution hyperparameters that bridge factor selection and portfolio returns: $TopN$, the number of selected stocks in the equal-weight portfolio, and $H$, the holding horizon in trading days. Figure~\ref{fig:heatmaps_main} reports the response surface on the primary asset pools while keeping the remaining evaluation protocol fixed.

The response surface varies systematically rather than noisily. Annualized return increases monotonically as the portfolio concentrates: on S\&P 500 it rises from 47.87\% at the default ($TopN{=}10$, $H{=}5$) to a peak of 176.03\% around $TopN{=}2$--$3$, and on CSI 300 from 40.57\% to 92.15\% at $TopN{=}2$. The only systematically weak region is the shortest horizon: at $H{=}1$, where daily full rotation lets transaction costs dominate, returns turn negative for $TopN \ge 7$ on S\&P 500 and $TopN \ge 5$ on CSI 300. Drawdowns follow the same horizon logic, peaking at $H{=}1$ (up to 24.26\% on S\&P 500 and 23.33\% on CSI 300) and shrinking as $H$ lengthens. The default configuration therefore sits in a mid-range plateau---positive across its whole neighborhood yet far below the surface peak---which is the expected signature of a code default fixed before evaluation rather than of a test-set optimum (Section~\ref{sec:dataset_and_factors}). We read the surface as evidence that the gating signal delivers positive returns across a broad range of execution rules, with the concentration gradient indicating that the top-ranked names under the selected factor sets carry the strongest signal.

\subsection{Statistical Robustness and Risk Attribution}
\label{sec:robustness_attribution}

\paragraph{Statistical inference.} We complement the point estimates above with formal uncertainty quantification (protocol and full tables in Appendix~\ref{app:statistical_robustness}). Circular block-bootstrap inference (10{,}000 resamples, block lengths 5/10/20 bracketing the 5-day holding period) yields 95\% AR confidence intervals excluding zero on three of four universes at block length 10 and on all four at block length 20 (on S\&P 500 the left tail grazes zero at blocks 5/10---reported as-is---a mechanical consequence of the single-year window containing the April 2025 tariff shock). Paired-difference intervals show \modelname significantly ahead of a majority of baselines on every universe at block lengths 10/20 (20/20 on CSI 1000 at every tier), and rolling 3-month sub-windows give per-window win rates of 78.3--100\%. The probabilistic Sharpe ratio is 0.96--0.996; the Deflated Sharpe Ratio is 0.39--0.85 under an honestly declared upper-bound trial count ($N{=}101$--$158$; 0.84--0.89 under the narrower $N{=}34$ family)---below 0.95, an inherent property of a single-year window under conservative multiple-testing correction---and the CSCV-based probability of backtest overfitting is 0.016--0.133. The candidate pool retains 79--91\% of its 2021 rank-IC level in the test year, i.e., it does not decay into the evaluation window.

\paragraph{Risk attribution.} Daily-return regressions (Appendix~\ref{app:exposure_capacity}) give market betas of only 0.12--0.24 ($R^2 \le 0.07$) with annualized intercept alphas of 32.0--59.7\% (Newey--West $t = 2.07$--$3.02$, 5\%-significant on all four universes): the returns are overwhelmingly alpha, not a leveraged market replay. Stripping a small-minus-large size proxy leaves residual intercept alphas of 39.8--42.2\% on the small-cap universes and 20.2--22.6\% on the large-cap ones, all 5\%-significant. Sector composition shows no single-industry concentration (top-5 industries $\le$43.5\% of average weights), and turnover is 20\% per day by construction of the slot-rotation execution.

\paragraph{Reasoning vs.\ keyword matching.} Three instruments (Appendix~\ref{app:semantic_vs_lexical}) separate mechanism-level semantic alignment from lexical matching. (i) A BM25 strawman fed \emph{exactly} the state text and profiles the LLM sees trails \modelname by 24.7--56.7\,pp in every universe (significant on three of the four; the Russell 2000 paired interval includes zero, reported as-is), with performance not even monotone in overlap volume. (ii) Under profile obfuscation---two independent rewrites, regime keywords audited to zero hits, mechanisms preserved---at least 70\% of full-profile performance survives on all four universes (retention 71--90\%; versions agree within 4\,pp), exactly where a keyword-matching account predicts collapse. (iii) Daily selection sets turn over 28--41\% with adjacent-day Jaccard 0.594--0.715 (versus $\approx$1.0 for a static keyword map and 0.03--0.13 for random draws): non-static, non-random dynamics---though set-level selections do not cluster by volatility regime, which we report as-is.

%% file: sections/07_conclusion.tex
\section{Conclusion}
\label{sec:conclusion}

This paper presents \modelname, an RL-aligned semantic gating framework for dynamic alpha screening. By matching factor rationales to market-state narratives and optimizing selections with realized-return GRPO, \modelname achieves the strongest results among the evaluated traditional quantitative strategies and zero-shot LLM baselines under our 12-month protocol, with out-of-universe generalization results on Russell 2000 and CSI 1000 reported in Section~\ref{sec:cross_universe_transfer}. The main contribution is a use-inspired formulation and system design for context-conditioned, second-stage factor reranking. The current evidence is limited to a single 12-month test window and a bounded candidate pool. Our statistical package mitigates but does not eliminate this: the Deflated Sharpe Ratio under a conservatively declared upper-bound trial count remains below the conventional 0.95 bar (Appendix~\ref{app:dsr_pbo}), and market-impact analysis bounds practical capacity to roughly $10^8$ CNY on CSI 300, $10^7$--$10^8$ CNY on CSI 1000, $\sim$$3\times10^7$ USD on S\&P 500, and only the million-USD scale on Russell 2000 (Appendix~\ref{app:capacity}). Broader claims about robustness, capacity, market impact, and deployment therefore require multi-year validation.

The intended positive impact is to make factor screening more adaptive, auditable, and explicit about the economic rationale behind selected signals, which can support better human review of systematic strategies. At the same time, automated factor-screening systems that concentrate on similar signals may amplify correlated trading, encourage over-reliance on backtested patterns, or contribute to systemic fragility if deployed without capacity controls. We therefore treat the release as a research artifact rather than a live trading service: restricted raw market/news feeds and proprietary LLM outputs are not redistributed, and responsible deployment should incorporate capacity monitoring, diversification constraints, independent validation, and human oversight. The computational cost of LLM training and inference also warrants consideration in production settings.

%% file: sections/09_01_appendix_data.tex
\section{Detailed Data Flow}
\label{sec:appendix_data_flow}

This section details the two data streams used in \framename: Price Market Data and News Data, with examples from both Chinese A-share and US markets. Section~\ref{sec:data_pipelines} outlines the construction process, Section~\ref{sec:data_samples} shows representative outputs, and Section~\ref{sec:appendix_prompt_sample} presents the prompt template. As discussed in Section~\ref{sec:methodology}, the critical reproducibility object is the task decomposition and interface design rather than the exact wording of a particular vendor model's output.

\subsection{Construction Methods}
\label{sec:data_pipelines}

\paragraph{Decision-Time Information Boundary.}
For each trading date, the state-construction pipeline uses only records available before the portfolio action for that date. The precise cutoffs are:

\begin{itemize}
    \item \textbf{Price data ($S^{\text{price}}_t$):} limited to the \emph{previous trading day's close} (day $t{-}1$). No same-day (day $t$) intraday data enters the state description, ensuring strict separation from the 09:31--10:00 VWAP execution window.
    \item \textbf{News data ($S^{\text{news}}_t$):} filtered by publication timestamp up to \emph{30 minutes before market open} on day $t$ (i.e., 09:00 for both Chinese A-share and U.S. markets). Items published after this cutoff are excluded.
\end{itemize}

\noindent Figure~\ref{fig:decision_timeline} illustrates the complete decision-day timeline.

\begin{figure}[htbp]
    \centering
    \small
    \setlength{\tabcolsep}{2pt}
    \begin{tabular}{@{}r@{\;\;}l@{}}
        \toprule
        \textbf{Time} & \textbf{Event} \\
        \midrule
        Day $t{-}1$ close & Price data cutoff: $S^{\text{price}}_t$ finalized \\
        \multicolumn{2}{@{}l@{}}{\quad\dotfill} \\
        Day $t$, 09:00 & News cutoff: $S^{\text{news}}_t$ finalized \\
        Day $t$, 09:00--09:30 & State $S_t$ constructed; prompt assembled; \\
         & \modelname outputs $\mathcal{A}_t$ \\
        Day $t$, 09:30 & Market open \\
        Day $t$, 09:31--10:00 & VWAP execution of slot $k = t \bmod H$ \\
        Day $t$--$(t{+}H{-}1)$ & Holding-period return measured \\
        \bottomrule
    \end{tabular}
    \caption{Decision-day timeline. The state $S_t$ is constructed entirely from day-$(t{-}1)$ closing prices and pre-09:00 news on day $t$. No data from the 09:31--10:00 execution window or the subsequent holding period enters the decision inputs.}
    \Description{Table showing the chronological sequence of events on each trading day, from previous-day close through VWAP execution and holding-period measurement.}
    \label{fig:decision_timeline}
\end{figure}

The execution return is then measured using the downstream VWAP protocol, but realized execution prices, post-decision intraday movements, and future holding-period returns are not provided to the state-construction or reasoning modules. This timestamp policy is applied together with the point-in-time universe construction described in Section~\ref{sec:dataset_and_factors}.

\subsubsection{Price Market Data Flow}
The Price Market Data module converts high-frequency trading data into semantic descriptions via a multimodal LLM. This appendix-level view corresponds to the market-state construction pipeline introduced in Section~\ref{sec:phase2_construction}, especially the asset-pool state description step in Section~\ref{sec:phase2_construction}, where raw market observations are mapped into a unified semantic interface for downstream reasoning.

\begin{itemize}
    \item Visual Encoding: We collect OHLC and volume data from Tushare (China) and Massive (US) up to the previous trading day's close, then render a 90-day K-line chart and a day-$(t{-}1)$ intraday chart.
    \item Multimodal Prompting: These charts, together with key statistics (e.g., amplitude, turnover, and price change), are fed to Claude 3.7 Sonnet.
    \item Semantic Extraction: The model returns a structured summary of support/resistance, trend signals, and divergence patterns.
\end{itemize}

The price stream uses only information available before the portfolio action; no future returns, realized execution outcomes, or post-decision data enter the state representation.

The price descriptions also standardize heterogeneous market data into a common language: U.S. and Chinese markets differ in trading calendars, liquidity patterns, and microstructure, but the semantic interface asks for comparable concepts (trend strength, volatility expansion, support/resistance, volume-price divergence). This makes the factor-selection prompt less dependent on exchange-specific raw formats while retaining market features relevant for short-horizon factor gating.

\subsubsection{News Data Flow}
The News Data module converts unstructured news into market-relevant narratives. Together with the price stream above, this module instantiates the multimodal market-state construction described in Section~\ref{sec:phase2_construction} and provides one of the context sources consumed by the reasoning model in Section~\ref{sec:alpha_r1}.

\begin{itemize}
    \item \textbf{Data Aggregation:} We collect financial news in the target window from Sina Finance (China) and Massive (US), retaining only items timestamped before the decision cutoff for the corresponding market.
    \item \textbf{LLM Summarization:} The model removes irrelevant or duplicate items and summarizes the remaining news into three dimensions: \textit{Macroeconomic Policy}, \textit{Industry Dynamics}, and \textit{Significant Corporate Events}.
\end{itemize}

The news stream plays a complementary role to the price stream. Price summaries describe what the market has already done, while news summaries help characterize the possible narrative drivers behind that movement. Duplicate articles, boilerplate market commentary, and items without an obvious connection to the tradable universe are filtered before the final summary is produced. The three summary dimensions are chosen to preserve different levels of regime information: macroeconomic policy captures broad risk appetite and discount-rate pressure, industry dynamics capture sector-level leadership or stress, and significant corporate events capture idiosyncratic catalysts that may affect cross-sectional dispersion.

\paragraph{Third-Party Data and Model Access.}
The market data, news feeds, and proprietary LLM services used in the pipeline remain subject to their respective provider licenses and terms of use. The released research resources therefore document the interface, prompts, temporal filtering rules, and evaluation protocol, while restricted raw feeds or proprietary model outputs must be obtained through the corresponding providers when exact reruns require them.

\begin{figure*}[htbp]
    \centering
    \fbox{
    \parbox{\dimexpr0.96\linewidth-2\fboxsep-2\fboxrule\relax}{
        \small
        \textbf{[Sample 1] Price Market Description ($S^{\text{price}}_{20250409}$)}
        \vspace{0.5em}
        \hrule
        \vspace{0.5em}

        \textbf{90-Day Market Trend Review:}
        Over the past 90 days, the major indices (Shanghai Composite, ChiNext, STAR 50) exhibited a synchronized cycle of "Rise (Late Nov--Mid Dec) $\rightarrow$ Adjustment $\rightarrow$ Rebound (Late Jan--Feb) $\rightarrow$ Decline (Mar--Apr)". Significant divergence in volatility was observed: the STAR 50 had the highest amplitude ($\sim$21\%), followed by ChiNext ($\sim$17\%) and Shanghai Composite ($\sim$9\%). Currently, all indices have closed below their initial levels from 90 days ago, indicating a downward shift in the market center of gravity. Volume analysis shows that growth indices (STAR 50, ChiNext) saw sustained volume amplification during the recent decline, indicating active capital turnover.

        \textbf{Intraday Characteristics (2025-04-08):}
        The indices displayed a highly synchronized "Morning Spike $\rightarrow$ Mid-day Consolidation $\rightarrow$ Late-session Rally" pattern:
        (1) Morning (9:30--9:45): Rapid surge to daily highs followed by a swift retracement; volume peaked at the open.
        (2) Mid-day (9:45--14:30): Range-bound oscillation with shrinking volume; ChiNext adjusted the most.
        (3) Late-session (14:30--15:00): Collective rally. The Shanghai Composite and STAR 50 showed significant gains with expanding volume, closing near daily highs (+1.58\% and +1.72\% respectively).
    }
    }
    \caption{Semantic description generated from price market data processed by Claude 3.7 Sonnet.}
    \Description{Text box containing a sample generated price market description with a 90-day trend review and intraday characteristics.}
    \label{fig:sample_price}
\end{figure*}

\begin{figure*}[htbp]
    \centering
    \fbox{
    \parbox{\dimexpr0.96\linewidth-2\fboxsep-2\fboxrule\relax}{
        \small
        \textbf{[Sample 2] News Data Summary ($S^{\text{news}}_{20250409}$)}
        \vspace{0.5em}
        \hrule
        \vspace{0.5em}

        \textbf{Corporate Violations \& Penalties:}
        (1) Aulion Electronics: Received CSRC penalty for misleading statements regarding the performance of key personnel in its Perovskite project. The company was fined 3 million RMB, and key executives were individually fined.
        (2) Hengxing Tech: Chairman and GM were fined for short-swing trading involving family members.

        \textbf{Restructuring \& Distress:}
        (1) Shimao Group: Received a liquidation petition filed by CCB (Asia) at the Hong Kong High Court involving $\sim$1.58 billion HKD. Stock fell 18.68\%.
        (2) *ST Bugao: Announced industrial investors (Baitu Group, etc.) for restructuring, triggering a stock rebound despite a daily drop.

        \textbf{Industry Dynamics:}
        (1) Gaming: NPPA approved 14 imported games (including titles from Tencent and Perfect World). The market expects stable approval cycles to drive sector growth.
        (2) Liquor: Rumors of Feitian Moutai wholesale prices dropping below 2600 RMB caused volatility; the sector remains weak despite stable actual terminal prices.
    }
    }
    \caption{Summarized financial news narratives generated from Chinese market news data processed by Claude 3.7 Sonnet.}
    \Description{Text box containing a sample Chinese market news summary organized by corporate violations, restructuring, and industry dynamics.}
    \label{fig:sample_news}
\end{figure*}

\subsection{Detailed Data Samples}
\label{sec:data_samples}

We present representative semantic data samples from the Chinese A-share market and the U.S. market under the same 2025-04-09 decision-date convention. These text boxes are concrete instances of the state abstraction described in Section~\ref{sec:phase2_construction} and are the actual inputs processed by the \modelname agent. They should therefore be read as audit examples of the same interface used in the experimental pipeline of Section~\ref{sec:experiments}, rather than as standalone qualitative illustrations.

\begin{figure*}[htbp]
    \centering
    \fbox{
    \parbox{\dimexpr0.96\linewidth-2\fboxsep-2\fboxrule\relax}{
        \small
        \textbf{[Sample 3] US Price Market Description ($S^{\text{price, US}}_{20250409}$)}
        \vspace{0.5em}
        \hrule
        \vspace{0.5em}

        \textbf{Market Overview:}
        The US market experienced a broad, high-volume decline. The growth-oriented Nasdaq-100 ETF (QQQ) underperformed the broader S\&P 500 ETF (SPY), exhibiting panic selling characteristics with a unilateral downward trend. Short sellers dominated the session as trading volume significantly exceeded typical levels.

        \textbf{Intraday Characteristics (2025-04-08):}
        (1) Morning (9:00–12:00): Indices opened higher but quickly retraced. QQQ hit an intraday high of \$443.14 and SPY \$524.98 before entering a volatile decline.
        (2) Mid-day (12:00–14:00): A rapid sell-off occurred post-noon. QQQ dropped from \$435.75 to \$426.38 between 12:00-13:00.
        (3) Late-session (14:00–16:00): Panic selling intensified. The final hour saw peak volume, with QQQ closing at \$416.24 (-5.00\%) and SPY at \$496.74 (-4.81\%), both near daily lows.
    }
    }
    \caption{Semantic description generated from US price market data processed by Claude 3.7 Sonnet. The quoted percentage figures denote intraday open-to-close moves of the decision-eve session (2025-04-08), not close-to-close daily returns.}
    \Description{Text box containing a sample U.S. price market description with a market overview and intraday characteristics.}
    \label{fig:sample_price_us}
\end{figure*}

\begin{figure*}[htbp]
    \centering
    \fbox{
    \parbox{\dimexpr0.96\linewidth-2\fboxsep-2\fboxrule\relax}{
        \small
        \textbf{[Sample 4] US News Data Summary ($S^{\text{news, US}}_{20250409}$)}
        \vspace{0.5em}
        \hrule
        \vspace{0.5em}

        \textbf{Corporate Regulation:}
        QNRX and TPST were halted pending news releases on the evening of April 8 (ET).

        \textbf{Industry Dynamics:}
        (1) Consumer Staples: Bank of America noted the sector's historical defensive nature during recessions but warned that high prices and weak volume growth could undermine resilience.
        (2) Analyst Ratings: Scotiabank adjusted targets for WM (raised to \$260) and others. Wells Fargo lowered targets for several financials/energy stocks including SHEL (to \$83) and SCHW (to \$87).

        \textbf{Macroeconomic Policy:}
        Tariff concerns sparked a sell-off in US equities and bonds. The 10-year Treasury yield rebound drove 30-year fixed mortgage rates up to 6.82\%, the largest single-day jump of the year.
    }
    }
    \caption{Summarized financial news narratives generated from US market news data processed by Claude 3.7 Sonnet.}
    \Description{Text box containing a sample U.S. market news summary organized by corporate regulation, industry dynamics, and macroeconomic policy.}
    \label{fig:sample_news_us}
\end{figure*}

\subsection{Detailed Prompt Sample}
\label{sec:appendix_prompt_sample}

This section presents the prompt template used for alpha factor selection. The template serves as an auditable interface contract: market-information slots, factor-description slots, and the XML output schema define the input-output specification, making the pipeline portable across model substitutions. In relation to Section~\ref{sec:alpha_r1}, Figure~\ref{fig:sample_prompt} should be read as the concrete interface consumed by the reasoning model after the data-construction steps illustrated in Figures~\ref{fig:sample_price}, \ref{fig:sample_news}, \ref{fig:sample_price_us}, and \ref{fig:sample_news_us}.

\begin{figure*}[htbp]
    \centering
    \fbox{
    \parbox{\dimexpr0.96\linewidth-2\fboxsep-2\fboxrule\relax}{
        \small
        \textbf{[Sample 5] Alpha Factor Selection Prompt Structure (Training Data Format)}
        \vspace{0.5em}
        \hrule
        \vspace{0.5em}

        \textbf{System Role Definition:}\\
        You are a senior quantitative investment expert, skilled in selecting the most suitable alpha factor combinations based on market environment. You need to analyze current market conditions and the characteristics of each factor to provide scientific and reasonable factor selection recommendations.

        \textbf{User Task Instruction:}\\
        Based on the following information, select the most suitable factor combinations for 2024-07-01's trading day for 5-day short-term strategy stock selection (rebalance one slot using 09:31--10:00 VWAP execution; hold that slot for $H{=}5$ trading days; aggregate across $H$ slots).

        \textbf{Target Date:} 2024-07-01

        \textbf{Market Environment Information:}\\
        • Previous Trading Day Closing Data (2024-06-28): [Market price data content...]\\
        • Previous Trading Day Market Analysis (2024-06-28): [Technical analysis content...]\\
        • Current Day Pre-Market News (2024-07-01): [Financial news content...]

        \textbf{Available Factor Descriptions:}\\
        • alpha001: [Detailed factor description including calculation method, historical performance...]\\
        • alpha002: [Detailed factor description...]\\
        • [Additional factors with complete descriptions...]

        \textbf{Analysis Framework:}\\
        1. Analyze each factor's nature and characteristics\\
        2. Evaluate factors' expected performance in current market\\
        3. Make final selection (maximum 10 factors)

        \textbf{Output Requirements:}\\
        • Provide detailed analytical reasoning first\\
        • Output XML-tagged factor list: {\textless}alpha\_list{\textgreater}{\textless}alpha id="alpha001"/{\textgreater}...{\textless}/alpha\_list{\textgreater}\\
        • Maximum 10 factors allowed\\
        • Skip selection if no factors expected to yield positive returns

        \textbf{Expected Response Format:}\\
        Factor analysis reasoning: [Detailed explanation of selection logic...]\\
        The most suitable factor selection for the current market is: {\textless}alpha\_list{\textgreater}{\textless}alpha id="alpha001"/{\textgreater}{\textless}alpha id="alpha003"/{\textgreater}{\textless}alpha id="alpha007"/{\textgreater}{\textless}/alpha\_list{\textgreater}
    }
    }
    \caption{Complete prompt structure for alpha factor selection training. This demonstrates the standardized format used to generate training data for \modelname, incorporating market data and factor descriptions.}
    \Description{Text box showing the alpha factor selection prompt template with system role, task instruction, market information, factor descriptions, analysis framework, and output requirements.}
    \label{fig:sample_prompt}
\end{figure*}

%% file: sections/09_02_appendix_judge.tex
\section{LLM-as-a-Judge Rubric}
\label{app:judge_rubric}

This appendix summarizes the rubric used by the judge model when assigning the consistency penalty in Equation~\ref{eq:consistency_eval}. The goal is not to reward eloquent writing, but to penalize reasoning traces that are disconnected from the decision context or that fail to justify the final factor set.

\paragraph{Evaluation Objective.}
For each response, the judge model receives the market state, the semantic factor descriptions, the generated reasoning, and the final selected factor list. It then assigns a penalty score in $[0, 10]$, where lower is better.

The judge receives the same decision context available to the reasoning model---no future returns or post-decision information---and outputs a scalar consistency penalty rather than a trading signal, alpha verifier, or proof that the generated rationale is the model's faithful internal computation.

\paragraph{Rubric Dimensions.}
\begin{itemize}
    \item \textbf{Logical contradic\-tions or reasoning inconsis\-tency:} whether the generated explanation conflicts with the provided market state, the stated factor mechanisms, or the final selected action.
    \item \textbf{Linguistic confusion:} whether the justification is so unclear or disorganized that the intended decision logic cannot be reliably interpreted.
    \item \textbf{Hallucinated or irrelevant content:} whether the response introduces unsupported claims, unrelated narratives, or statements that are not grounded in the prompt context.
    \item \textbf{Repetition or redundancy:} whether the model repeats the same point without adding decision-relevant information.
    \item \textbf{Other irrationalities:} whether the response contains obvious factual inconsistencies, malformed reasoning steps, or nonsensical statements.
\end{itemize}

These dimensions are related but not interchangeable. A logical contradiction occurs when the response uses the provided evidence in an inconsistent way, such as describing a risk-off session while justifying factors solely by calm-market assumptions. Hallucinated content refers to claims that are not grounded in the prompt, such as inventing a news catalyst or a factor property that was never provided. Irrelevance is milder than hallucination: the statement may be true in general, but it does not help justify the selected factor set under the current state. Redundancy is penalized only when repetition reduces decision support, not when a response restates a key point briefly for clarity. Linguistic confusion is reserved for cases where the reasoning cannot be reliably parsed into a market view, factor interpretation, and action justification.

\paragraph{Scoring Interpretation.}
A score near 0 indicates that the reasoning is internally consistent, context-grounded, and aligned with the final action. Mid-range penalties correspond to noticeable but non-fatal issues such as local contradictions or mild redundancy. High penalties are assigned when the reasoning is largely disconnected from the market evidence or when the action cannot be justified from the generated explanation. These scores measure the quality of the rationale relative to the prompt context; they do not certify that the selected factors will be profitable or that the rationale is economically complete.

\paragraph{Manual Calibration.}
To verify directional reliability, three finance master's reviewers independently scored a calibration subset under the same $0$--$10$ rubric, blind to model identity and realized returns. The judge showed high directional agreement with the aggregated human reference, confirming its suitability as a consistency regularizer.

\paragraph{Role in Training.}
The judge penalty is an auxiliary shaping term that rescales the market-based reward; realized portfolio performance remains the dominant training signal.

\begin{figure*}[htbp]
    \centering
    \begin{subfigure}[b]{0.49\textwidth}
        \centering
        \includegraphics[width=\textwidth]{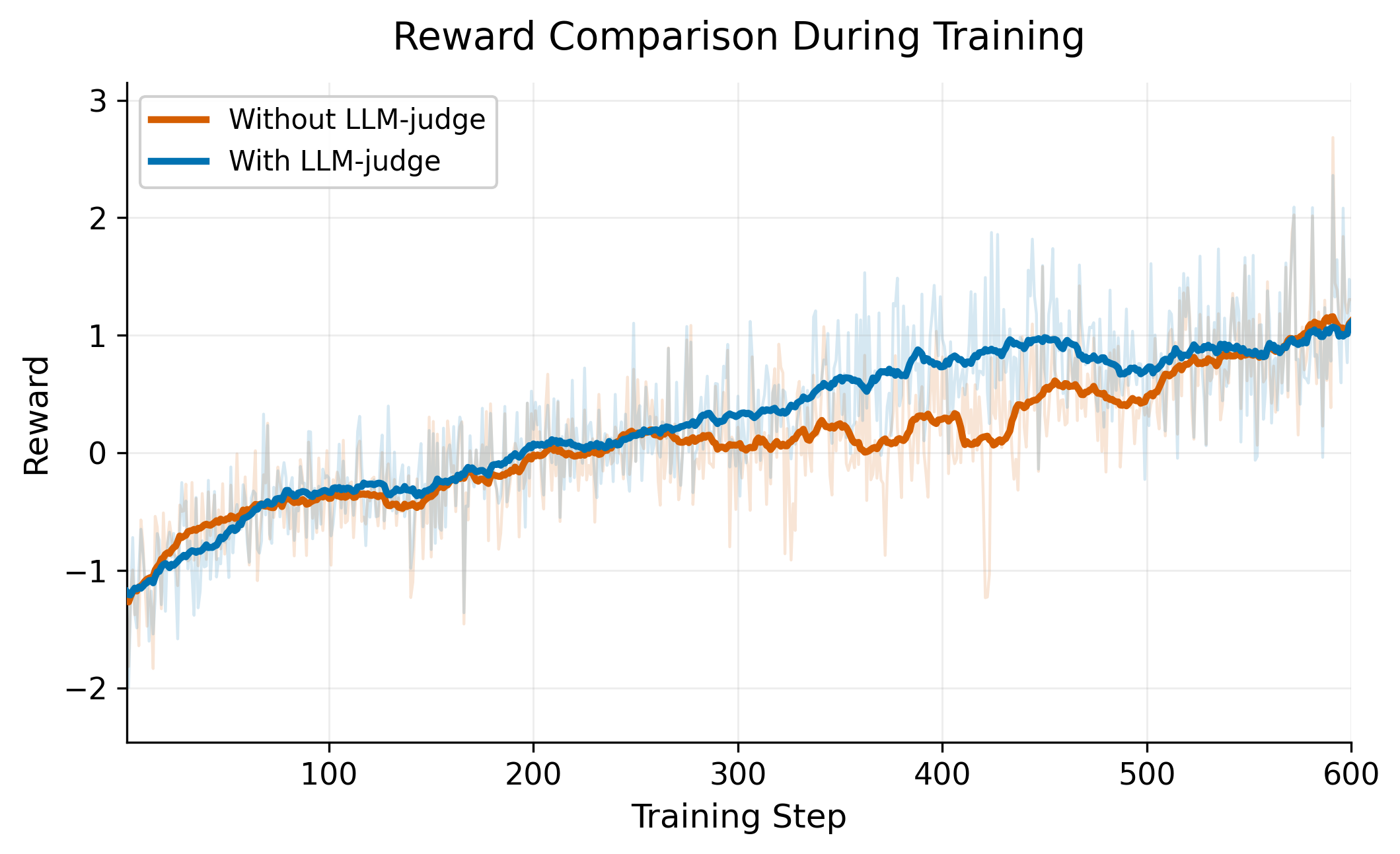}
        \caption{Training reward}
        \label{fig:judge_reward_comparison}
    \end{subfigure}
    \hfill
    \begin{subfigure}[b]{0.49\textwidth}
        \centering
        \includegraphics[width=\textwidth]{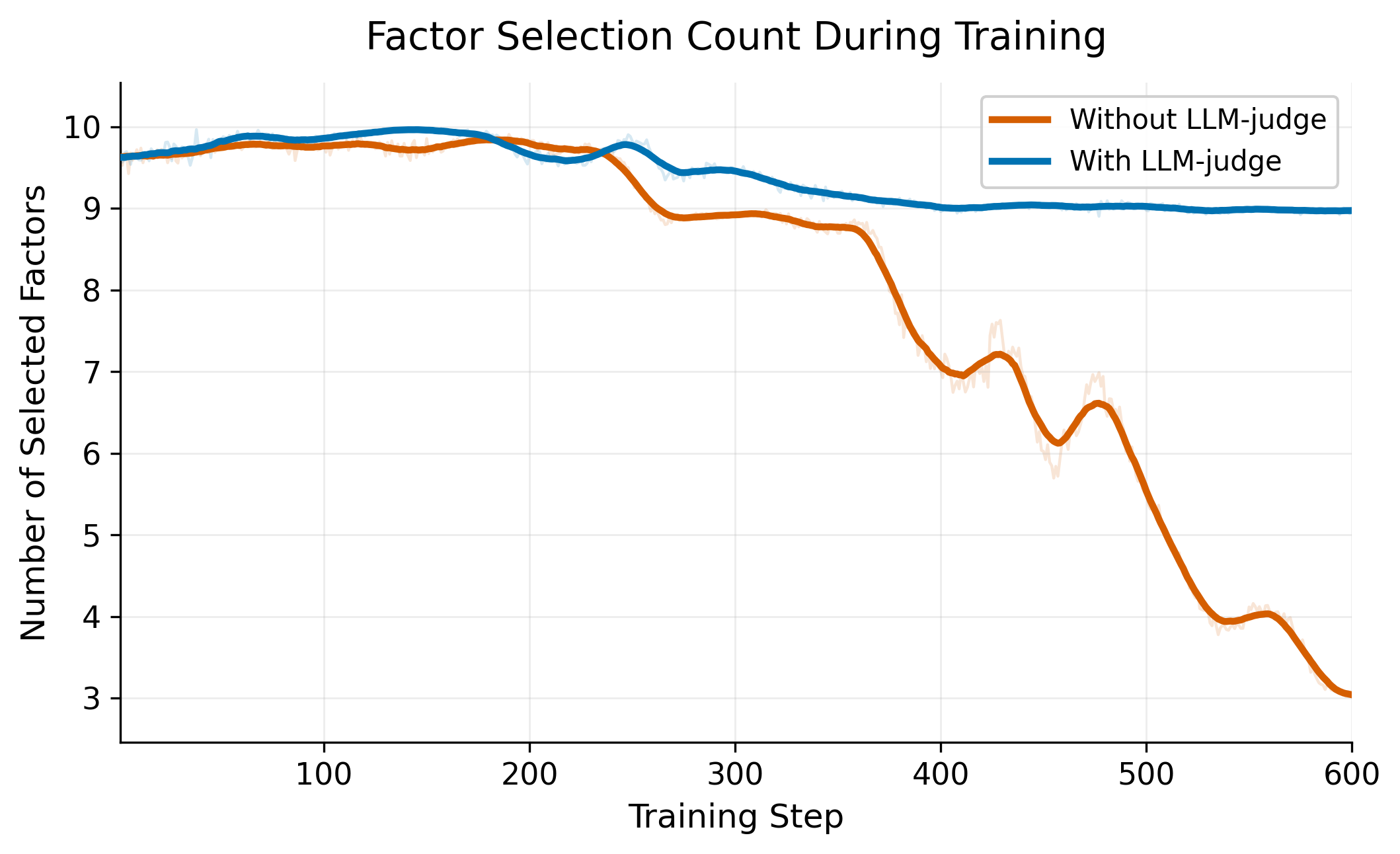}
        \caption{Selected-factor count}
        \label{fig:judge_factor_count_comparison}
    \end{subfigure}
    \caption{Training dynamics with and without the LLM-as-judge consistency regularizer. Panel (a) reports scalar training reward, and panel (b) tracks the selected-factor count, showing that the no-judge run collapses to a much smaller factor set. The two reward curves are not directly comparable: the no-judge run omits the consistency adjustment, so its higher raw reward does not indicate better downstream performance.}
    \Description{Two training curves comparing scalar reward and selected-factor count with and without the LLM-as-judge regularizer.}
    \label{fig:judge_training_dynamics}
\end{figure*}

\paragraph{Training Dynamics without the Judge Term.}
Figure~\ref{fig:judge_training_dynamics} compares training dynamics with and without the judge. Removing the judge causes the selected-factor count to collapse from roughly ten to about two, indicating degenerate optimization. Because this diagnostic run was not evaluated downstream, we make no claim that the judge term improves test-period returns; its role is to prevent the policy from collapsing onto minimal, rationale-free selections during training.

%% file: sections/09_03_appendix_metrics.tex
\section{Performance Evaluation Metrics}
\label{sec:appendix_metrics}

This appendix defines the standard metric suite used to evaluate all strategies. Let $V_t$ be the portfolio value at time $t$ and $R_t = V_t / V_{t-1} - 1$ the daily return.

\subsection{Metric Computation Protocol}

All metrics are computed from daily NAV curves produced by the out-of-sample 2025 evaluation protocol of Section~\ref{sec:experiments}. The annualization factor is $K=252$. Excess-return metrics use annual risk-free rates of 4.5\% (U.S.) and 1.5\% (China). The IR benchmark is the Buy \& Hold NAV curve for each asset pool. These definitions underpin the main-results, cross-universe, ablation, and gating-baseline tables reported in Section~\ref{sec:experiments}, so the appendix metric suite should be read as the common evaluation layer shared across all major empirical comparisons.

\subsection{Metric Definitions}

\paragraph{Cumulative Return (CR)}
\[
\mathrm{CR} = \frac{V_T}{V_0} - 1.
\]

\paragraph{Annualized Return (AR)}
\[
\mathrm{AR} = \left(\frac{V_T}{V_0}\right)^{\frac{K}{T}} - 1,
\]
where $T$ is the total number of trading periods in the evaluation.

\paragraph{Sharpe Ratio (SR)}
Let $R_e = R_p - R_f$ be the daily excess return with mean $\mu_e$ and standard deviation $\sigma_e$.
\[
\mathrm{SR}_{\text{period}} = \frac{\mu_e}{\sigma_e}, \qquad \mathrm{SR}_{\text{ann}} = \mathrm{SR}_{\text{period}} \cdot \sqrt{K}.
\]

\paragraph{Maximum Drawdown (MDD)}
\[
\mathrm{MDD} = \max_{t \in [1, T]} \left( \frac{\max_{\tau \in [1, t]} V_\tau - V_t}{\max_{\tau \in [1, t]} V_\tau} \right).
\]
Reported as a positive magnitude; lower is better.

\paragraph{Annualized Volatility (Vol)}
\[
\mathrm{Vol}_{\text{ann}} = \sigma(R_t) \sqrt{K}.
\]

\paragraph{Downside Deviation and Sortino Ratio}
Downside deviation uses the market-specific risk-free rate as the minimum acceptable return. Let $R_f$ be the daily risk-free rate (annual 4.5\% for U.S. markets, 1.5\% for Chinese markets, divided by $K$):
\[
\sigma_{\text{down}} = \sqrt{K} \cdot \sigma(\min(R_t - R_f, 0)), \qquad \mathrm{Sortino} = \frac{K \cdot \mathbb{E}[R_t - R_f]}{\sigma_{\text{down}}}.
\]

\paragraph{Calmar Ratio}
\[
\mathrm{Calmar} = \frac{\mathrm{AR}}{\mathrm{MDD}}.
\]

\paragraph{Information Ratio (IR)}
Let $R_{a,t}=R_t-R_{b,t}$ denote active return over the Buy \& Hold benchmark:
\[
\mathrm{IR} = \frac{\mathbb{E}[R_{a,t}]}{\sigma(R_{a,t})}\sqrt{K}.
\]

\paragraph{Daily Win Rate}
Fraction of trading days with positive return:
\[
\mathrm{WinRate} = \frac{1}{T}\sum_{t=1}^{T}\mathbf{1}\{R_t>0\}.
\]

\paragraph{Return Distribution Statistics}
We report best daily return, worst daily return, skewness, and kurtosis in the machine-readable metric output to summarize distributional shape and tail behavior. These distributional statistics complement the return diagnostics in Appendix~\ref{app:return_distribution_analysis}.

\paragraph{Monthly and Quarterly Return Aggregation}
For a calendar period $\mathcal{P}$, the compounded period return is:
\[
R_{\mathcal{P}} = \prod_{t \in \mathcal{P}}(1+R_t) - 1.
\]
Monthly and quarterly decomposition figures use this compounded definition rather than summing daily returns.

\paragraph{Complete Metric Tables}
Table~\ref{tab:extended_metrics_table} reports the complete metric set across Panel~A (main and cross-universe), Panel~B (ablation), and Panel~C (gating baselines). Bold marks the best value within each panel and asset pool. Strong AR accompanied by high SR, Sortino, Calmar, and positive IR indicates that the return is not merely compensation for extreme volatility or benchmark exposure. Together with the return-timing and drawdown figures in Appendix~\ref{app:return_distribution_analysis}, these tables connect the scalar summaries in Section~\ref{sec:experiments} to their underlying NAV behavior and risk profile.

\begin{table*}[htbp]
    \centering
    \caption{Grouped NAV-derived risk and return metrics. Metrics are computed from daily NAV curves over the 2025 test period. SR is the annualized excess Sharpe ratio using the market-specific risk-free rate, and MDD is reported as a positive drawdown magnitude. Bold entries mark the best value within the same panel and asset pool; Buy \& Hold is a passive benchmark, so its strategy Win Rate is not reported.}
    \label{tab:extended_metrics_table}
    \begingroup
    \tablefontsize
    \setlength{\tabcolsep}{\tablecolsep}
    \renewcommand{\arraystretch}{0.92}
    \fittextwidth{
    \begin{tabular}{llccccccccc}
        \toprule
        \textbf{Market} & \textbf{Method} & \textbf{CR (\%)} & \textbf{AR (\%)} & \textbf{SR} & \textbf{MDD (\%)} & \textbf{Vol (\%)} & \textbf{Sortino} & \textbf{Calmar} & \textbf{IR} & \textbf{Win Rate (\%)} \\
        \midrule
        \multicolumn{11}{l}{\textbf{Panel A: Main and Cross-Universe Results}} \\
        \midrule
        \multirow{11}{*}{S\&P 500} & Buy \& Hold & 19.00 & 19.34 & 0.80 & 18.75 & 18.76 & 1.30 & 1.03 & 0.00 & -- \\
         & PCA & 7.84 & 7.98 & 0.27 & 17.30 & 18.06 & 0.67 & 0.46 & -0.40 & 50.81 \\
         & XGBoost & 3.43 & 3.49 & 0.03 & 18.45 & 18.10 & 0.38 & 0.19 & -0.56 & 51.21 \\
         & LightGBM & -5.33 & -5.42 & -0.43 & 20.93 & 19.21 & -0.24 & -0.26 & -0.89 & 49.60 \\
         & A2C & 10.64 & 10.82 & 0.40 & 17.70 & 18.89 & 0.86 & 0.61 & -0.28 & 51.21 \\
         & PPO & 7.56 & 7.68 & 0.25 & 14.97 & \textbf{17.40} & 0.68 & 0.51 & -0.42 & 54.84 \\
         & Gemini 2.5 Pro & 13.99 & 14.23 & 0.55 & 17.01 & 19.40 & 1.05 & 0.84 & -0.16 & 53.23 \\
         & Claude 3.7 Sonnet & 10.74 & 10.92 & 0.40 & 18.88 & 19.16 & 0.83 & 0.58 & -0.27 & 51.61 \\
         & DeepSeek-R1 & 21.56 & 21.94 & 0.93 & \textbf{14.36} & 18.37 & 1.55 & 1.53 & 0.08 & 53.63 \\
         & Qwen3-8B & 12.63 & 12.85 & 0.47 & 19.52 & 20.58 & 0.87 & 0.66 & -0.20 & 54.84 \\
         & Alpha-R1 & \textbf{46.95} & \textbf{47.87} & \textbf{1.62} & 16.91 & 23.06 & \textbf{2.57} & \textbf{2.83} & \textbf{0.81} & \textbf{56.85} \\
        \midrule
        \multirow{11}{*}{CSI 300} & Buy \& Hold & 21.19 & 22.16 & 1.31 & 10.49 & 14.96 & 1.65 & 2.11 & 0.00 & -- \\
         & PCA & 2.82 & 2.93 & 0.17 & 14.46 & \textbf{14.72} & 0.39 & 0.20 & -0.92 & 49.17 \\
         & XGBoost & 8.62 & 8.99 & 0.50 & 16.26 & 17.11 & 0.88 & 0.55 & -0.53 & 51.65 \\
         & LightGBM & 17.65 & 18.44 & 1.05 & 14.92 & 15.99 & 1.51 & 1.24 & -0.17 & 51.24 \\
         & A2C & 21.96 & 22.96 & 1.20 & 14.86 & 17.24 & 1.85 & 1.55 & 0.05 & 53.31 \\
         & PPO & 14.33 & 14.96 & 0.81 & 12.95 & 17.31 & 1.35 & 1.16 & -0.28 & 50.41 \\
         & Gemini 2.5 Pro & 15.60 & 16.29 & 0.90 & 14.01 & 16.67 & 1.47 & 1.16 & -0.24 & \textbf{53.72} \\
         & Claude 3.7 Sonnet & 9.71 & 10.13 & 0.57 & 14.49 & 16.73 & 0.99 & 0.70 & -0.50 & 49.59 \\
         & DeepSeek-R1 & 14.04 & 14.66 & 0.81 & 14.60 & 16.70 & 1.29 & 1.00 & -0.32 & 51.24 \\
         & Qwen3-8B & 14.78 & 15.44 & 0.79 & 14.38 & 18.36 & 1.34 & 1.07 & -0.24 & 52.07 \\
         & Alpha-R1 & \textbf{38.68} & \textbf{40.57} & \textbf{2.23} & \textbf{6.58} & 15.10 & \textbf{4.40} & \textbf{6.17} & \textbf{0.73} & 52.48 \\
        \midrule
        \multirow{11}{*}{Russell 2000} & Buy \& Hold & 13.53 & 13.76 & 0.48 & 23.79 & 23.09 & 1.00 & 0.58 & 0.00 & -- \\
         & PCA & 13.80 & 14.04 & 0.47 & 28.15 & 25.19 & 0.94 & 0.50 & 0.02 & 50.00 \\
         & XGBoost & 0.51 & 0.52 & -0.04 & 27.00 & 24.26 & 0.20 & 0.02 & -0.39 & 52.02 \\
         & LightGBM & -8.30 & -8.43 & -0.37 & 36.46 & 26.44 & -0.26 & -0.23 & -0.61 & 48.79 \\
         & A2C & 54.65 & 55.74 & 1.91 & \textbf{16.46} & \textbf{22.11} & 3.08 & 3.39 & 1.02 & 57.26 \\
         & PPO & -25.48 & -25.83 & -1.12 & 42.99 & 27.40 & -1.28 & -0.60 & -1.24 & 48.79 \\
         & Gemini 2.5 Pro & 19.60 & 19.95 & 0.64 & 29.24 & 27.21 & 1.17 & 0.68 & 0.20 & 53.23 \\
         & Claude 3.7 Sonnet & 29.60 & 30.14 & 0.72 & 27.92 & 43.57 & 1.27 & 1.08 & 0.44 & 53.63 \\
         & DeepSeek-R1 & 29.95 & 30.50 & 0.93 & 31.12 & 28.23 & 1.46 & 0.98 & 0.45 & 52.82 \\
         & Qwen3-8B & -9.27 & -9.41 & -0.43 & 32.97 & 25.59 & -0.36 & -0.29 & -0.69 & 50.40 \\
         & Alpha-R1 & \textbf{78.85} & \textbf{80.54} & \textbf{2.46} & 17.67 & 23.36 & \textbf{3.83} & \textbf{4.56} & \textbf{1.50} & \textbf{60.08} \\
        \midrule
        \multirow{11}{*}{CSI 1000} & Buy \& Hold & 31.02 & 32.49 & 1.33 & 16.87 & 21.89 & 1.50 & 1.93 & 0.00 & -- \\
         & PCA & 0.60 & 0.63 & 0.06 & 20.99 & 20.21 & 0.18 & 0.03 & -1.12 & 50.00 \\
         & XGBoost & 0.69 & 0.71 & 0.05 & 22.24 & 18.94 & 0.17 & 0.03 & -1.16 & 50.83 \\
         & LightGBM & -4.63 & -4.82 & -0.23 & 20.88 & 19.58 & -0.20 & -0.23 & -1.37 & 48.76 \\
         & A2C & -7.72 & -8.03 & -0.43 & 22.90 & \textbf{18.86} & -0.49 & -0.35 & -1.56 & 48.35 \\
         & PPO & 14.63 & 15.28 & 0.72 & 20.39 & 20.62 & 1.06 & 0.75 & -0.58 & 53.72 \\
         & Gemini 2.5 Pro & 0.47 & 0.49 & 0.06 & 28.37 & 21.71 & 0.17 & 0.02 & -1.09 & 55.37 \\
         & Claude 3.7 Sonnet & -1.85 & -1.92 & 0.08 & 23.86 & 35.44 & 0.18 & -0.08 & -0.68 & 49.59 \\
         & DeepSeek-R1 & -6.21 & -6.45 & -0.27 & 30.65 & 21.71 & -0.25 & -0.21 & -1.41 & 53.31 \\
         & Qwen3-8B & 13.87 & 14.48 & 0.61 & 23.02 & 25.05 & 0.96 & 0.63 & -0.47 & 51.24 \\
         & Alpha-R1 & \textbf{69.76} & \textbf{73.52} & \textbf{2.80} & \textbf{13.75} & 19.89 & \textbf{3.90} & \textbf{5.35} & \textbf{1.05} & \textbf{58.68} \\
        \bottomrule
    \end{tabular}
    }
    \endgroup
\end{table*}

\begin{table*}[htbp]
    \centering
    \caption{Grouped NAV-derived risk and return metrics. Metrics are computed from daily NAV curves over the 2025 test period. SR is the annualized excess Sharpe ratio using the market-specific risk-free rate, and MDD is reported as a positive drawdown magnitude. Bold entries mark the best value within the same panel and asset pool; Buy \& Hold is a passive benchmark, so its strategy Win Rate is not reported. (continued)}
    \label{tab:extended_metrics_table_cont}
    \begingroup
    \tablefontsize
    \setlength{\tabcolsep}{\tablecolsep}
    \renewcommand{\arraystretch}{0.92}
    \fittextwidth{
    \begin{tabular}{llccccccccc}
        \toprule
        \textbf{Market} & \textbf{Method} & \textbf{CR (\%)} & \textbf{AR (\%)} & \textbf{SR} & \textbf{MDD (\%)} & \textbf{Vol (\%)} & \textbf{Sortino} & \textbf{Calmar} & \textbf{IR} & \textbf{Win Rate (\%)} \\
        \midrule
        \multicolumn{11}{l}{\textbf{Panel B: Ablation Study}} \\
        \midrule
        \multirow{6}{*}{S\&P 500} & Alpha-R1 (Full) & \textbf{46.95} & \textbf{47.87} & \textbf{1.62} & \textbf{16.91} & 23.06 & \textbf{2.57} & \textbf{2.83} & \textbf{0.81} & \textbf{56.85} \\
         & Buy \& Hold & 19.00 & 19.34 & 0.80 & 18.75 & \textbf{18.76} & 1.30 & 1.03 & 0.00 & -- \\
         & w/o Market Price & 28.31 & 28.83 & 1.02 & 17.59 & 23.18 & 1.74 & 1.64 & 0.31 & 54.44 \\
         & w/o News & 33.99 & 34.63 & 1.20 & 17.60 & 23.33 & 2.01 & 1.97 & 0.47 & 55.24 \\
         & w/o Factor Semantics & 39.69 & 40.45 & 1.37 & 17.16 & 23.58 & 2.25 & 2.36 & 0.63 & 55.65 \\
         & w/o RL Optimization & 12.63 & 12.85 & 0.47 & 19.52 & 20.58 & 0.87 & 0.66 & -0.20 & 54.84 \\
        \midrule
        \multirow{6}{*}{CSI 300} & Alpha-R1 (Full) & \textbf{38.68} & \textbf{40.57} & \textbf{2.23} & 6.58 & 15.10 & \textbf{4.40} & \textbf{6.17} & \textbf{0.73} & 52.48 \\
         & Buy \& Hold & 21.19 & 22.16 & 1.31 & 10.49 & \textbf{14.96} & 1.65 & 2.11 & 0.00 & -- \\
         & w/o Market Price & 26.68 & 27.93 & 1.55 & 7.21 & 15.76 & 2.95 & 3.87 & 0.24 & \textbf{52.89} \\
         & w/o News & 30.36 & 31.79 & 1.75 & \textbf{6.51} & 15.62 & 3.45 & 4.89 & 0.39 & 52.07 \\
         & w/o Factor Semantics & 34.03 & 35.67 & 1.94 & \textbf{6.51} & 15.55 & 4.01 & 5.48 & 0.54 & 50.83 \\
         & w/o RL Optimization & 14.78 & 15.44 & 0.79 & 14.38 & 18.36 & 1.34 & 1.07 & -0.24 & 52.07 \\
        \midrule
        \multicolumn{11}{l}{\textbf{Panel C: Factor-Gating Baseline Comparison}} \\
        \midrule
        \multirow{10}{*}{S\&P 500} & Alpha-R1 & \textbf{46.95} & \textbf{47.87} & \textbf{1.62} & 16.91 & 23.06 & \textbf{2.57} & \textbf{2.83} & \textbf{0.81} & \textbf{56.85} \\
         & Buy \& Hold & 19.00 & 19.34 & 0.80 & 18.75 & 18.76 & 1.30 & 1.03 & 0.00 & -- \\
         & Random & 5.27 & 5.35 & 0.13 & 15.48 & \textbf{17.68} & 0.50 & 0.35 & -0.50 & 51.61 \\
         & Rolling IC & 15.04 & 15.30 & 0.61 & 14.78 & 18.96 & 1.10 & 1.04 & -0.12 & 53.23 \\
         & Top-k & 1.31 & 1.33 & -0.07 & 15.60 & 19.07 & 0.22 & 0.09 & -0.62 & 52.02 \\
         & Greedy & -12.44 & -12.62 & -0.82 & 22.50 & 19.66 & -0.72 & -0.56 & -1.19 & 45.56 \\
         & Beam & 16.31 & 16.59 & 0.61 & 16.20 & 21.45 & 1.13 & 1.02 & -0.06 & 54.03 \\
         & Lasso & 0.49 & 0.50 & -0.12 & 18.88 & 18.80 & 0.16 & 0.03 & -0.67 & 50.00 \\
         & MLP & 16.25 & 16.53 & 0.69 & 14.67 & 18.10 & 1.19 & 1.13 & -0.10 & 51.61 \\
         & Transformer & 20.21 & 20.57 & 0.86 & \textbf{14.29} & 18.53 & 1.46 & 1.44 & 0.04 & 55.24 \\
        \midrule
        \multirow{10}{*}{CSI 300} & Alpha-R1 & \textbf{38.68} & \textbf{40.57} & \textbf{2.23} & \textbf{6.58} & 15.10 & \textbf{4.40} & \textbf{6.17} & \textbf{0.73} & \textbf{52.48} \\
         & Buy \& Hold & 21.19 & 22.16 & 1.31 & 10.49 & 14.96 & 1.65 & 2.11 & 0.00 & -- \\
         & Random & 0.69 & 0.75 & 0.05 & 14.51 & 18.32 & 0.18 & 0.05 & -0.94 & 51.08 \\
         & Rolling IC & 10.69 & 11.15 & 0.63 & 13.84 & 16.62 & 0.99 & 0.81 & -0.47 & 50.41 \\
         & Top-k & -9.41 & -10.22 & -0.44 & 19.55 & 22.37 & -0.47 & -0.52 & -1.23 & 50.65 \\
         & Greedy & 22.62 & 24.92 & 1.17 & 11.26 & 19.29 & 1.87 & 2.21 & 0.13 & 50.65 \\
         & Beam & 20.91 & 23.01 & 1.09 & 11.02 & 19.35 & 1.69 & 2.09 & 0.05 & 50.65 \\
         & Lasso & 6.15 & 6.42 & 0.46 & 8.75 & \textbf{11.87} & 0.84 & 0.73 & -0.89 & 50.83 \\
         & MLP & -5.66 & -6.16 & -0.30 & 16.61 & 19.88 & -0.29 & -0.37 & -1.20 & 51.52 \\
         & Transformer & 7.99 & 8.75 & 0.45 & 14.48 & 19.26 & 0.78 & 0.60 & -0.54 & 50.22 \\
        \bottomrule
    \end{tabular}
    }
    \endgroup
\end{table*}

The extended metrics confirm that \modelname's advantage extends across risk-adjusted, active-return, and drawdown dimensions.

%% file: sections/09_04_appendix_supplementary_figures.tex
\section{Supplementary Results and Figures}
\label{app:supplementary_figures}

This appendix collects supplementary results and visualizations for the experiments in Section~\ref{sec:experiments}. All figures use the same 2025 out-of-sample NAV curves and trading protocol.

\subsection{Return Distribution Analysis}
\label{app:return_distribution_analysis}

This section provides NAV-level return timing, distribution, and drawdown diagnostics.

\begin{figure*}[htbp]
    \centering
    \includegraphics[width=0.96\textwidth]{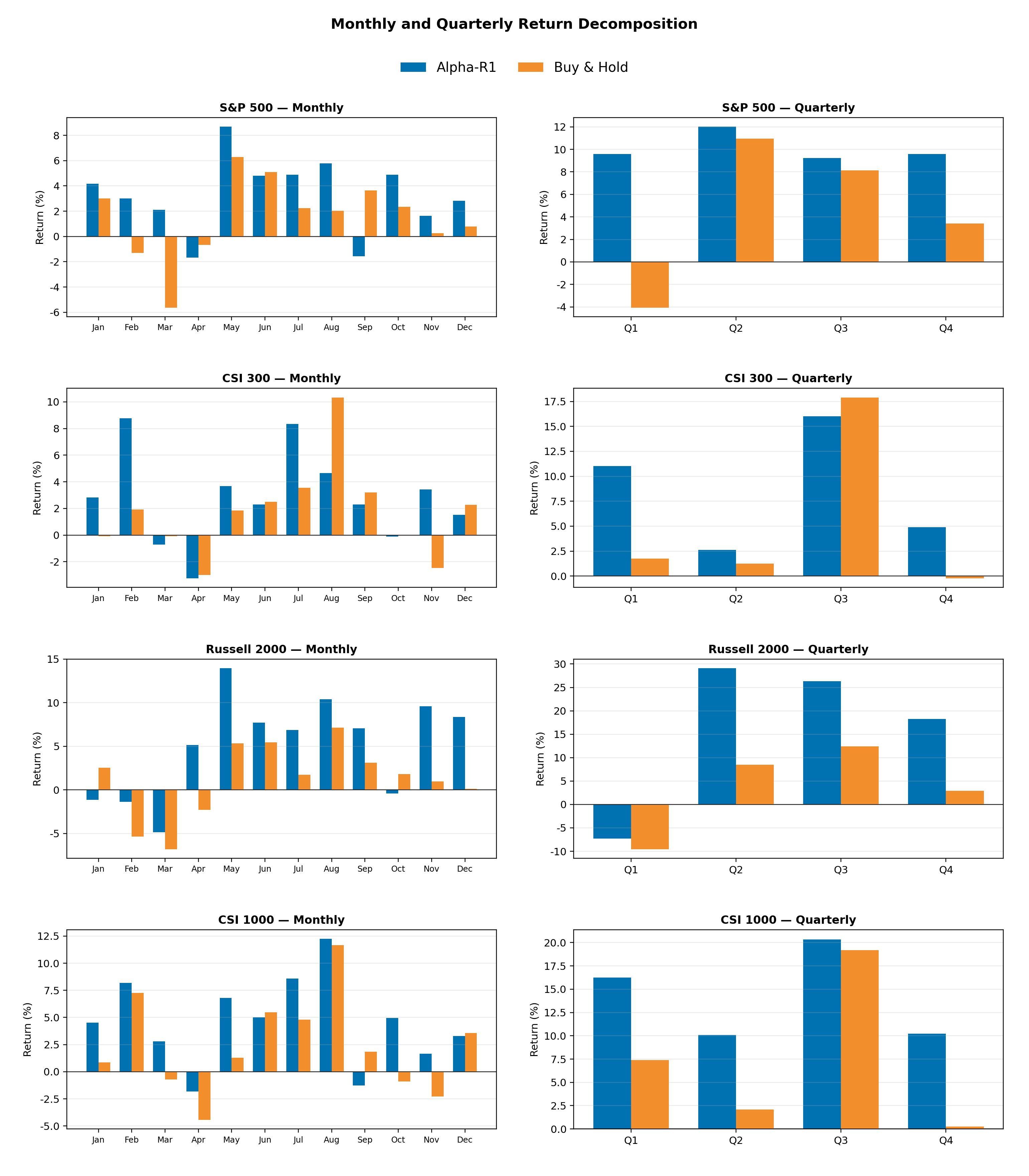}
    \caption{Monthly (left) and quarterly (right) return decomposition for \modelname and Buy \& Hold across all four asset pools. Returns are compounded from daily NAV returns, so each bar measures realized wealth growth within the corresponding calendar period.}
    \Description{Bar charts comparing monthly and quarterly compounded returns for \modelname and Buy \& Hold across the four evaluated asset pools.}
    \label{fig:risk_monthly_quarterly_bar_main}
\end{figure*}

\paragraph{Monthly and Quarterly Return Decomposition}
Figure~\ref{fig:risk_monthly_quarterly_bar_main} reports compounded monthly and quarterly returns for \modelname and Buy \& Hold across all four asset pools. The monthly view (left column) shows whether annual performance is spread across the testing horizon or concentrated in a small number of calendar months, while the quarterly view (right column) provides a coarser comparison of active gains against the market benchmark.

\begin{figure*}[htbp]
    \centering
    \includegraphics[width=\textwidth]{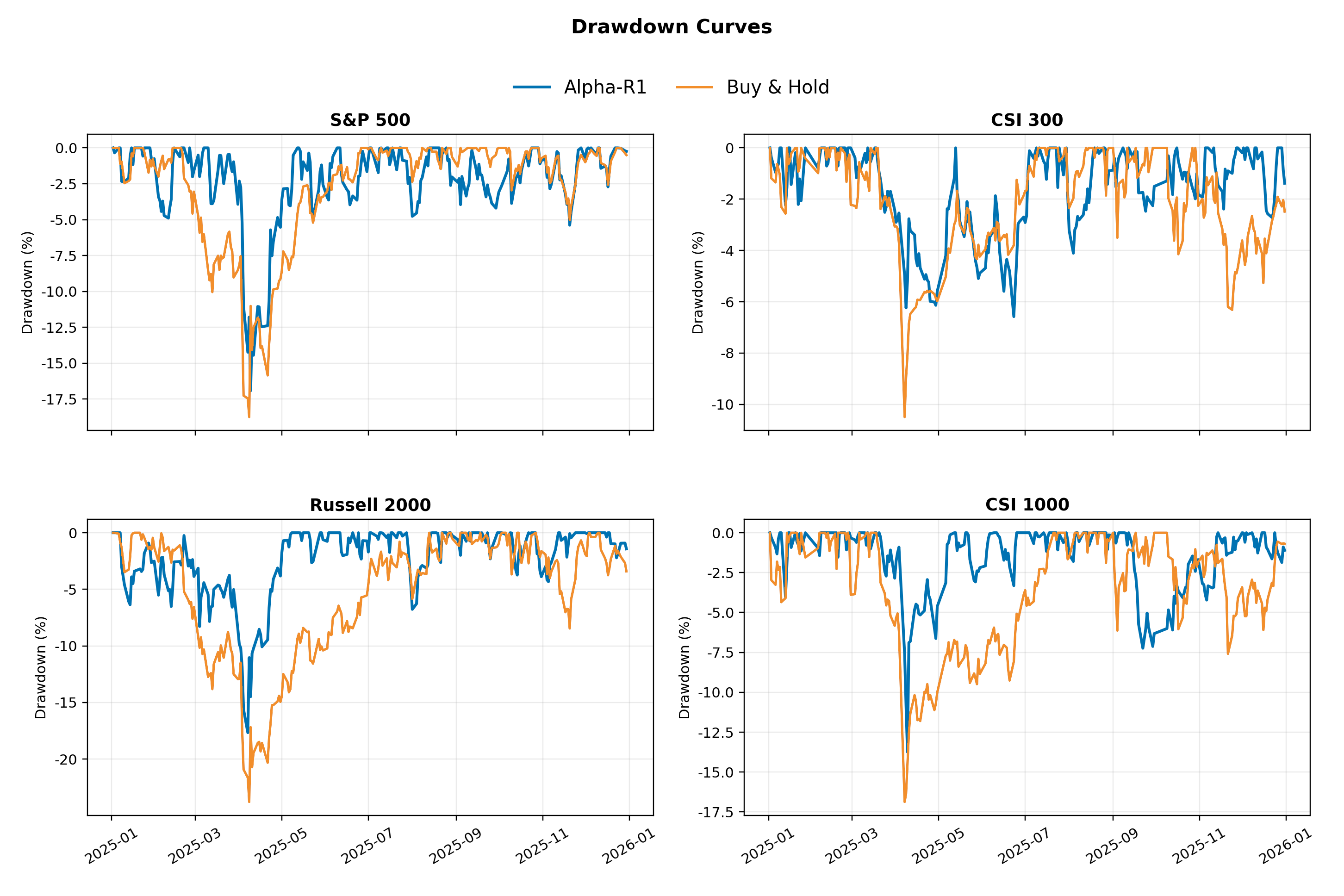}
    \caption{Drawdown curves for \modelname and Buy \& Hold across all four asset pools. Each curve reports the percentage decline from the running NAV peak, making it possible to compare not only the maximum drawdown magnitude but also the timing and persistence of underwater periods.}
    \Description{Line charts showing drawdown paths for \modelname and Buy \& Hold across the four evaluated asset pools.}
    \label{fig:risk_drawdown_curves_main}
\end{figure*}

\paragraph{Drawdown Path}
Figure~\ref{fig:risk_drawdown_curves_main} plots the full drawdown path rather than only the scalar MDD value. This view shows the timing and persistence of peak-to-trough losses throughout the 2025 test period.

\subsection{Ablation Trajectory Analysis}

\begin{figure*}[htbp]
    \centering
    \begin{subfigure}[b]{0.49\textwidth}
        \centering
        \includegraphics[width=\textwidth]{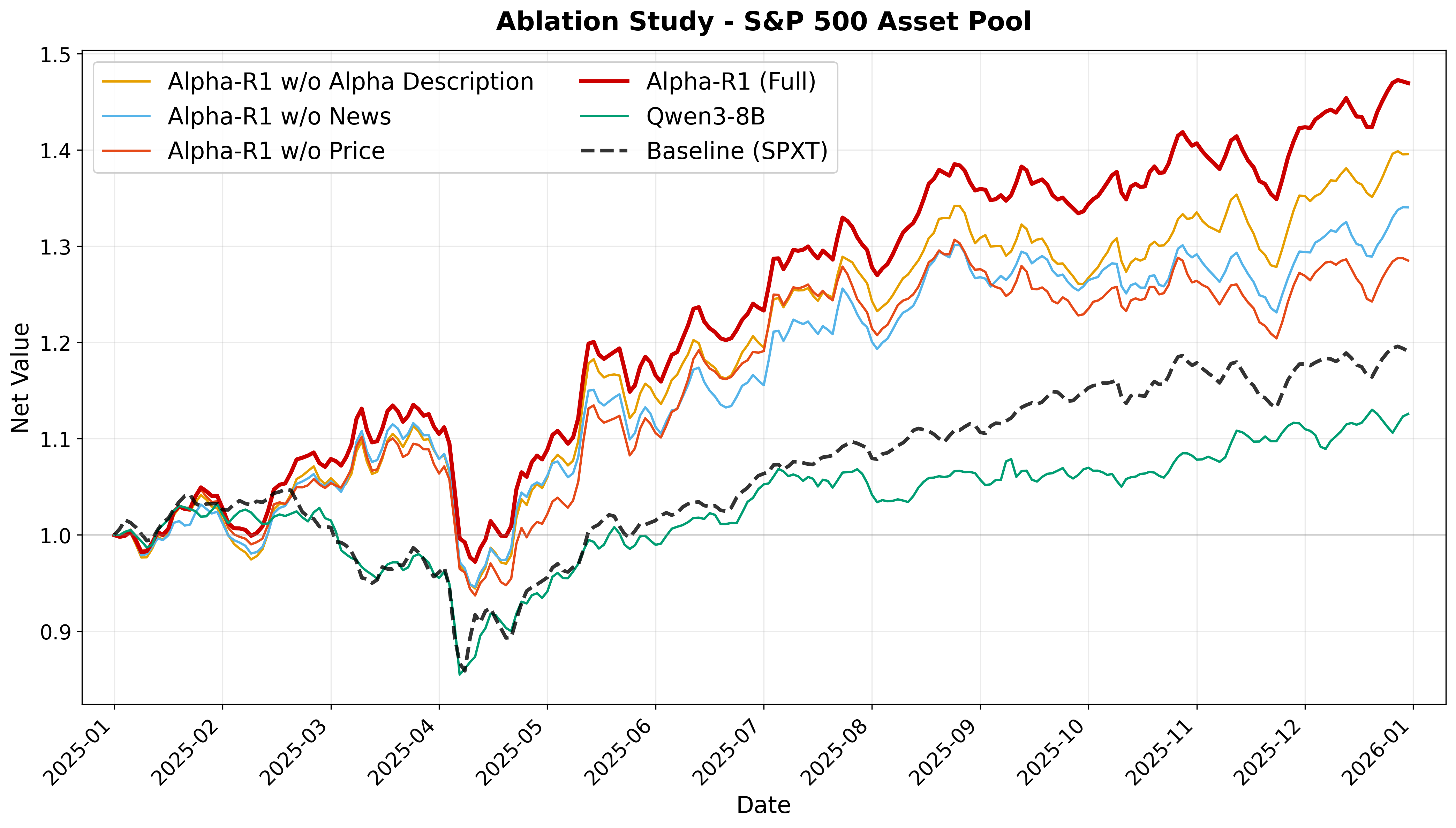}
        \caption{S\&P 500}
        \label{fig:ablation_curves_sp500}
    \end{subfigure}
    \hfill
    \begin{subfigure}[b]{0.49\textwidth}
        \centering
        \includegraphics[width=\textwidth]{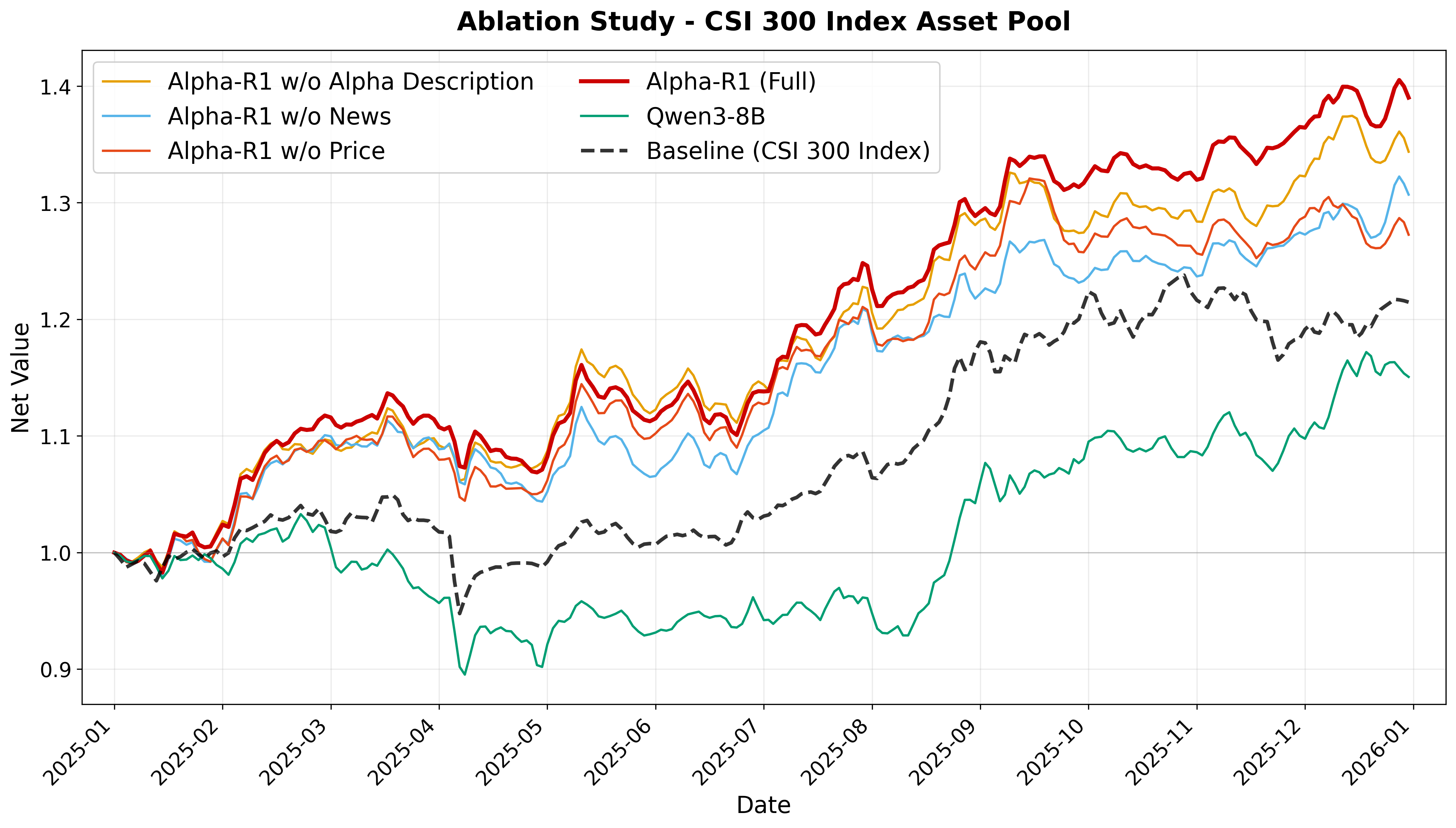}
        \caption{CSI 300}
        \label{fig:ablation_curves_csi300}
    \end{subfigure}
    \caption{Ablation cumulative net value curves over the 12-month 2025 testing set. All variants are evaluated under the same downstream trading protocol, so the gaps isolate the contribution of market-price inputs, news inputs, factor semantics, and RL optimization.}
    \Description{Two cumulative net value line charts comparing the full \modelname configuration with ablation variants on S\&P 500 and CSI 300.}
    \label{fig:ablation_curves}
\end{figure*}

This subsection expands the ablation results discussed in Section~\ref{sec:experiments}, especially the component-level comparisons summarized in Table~\ref{tab:ablation}. Readers should focus on whether the relative gaps reported in the table remain persistent across the 2025 horizon rather than arising from a few isolated dates. In particular, Figure~\ref{fig:ablation_curves} visually reinforces the conclusion from the ablation analysis that removing RL optimization or factor semantics weakens cumulative growth more materially than removing any single auxiliary market-information stream.

\subsection{Factor-Gating Baseline Trajectories}

\begin{figure*}[htbp]
    \centering
    \begin{subfigure}[b]{0.49\textwidth}
        \centering
        \includegraphics[width=\textwidth]{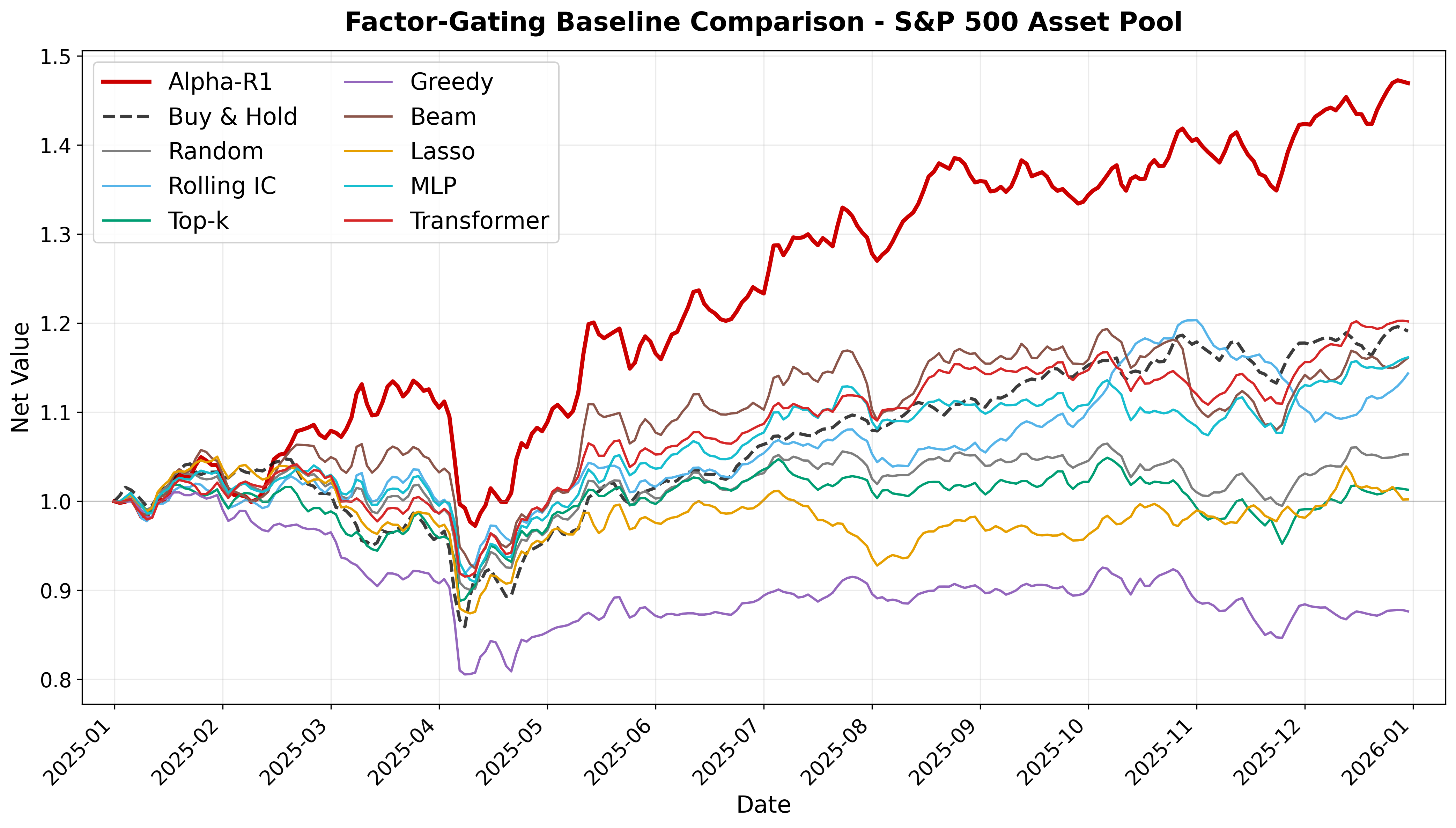}
        \caption{S\&P 500}
    \end{subfigure}
    \hfill
    \begin{subfigure}[b]{0.49\textwidth}
        \centering
        \includegraphics[width=\textwidth]{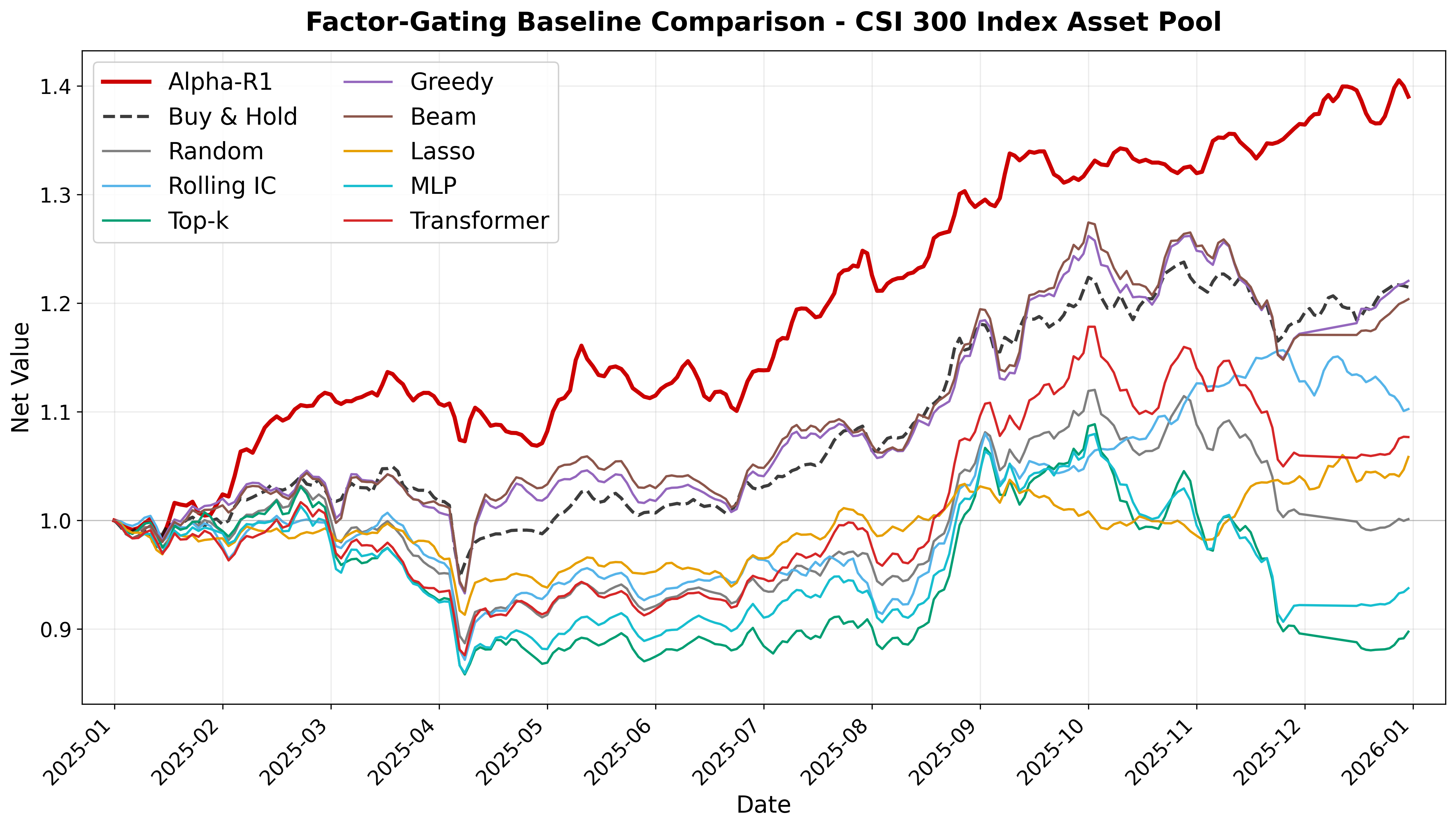}
        \caption{CSI 300}
    \end{subfigure}
    \caption{Factor-gating baseline comparison curves over the 12-month 2025 testing set. The figure compares \modelname against heuristic rules, search-based gating, and learned gating models under the same downstream execution protocol, alongside the Buy \& Hold benchmark.}
    \Description{Two cumulative net value line charts comparing \modelname with heuristic, search-based, learned, and Buy \& Hold baselines on S\&P 500 and CSI 300.}
    \label{fig:heuristic_curves}
\end{figure*}

This subsection provides the trajectory-level counterpart to the quantitative comparison in Table~\ref{tab:gating_comparison} and the discussion in Section~\ref{sec:experiments}. The main reading objective is to compare not only terminal performance but also the timing, persistence, and recovery profile of each gating strategy under the shared downstream protocol. Figure~\ref{fig:heuristic_curves} therefore serves as the time-series expansion of the gating-baseline results, illustrating why the semantic conditioning advantage described in the main text is not reducible to a single favorable subperiod.

\begin{table*}[!t]
    \centering
    \caption{Positioning of \modelname relative to adjacent method families. The contribution is complementary to LLM alpha \emph{mining}---a mined factor zoo is precisely the kind of input our second stage consumes---and distinct in both state representation and alignment signal from RL-finance LLMs and numerical-state RL portfolio managers.}
    \label{tab:positioning}
    \begingroup
    \tablefontsize
    \setlength{\tabcolsep}{\tablecolsep}
    \fittextwidth{
    \begin{tabular}{@{}L{0.19\linewidth}L{0.22\linewidth}L{0.26\linewidth}L{0.26\linewidth}@{}}
        \toprule
        \textbf{Family} & \textbf{Task} & \textbf{State representation} & \textbf{Alignment signal} \\
        \midrule
        LLM alpha mining~\cite{wang2024llmfactorextractingprofitablefactors,Wang2025AlphaGPT,Tang2025AlphaAgentLA,Shi2024AlphaForgeAF,shi2026alphajungle,li2025rdagentquant} & Factor generation / discovery & Research context, factor and performance text & Backtest feedback on generated factors \\
        RL-finance LLMs~\cite{xiao2025tradingr1financialtradingllm,liu2025finr1largelanguagemodel,deng2026alphaquanter} & End-to-end trading decisions, financial QA & Market text, trade context & Price-action or answer-correctness rewards \\
        RL portfolio managers~\cite{liu2020finrl,lee2020maps,ye2020sarl} & Per-asset weight allocation & Numerical price / indicator states & Return of the weight allocation \\
        \midrule
        \textbf{\modelname (Ours)} & Context-conditioned second-stage factor subset selection & Semantic market state $+$ factor semantic profiles & Realized portfolio return of the selected factor subset \\
        \bottomrule
    \end{tabular}
    }
    \endgroup
\end{table*}

\subsection{Positioning Relative to Adjacent Method Families}
\label{app:positioning}

Table~\ref{tab:positioning} provides the task-level positioning of \modelname summarized in Section~\ref{sec:related_work}.

%% file: sections/09_05_appendix_case_study.tex
\section{Interpretability Case Study}
\label{app:case_study}

To illustrate how \modelname connects market semantics to factor screening, we present a representative U.S. case from 2025-04-09 and a compact multi-case summary. These serve as hu\-man-read\-able audit artifacts that illustrate contextual consistency, not as mecha\-nism-faith\-ful explanations of internal computation.

\begin{figure*}[htbp]
    \centering
    \fbox{
    \parbox{\dimexpr0.96\linewidth-2\fboxsep-2\fboxrule\relax}{
        \small
        \textbf{[Case Study] Risk-Off Selloff in U.S. Equities (2025-04-09)}
        \vspace{0.5em}
        \hrule
        \vspace{0.5em}

        \textbf{1. Market Perception (Input):}
        The model analyzes the intraday price action together with pre-market news, identifying a \emph{risk-off} session in which U.S. index ETFs sell off sharply and growth-heavy exposures underperform broader benchmarks. It characterizes the current regime as \emph{panic selling} with \emph{short-horizon volatility expansion}. The important input signal is the joint pattern: directional weakness, elevated intraday range, high-volume repricing, and a news backdrop that reinforces defensive positioning rather than a routine low-volatility pullback.

        \vspace{0.5em}
        \textbf{2. Reasoning \& Factor Alignment:}
        \begin{itemize}
            \item \textbf{Logic A (Selected):} The model prioritizes short-horizon price/volume dislocation signals together with intraday range-sensitive logic; across repeated runs, this family of signals remains the most stable response to the regime. This includes factors whose interpretation is tied to close-versus-VWAP pressure, abnormal intraday movement, and price-volume interaction during rapid repricing.

            \textit{Reasoning:} ``Under directional pressure and order-flow imbalance, close-versus-VWAP dislocation becomes more informative for short-horizon price discovery. Intraday price-range structure is also more revealing when volatility expands and the market is repricing rapidly.''

            \item \textbf{Logic B (Downweighted):} The model downweights factor logic that is better suited to stable, lower-volatility, or longer-horizon trend environments. These signals may remain useful in calmer regimes, but their economic rationale is less aligned with a session dominated by forced de-risking and short-horizon liquidity pressure.

            \textit{Reasoning:} ``During a panic-driven selloff, longer-horizon structural signals are less aligned with the immediate repricing dynamics and may introduce noise relative to short-horizon price/volume dislocation signals.''
        \end{itemize}

        \vspace{0.5em}
        \textbf{3. Outcome:}
        The reasoning module emphasizes short-horizon price/volume dislocation signals during a U.S. panic selloff and de-emphasizes factor logic that is more compatible with stable trend continuation or lower-volatility regimes. This case shows that \modelname does not merely rank factors by recent standalone performance. Instead, it reasons over the compatibility between the current market narrative and the economic logic of each factor, then activates factors whose mechanisms are most consistent with the inferred regime. The case is therefore qualitatively consistent with two quantitative results: the degradation of the ``w/o Factor Semantics'' variant, which removes economic descriptions from the gating input, and the weaker performance of static or non-semantic gating baselines, which cannot explicitly condition factor activation on the current narrative.
    }
    }
    \caption{Structured interpretability case summary for the U.S. 2025-04-09 decision. The example illustrates how market semantics, such as risk-off pressure and short-horizon volatility expansion, can be mapped to factor logic based on price/volume dislocation and intraday range sensitivity, while less compatible stable-regime logic is downweighted.}
    \Description{Text box summarizing a held-out U.S. case study with market perception, selected and downweighted factor logic, and the resulting interpretation.}
    \label{fig:case_summary}
\end{figure*}

\paragraph{Multi-Case Pattern Summary.}
To reduce reliance on a single anecdotal example, Table~\ref{tab:multi_case_patterns} summarizes theme--factor alignment patterns across six additional held-out U.S. dates. The table serves as a qualitative consistency check: the market theme describes the regime inferred from inputs, the downweighted logic summarizes the factor family judged less compatible with that regime, and the observed pattern records how the model is presented with market semantics. Across the six cases, the summary indicates distinct factor families---short-hori\-zon dislocation, mean-rever\-sion, trend\slash momentum, quality, and deep-value---rather than a single default signal type. This appendix should therefore be read jointly with the degradation of the ``w/o Factor Semantics'' variant in Table~\ref{tab:ablation} and the broader comparison against non-semantic gating rules in Table~\ref{tab:gating_comparison}, rather than as a free-standing anecdotal narrative.

\begin{table*}[htbp]
    \centering
    \tablefontsize
    \setlength{\tabcolsep}{\tablecolsep}
    \renewcommand{\arraystretch}{1.08}
    \renewcommand{\tabularxcolumn}[1]{m{#1}}
    \caption{Compact multi-case summary across held-out U.S. test dates, illustrating how \modelname conditions factor selection on the inferred market regime. Distinct regimes activate distinct factor families---short-horizon dislocation, mean-reversion, trend/momentum, quality, and deep-value---rather than defaulting to a single signal type.}
    \label{tab:multi_case_patterns}
    \begin{tabularx}{\textwidth}{@{}>{\centering\arraybackslash}m{1.25cm}>{\raggedright\arraybackslash}m{2.55cm}>{\raggedright\arraybackslash}m{2.55cm}Y@{}}
        \toprule
        \multicolumn{1}{c}{Date} & \multicolumn{1}{c}{Market Theme} & \multicolumn{1}{c}{Downweighted Logic} & \multicolumn{1}{c}{Observed Pattern} \\
        \midrule
        20250304 & Broad selloff with high-volatility downside pressure and weak risk appetite & Lower-volatility or longer-horizon structural logic that assumes smoother continuation & The model prioritizes short-horizon price/volume dislocation signals. Abnormal range expansion and volume imbalance make immediate liquidity-sensitive factors more relevant than slow-moving structural signals under forced repricing. \\
        20250409 & U.S. panic selloff with growth-heavy exposures underperforming the broad market & Stable-trend or lower-volatility logic whose payoff depends on orderly continuation & Risk-off conditions activate close-versus-VWAP and intraday range signals, consistent with the detailed case study above. Stable-regime factors are downweighted because their assumptions are mismatched to a session driven by forced de-risking. \\
        20250519 & Intraday rebound with noisy event-driven flow after earlier pressure & Trend-following momentum logic that would extrapolate the prior downward direction rather than capture the reversal & The model shifts to mean-reversion and reversal-sensitive signals. It treats the rebound as an oversold recovery rather than a trend initiation, favoring factors tied to short-term price normalization and bounce-detection over directional momentum. \\
        20250520 & U.S. up day with technology leadership stronger than broad benchmarks & Contrarian or value-oriented logic that would fade leadership strength in favor of laggard rotation & The model activates momentum and trend-confirmation signals tied to relative strength. With clear sector leadership and broad participation, it favors factors that reinforce directional persistence over contrarian or defensive screens. \\
        20251107 & Volatile downtrend with elevated uncertainty and unstable recovery attempts & High-beta or speculative growth logic that depends on stable risk appetite, which is absent under elevated uncertainty & The model pivots to quality and fundamental-strength factors, emphasizing earnings stability, low leverage, and resilient margins. Under persistent uncertainty, it selects factors whose economic rationale is tied to balance-sheet durability rather than directional timing. \\
        20251121 & Growth-led weakness under a renewed risk-off tone and fragile sentiment & Growth-oriented or high-multiple logic that depends on expansion narratives, which fragile sentiment cannot sustain & The model favors deep-value and dividend-yield factors emphasizing margin-of-safety. With growth de-rating under way, it selects factors anchored to tangible valuation support and income durability rather than to short-horizon price dynamics. \\
        \bottomrule
    \end{tabularx}
\end{table*}

%% file: sections/09_06_appendix_method_details.tex
\section{Additional Method Details}
\label{app:method_details}

This appendix provides the execution details summarized in Section~\ref{sec:execution_mechanism} and the numerical GRPO hyperparameters used for training.

\subsection{GRPO Hyperparameters}
\label{app:grpo_hyperparams}

The full GRPO objective is given in Section~\ref{sec:grpo}. Table~\ref{tab:grpo_hyperparams} lists the corresponding numerical hyperparameters.

\begin{table}[htbp]
    \centering
    \caption{GRPO hyperparameters used for \modelname.}
    \label{tab:grpo_hyperparams}
    \begin{tabular}{@{}ll@{}}
        \toprule
        \textbf{Parameter} & \textbf{Value} \\
        \midrule
        Group size $G$ & 8 \\
        Clipping $\epsilon$ & 0.2 \\
        KL coefficient $\beta$ & 0.001 \\
        Learning rate & $1 \times 10^{-6}$ \\
        LR warmup ratio & 0.005 \\
        Batch size (train) & 64 \\
        Total epochs & 2 \\
        Max prompt length & 32,768 tokens \\
        Max response length & 8,192 tokens \\
        Rollout temperature & 1.0 \\
        Optimizer & AdamW \\
        Reference model $\pi_{\text{ref}}$ & Qwen3-8B (frozen pre-trained backbone) \\
        \bottomrule
    \end{tabular}
\end{table}

\subsection{Fixed Linear Return Model}
\label{app:linear_model_spec}

The fixed linear model converts the selected factor set $\mathcal{A}_t$ into cross-sectional stock scores. We provide the full specification here to clarify the role of this component relative to the semantic gating mechanism.

\paragraph{Estimation.}
The coefficients $\{\beta_i\}$ are estimated via ordinary least squares (OLS) on a pooled cross-sectional panel over the historical window 2020.01.01--2023.12.31. The dependent variable is the forward $H$-day stock return. All factor values are cross-sectionally standardized (zero mean, unit variance) and winsorized at the 1st and 99th percentiles on each trading day before entering the regression. Missing factor values are filled with the cross-sectional median. The same coefficient set is used for all stocks within each asset pool (i.e., coefficients are not estimated separately per sector or size group). The intercept $\beta_0$ does not affect cross-sectional ranking and is retained only for numerical stability.

\paragraph{Inference-Time Scoring.}
At each decision time $t$, for each stock, predicted returns are computed using Equation~\ref{eq:predicted_return}: only factors in $\mathcal{A}_t$ contribute; unselected factors have zero weight. Stocks are then ranked by predicted return, and the top $N$ form the equal-weighted portfolio. Because the coefficients are fixed, the \emph{only} degree of freedom at decision time is the factor selection $\mathcal{A}_t$ produced by \modelname. This design ensures that observed performance differences across methods are attributable to the factor-selection mechanism rather than to differences in the scoring model.

\subsection{Portfolio Construction and Execution}
\label{app:execution_details}

To mitigate turnover costs, we divide capital into $H$ independent sub-portfolios (slots), where $H$ is the holding period. On each trading day $t$, only slot $k = t \pmod H$ is rebalanced: existing positions in slot $k$ are liquidated and replaced with equal-weighted positions in the top $N$ stocks ranked by the predicted returns implied by the current factor set $\mathcal{A}_t$:
\begin{equation}
    P_{t, k} = \text{Top}_N(\text{Rank}(\text{LinearModel}(\mathcal{A}_t))).
\end{equation}
Here, $P_{t,k}$ denotes the holdings of the $k$-th slot after rebalancing on day $t$, while the remaining $H-1$ slots are left unchanged. This design smooths the equity curve and reduces the average daily turnover rate to $1/H$.

Rather than assuming execution at the opening price, we approximate fills using a VWAP model computed from the first 30 minutes of the session (09:31--10:00). For stock $s$, the execution price $\hat{P}_{s,t}$ is
\begin{equation}
    \hat{P}_{s,t} = \frac{\sum_{i=1}^{30} \left( Price_{s,t,i} \times Volume_{s,t,i} \right)}{\sum_{i=1}^{30} Volume_{s,t,i}}.
\end{equation}
We further enforce three practical constraints: (1) buy orders are rejected if a stock is locked at the upper price limit and sell orders are deferred if it is locked at the lower price limit during the execution window; (2) newly listed stocks are excluded on their first trading day; and (3) a transaction fee of 0.1\% (10 bps) is charged on both buys and sells. The daily portfolio return $R_t$ is computed as the aggregate return across all $H$ slots.

\paragraph{Long-Only Strategy.}
The strategy is strictly long-only: no short positions are taken. When fewer than $N$ valid stocks are available in a given slot due to limit locks, trading suspensions, or first-day listing exclusions, the unfilled portion of that slot is held in cash (earning the risk-free rate) until the next rebalancing date for that slot. Active returns reported throughout the paper should therefore be interpreted as long-only active selection relative to the benchmark, not as a dollar-neutral alpha strategy.

\paragraph{On Overlapping Holding Periods.}
We do not average overlapping $H$-day signal returns: capital is partitioned into $H$ independent sleeves, only one sleeve is rebalanced per day (daily turnover $1/H$), each sleeve is charged real transaction fees, and the reported curve is the NAV of this single realizable portfolio---exactly what an investor running this strategy would have realized. No ``overlap correction'' therefore applies to the return stream itself. Where overlap does matter is statistical inference on Sharpe-type ratios: the induced autocorrelation is handled at the inference layer by the block bootstrap in Appendix~\ref{app:bootstrap_protocol}, with block lengths bracketing the holding period.

%% file: sections/09_07_appendix_experimental_protocol.tex
\section{Experimental Protocol Details}
\label{app:experimental_protocol}

This appendix provides the anti-leakage enforcement, constituent-membership handling, and training/inference configuration details summarized in Section~\ref{sec:dataset_and_factors}.

\subsection{Anti-Leakage Protocol and Factor Construction}
\label{app:anti_leakage}

Experiments are conducted with strict anti-leakage protocols. The factor-synthesis coefficients $\beta_i$ are estimated from historical data spanning 2020--2023. During training (2024.07.01--2024.12.31), we apply randomized factor augmentation and sample 40 factors per episode to encourage robust regime-to-fac\-tor matching rather than memo\-ri\-za\-tion of a fixed or\-der\-ing. During testing (2025.01.01--2025.12.31), the candidate pool is fixed to the top factors ranked by Mean Rank IC computed over the historical factor-back\-testing window 2020.01.01--2023.12.31. The same window is used to construct the factor performance summaries $P_i$ and the semantic factor profiles $\alpha_{\text{des}, i}$, both of which are frozen before the 2025 evaluation period begins. This protocol reflects the two-stage design used throughout the paper: coarse pre-screening first narrows the broader factor universe to a pre-screened candidate pool, after which \modelname performs semantic reranking within that bounded set. We therefore interpret the reported results as evidence for the second-stage semantic gating mechanism under a controlled candidate budget, rather than as evidence that an LLM can already reason directly over a much larger factor zoo without degradation. The full training corpus contains approximately 96,000 samples constructed from the Alpha101 factor universe.

\subsection{Point-in-Time Constituent Membership}
\label{app:point_in_time}

All four asset pools are constructed using point-in-time constituent membership available as of each decision date. We do not apply a fixed end-of-2025 or otherwise ex-post constituent list retroactively. Index additions, deletions, trading suspensions, ST status in China, newly listed stocks, and delistings are handled according to the security status available on the corresponding trading date, so the investable universe is defined by the historical membership and tradability information available at portfolio formation time.

\subsection{Training and Inference Configuration}
\label{app:training_config}

We train \modelname on 64 H800 GPUs for approximately 120 hours, including the reinforcement learning stage and distributed rollout collection. For inference, \modelname and the baseline LLMs use greedy decoding with \texttt{temperature}=0. We additionally set \texttt{top\_p}=0.7 in the API configuration; under greedy decoding the top-$p$ threshold is not binding, since the single highest-probability token is always selected. Table~\ref{tab:model_cutoffs} lists every LLM used in the pipeline with its served identifier and provider-documented pre-training cutoff. All components of \modelname's own decision chain have cutoffs no later than 2024-11: the RL backbone Qwen3-8B (2024-09) is a frozen, locally hosted open-weight model rather than a live API; the state-synthesis and factor-profiling module is pinned to the dated, immutable snapshot \texttt{claude-3-7-sonnet-20250219} (2024-11), so the served weights cannot be silently updated during the evaluation window; and the consistency judge is Claude 3.5 Haiku (2024-07). Benchmark LLMs are evaluated zero-shot under the same timestamp-filtered inputs: DeepSeek-R1 (2024-11; served snapshot \texttt{deepseek-r1-250528}) and Gemini 2.5 Pro (provider-documented cutoff 2025-01). Gemini's one-month-later cutoff can only advantage the baseline, so the relative gap we report is conservative; no benchmark LLM output enters \modelname's inputs. All API calls were made between January and December 2025. For proprietary models, the stated cutoffs are taken from official provider documentation and cannot be independently verified beyond the provider's disclosure; if future evidence reveals that any model had access to 2025 information through its pretraining corpus, the anti-leakage guarantee for the affected module would be weakened. At test time, these modules operate only on information available on or before the decision date and do not use external retrieval, tool calls, future labels, or any information timestamped after the decision date. The decision state is formed before the portfolio action for that date; data produced after portfolio formation or during the realized holding period is excluded from the state description.

\begin{table}[htbp]
    \centering
    \caption{LLMs used in the pipeline: served identifiers and provider-documented pre-training cutoffs. \modelname's own decision chain (backbone, state synthesis and factor profiling, judge) has cutoffs no later than 2024-11; benchmark LLMs are zero-shot baselines evaluated under the same timestamp-filtered inputs and never feed \modelname's inputs.}
    \label{tab:model_cutoffs}
    \begingroup
    \tablefontsize
    \setlength{\tabcolsep}{\tablecolsep}
    \fitcolumnwidth{
    \begin{tabular}{@{}llll@{}}
        \toprule
        \textbf{Role} & \textbf{Model} & \textbf{Served identifier} & \textbf{Cutoff} \\
        \midrule
        RL backbone (also zero-shot baseline) & Qwen3-8B & locally hosted open-weight & 2024-09 \\
        State synthesis \& factor profiling & Claude 3.7 Sonnet & \texttt{claude-3-7-sonnet-20250219} (dated snapshot) & 2024-11 \\
        Consistency judge & Claude 3.5 Haiku & API & 2024-07 \\
        Zero-shot benchmark & DeepSeek-R1 & \texttt{deepseek-r1-250528} & 2024-11 \\
        Zero-shot benchmark & Gemini 2.5 Pro & API & 2025-01 \\
        \bottomrule
    \end{tabular}
    }
    \endgroup
\end{table}

Checkpoint selection uses the training-period reward only, and the reported execution configuration ($H{=}5$, $TopN{=}10$) is the code default fixed before evaluation (Section~\ref{sec:dataset_and_factors}). The 64-GPU figure is a one-time training expense; at inference the system is a single 8B model with deterministic decoding. Measured from the serving logs (vLLM, single GPU), one daily decision per universe consumes $\approx$15.6K input and $\approx$1.85K output tokens and completes in $\approx$15 seconds---the full four-universe year of daily decisions totals $\approx$4.2 GPU-hours, i.e., a recurring cost comparable to a handful of ordinary API calls per day.

\subsection{Forward-Looking-Content Audit}
\label{app:input_audit}

To close the temporal-leakage question empirically, we audited every text input the decision pipeline consumes---2{,}690 files spanning 2024-01$\,\to\,$2025-12, covering for each decision the day-$t{-}1$ close summary, the day-$t$ 09:00 pre-open news, and the day-$t{-}1$ synthesized index-analysis report. The protocol combined 100\% rule-based scanning for date mentions later than the file date with manual triage of all 632 flagged mentions. Every flag resolves to publicly scheduled events, scanner artifacts, or cosmetic VLM header stamps; the single year-typo found cites a level matching the \emph{past} trading day exactly (6{,}579.01 = CSI 1000 close on 2024-11-11), and the single cross-dated news reference embeds the decision day's own lawful 09:00 pre-open news. No post-decision information was found in any file.

%% file: sections/09_08_appendix_statistical_robustness.tex
\section{Statistical Robustness}
\label{app:statistical_robustness}

This appendix reports the statistical-inference layer for the results in Section~\ref{sec:experiments}: block-bootstrap confidence intervals, paired-difference tests against every baseline, rolling sub-window evaluation within the test year, Sharpe-ratio multiple-testing statistics (PSR/DSR/PBO), and a factor-level rank-IC decay check.

\subsection{Block-Bootstrap Inference Protocol}
\label{app:bootstrap_protocol}

All reported curves are the NAV of a single realizable slot-rota\-tion portfolio ($H$ independent sleeves, one re\-bal\-anced per day, fees charged per sleeve; Appendix~\ref{app:execution_details}); there is no averaged-overlapping-signal object to correct at the return level. The 5-day holding period does, however, induce autocorrelation in daily returns, which we handle at the inference layer with a circular moving-block bootstrap~\cite{politis1994stationary}: $B=10{,}000$ resamples with block lengths $L\in\{5,10,20\}$ bracketing the holding period $H=5$, applied to daily returns with annualization over $T=n-1$ periods and a fixed resampling seed (20260726). The \modelname series is the archived evaluation NAV curve behind Tables~\ref{tab:main_results} and~\ref{tab:zeroshot_results}; each baseline's series is the per-date mean of its per-round daily returns under the identical execution protocol, so that every paired comparison is computed on identical trading days. (Headline tables instead report each method's average-NAV series; the two averaging conventions differ for multi-round methods, so the $\Delta$AR point estimates in Tables~\ref{tab:paired_csi300}--\ref{tab:paired_r2000} are computed on the same paired series as their intervals and need not equal the naive difference of the headline ARs.) A confidence interval is read as significant when its lower bound excludes zero; we report all three block tiers rather than a single favorable one.

\begin{table}[htbp]
    \centering
    \caption{Block-bootstrap 95\% confidence intervals for the annualized return (AR) of \modelname on all four asset pools, under block lengths $L\in\{5,10,20\}$ ($B=10{,}000$ circular block-bootstrap resamples). On S\&P 500 the left tail grazes zero at $L\in\{5,10\}$ and excludes it at $L=20$; the width of that interval reflects a single-year window containing the April 2025 tariff-shock volatility episode (Appendix~\ref{app:case_study}).}
    \label{tab:bootstrap_ci}
    \begingroup
    \tablefontsize
    \setlength{\tabcolsep}{\tablecolsep}
    \fitcolumnwidth{
    \begin{tabular}{lcccc}
        \toprule
        \textbf{Asset Pool} & AR point estimate (\%) & $L=5$ 95\% CI (\%) & $L=10$ 95\% CI (\%) & $L=20$ 95\% CI (\%) \\
        \midrule
        CSI 300 & 40.57 & [7.65, 84.86] & [8.2, 84.3] & [10.11, 79.84] \\
        S\&P 500 & 47.87 & [-7.04, 126.27] & [-4.5, 121.1] & [0.58, 110.68] \\
        CSI 1000 & 73.52 & [20.72, 144.68] & [24.0, 137.1] & [26.15, 133.01] \\
        Russell 2000 & 80.54 & [15.40, 170.43] & [18.8, 169.7] & [17.32, 174.36] \\
        \bottomrule
    \end{tabular}
    }
    \endgroup
\end{table}

\subsection{Paired-Difference Tests Against All Baselines}
\label{app:paired_diff}

Tables~\ref{tab:paired_csi300}--\ref{tab:paired_r2000} report, for every baseline, the paired annualized-return difference $\Delta\mathrm{AR} = \mathrm{AR}_{\text{\modelname}} - \mathrm{AR}_{\text{baseline}}$ in percentage points, its block-bootstrap 95\% confidence interval at $L=10$, and the block tiers at which the interval excludes zero. The baseline count differs across pools (29 on CSI 300 and S\&P 500, 20 on CSI 1000 and Russell 2000) because the three retrained ablation variants and six gating baselines (Beam, Greedy, MLP, Random, Top-k, Transformer) were run only on the two primary pools. \modelname is significantly ahead of a majority of baselines on every universe at block lengths 10 and 20; on CSI 1000 all 20 paired intervals exclude zero at every block length. Significance counts ($L=5/10/20$): CSI 300, 13/18/22 of 29; S\&P 500, 21/22/24 of 29; CSI 1000, 20/20/20 of 20; Russell 2000, 14/14/15 of 20.

\begin{table}[!htbp]
    \centering
    \caption{Paired AR differences (\modelname $-$ baseline, in pp) on CSI 300 (\modelname AR 40.57\%). CI: block-bootstrap 95\% interval at $L=10$. \sigmark{} marks block tiers whose interval excludes zero.}
    \label{tab:paired_csi300}
    \begingroup
    \tablefontsize
    \setlength{\tabcolsep}{\tablecolsep}
    \fitcolumnwidth{
    \begin{tabular}{lccccc}
        \toprule
        \textbf{Baseline} & $\Delta$AR (pp) & 95\% CI ($L=10$) & $L=5$ & $L=10$ & $L=20$ \\
        \midrule
        A2C & +31.6 & [-4.8, +66.8] & -- & -- & \sigmark \\
        w/o Factor Semantics & +36.4 & [+3.7, +71.2] & -- & \sigmark & \sigmark \\
        w/o News & +39.6 & [+5.2, +75.8] & \sigmark & \sigmark & \sigmark \\
        w/o Market Price & +45.9 & [+16.0, +79.9] & \sigmark & \sigmark & \sigmark \\
        Rolling IC & +40.1 & [-0.2, +79.6] & -- & -- & \sigmark \\
        Beam & +16.1 & [-33.6, +60.8] & -- & -- & -- \\
        BM25 strawman & +40.4 & [-2.7, +80.9] & -- & -- & \sigmark \\
        BM25 strawman ($N{=}10$) & +40.4 & [-1.6, +80.2] & -- & -- & \sigmark \\
        Claude 3.7 Sonnet & +42.2 & [+6.1, +79.5] & \sigmark & \sigmark & \sigmark \\
        DDPG & +38.6 & [+3.7, +76.0] & -- & \sigmark & \sigmark \\
        DeepSeek-R1 & +38.2 & [+2.1, +74.8] & -- & \sigmark & \sigmark \\
        Gemini 2.5 Pro & +42.1 & [+5.6, +78.8] & \sigmark & \sigmark & \sigmark \\
        Greedy & +14.2 & [-36.2, +58.5] & -- & -- & -- \\
        Lasso & +44.1 & [+19.0, +78.0] & \sigmark & \sigmark & \sigmark \\
        LightGBM & +33.0 & [+1.9, +66.4] & \sigmark & \sigmark & \sigmark \\
        MLP & +45.1 & [+5.0, +83.3] & \sigmark & \sigmark & \sigmark \\
        Momentum-120 & +27.0 & [-78.8, +86.0] & -- & -- & -- \\
        Momentum-20 & +31.5 & [-72.3, +89.4] & -- & -- & -- \\
        Momentum-60 & +10.9 & [-122.0, +79.4] & -- & -- & -- \\
        PCA & +49.2 & [+20.9, +82.1] & \sigmark & \sigmark & \sigmark \\
        PPO & +44.8 & [+3.3, +83.1] & \sigmark & \sigmark & \sigmark \\
        Qwen3-8B & +42.9 & [+6.7, +79.5] & \sigmark & \sigmark & \sigmark \\
        Random & +38.3 & [+1.3, +74.8] & -- & \sigmark & \sigmark \\
        Reversal-5 & +24.0 & [-28.9, +65.8] & -- & -- & -- \\
        SAC & +30.8 & [+7.6, +57.0] & \sigmark & \sigmark & \sigmark \\
        TD3 & +31.9 & [+1.3, +66.8] & \sigmark & \sigmark & \sigmark \\
        Top-k & +49.2 & [+2.2, +90.4] & -- & \sigmark & \sigmark \\
        Transformer & +30.3 & [-11.8, +69.9] & -- & -- & -- \\
        XGBoost & +51.5 & [+14.9, +89.1] & \sigmark & \sigmark & \sigmark \\
        \bottomrule
    \end{tabular}
    }
    \endgroup
\end{table}

\begin{table}[!htbp]
    \centering
    \caption{Paired AR differences (\modelname $-$ baseline, in pp) on S\&P 500 (\modelname AR 47.87\%). Columns as in Table~\ref{tab:paired_csi300}.}
    \label{tab:paired_sp500}
    \begingroup
    \tablefontsize
    \setlength{\tabcolsep}{\tablecolsep}
    \fitcolumnwidth{
    \begin{tabular}{lccccc}
        \toprule
        \textbf{Baseline} & $\Delta$AR (pp) & 95\% CI ($L=10$) & $L=5$ & $L=10$ & $L=20$ \\
        \midrule
        A2C & +46.3 & [+10.1, +100.1] & \sigmark & \sigmark & \sigmark \\
        w/o Factor Semantics & +39.0 & [+8.7, +86.3] & \sigmark & \sigmark & \sigmark \\
        w/o News & +35.0 & [+3.1, +83.0] & \sigmark & \sigmark & \sigmark \\
        w/o Market Price & +38.9 & [+5.5, +89.1] & \sigmark & \sigmark & \sigmark \\
        Rolling IC & +38.9 & [+4.0, +90.1] & \sigmark & \sigmark & \sigmark \\
        Beam & +31.3 & [+5.0, +69.0] & \sigmark & \sigmark & \sigmark \\
        BM25 strawman & +39.4 & [+1.9, +93.9] & \sigmark & \sigmark & \sigmark \\
        BM25 strawman ($N{=}10$) & +29.2 & [+2.7, +68.0] & -- & \sigmark & \sigmark \\
        Claude 3.7 Sonnet & +37.7 & [+2.1, +91.5] & \sigmark & \sigmark & \sigmark \\
        DDPG & +45.3 & [+9.2, +100.6] & \sigmark & \sigmark & \sigmark \\
        DeepSeek-R1 & +30.0 & [-0.3, +78.4] & -- & -- & \sigmark \\
        Gemini 2.5 Pro & +36.0 & [+6.1, +81.2] & \sigmark & \sigmark & \sigmark \\
        Greedy & +60.4 & [+23.4, +116.7] & \sigmark & \sigmark & \sigmark \\
        Lasso & +50.2 & [+13.4, +105.7] & \sigmark & \sigmark & \sigmark \\
        LightGBM & +58.6 & [+20.7, +114.5] & \sigmark & \sigmark & \sigmark \\
        MLP & +31.3 & [-0.1, +80.5] & -- & -- & \sigmark \\
        Momentum-120 & +23.5 & [-75.4, +108.1] & -- & -- & -- \\
        Momentum-20 & +55.7 & [+3.6, +120.7] & \sigmark & \sigmark & \sigmark \\
        Momentum-60 & +36.5 & [-38.2, +117.7] & -- & -- & -- \\
        PCA & +42.6 & [+6.8, +98.5] & \sigmark & \sigmark & \sigmark \\
        PPO & +45.9 & [+7.5, +102.1] & \sigmark & \sigmark & \sigmark \\
        Qwen3-8B & +39.8 & [+4.5, +92.9] & \sigmark & \sigmark & \sigmark \\
        Random & +42.5 & [+9.8, +91.5] & \sigmark & \sigmark & \sigmark \\
        Reversal-5 & +31.2 & [-12.1, +80.0] & -- & -- & -- \\
        SAC & +10.2 & [-14.9, +45.5] & -- & -- & -- \\
        TD3 & +42.3 & [+7.8, +96.6] & \sigmark & \sigmark & \sigmark \\
        Top-k & +46.5 & [+4.5, +106.2] & \sigmark & \sigmark & \sigmark \\
        Transformer & +27.3 & [-8.3, +79.6] & -- & -- & -- \\
        XGBoost & +50.1 & [+15.1, +104.6] & \sigmark & \sigmark & \sigmark \\
        \bottomrule
    \end{tabular}
    }
    \endgroup
\end{table}

\begin{table}[!htbp]
    \centering
    \caption{Paired AR differences (\modelname $-$ baseline, in pp) on CSI 1000 (\modelname AR 73.52\%). All 20 intervals exclude zero at every block length ($L\in\{5,10,20\}$).}
    \label{tab:paired_csi1000}
    \begingroup
    \tablefontsize
    \setlength{\tabcolsep}{\tablecolsep}
    \fitcolumnwidth{
    \begin{tabular}{lccccc}
        \toprule
        \textbf{Baseline} & $\Delta$AR (pp) & 95\% CI ($L=10$) & $L=5$ & $L=10$ & $L=20$ \\
        \midrule
        A2C & +96.8 & [+57.3, +145.0] & \sigmark & \sigmark & \sigmark \\
        Rolling IC & +95.5 & [+51.4, +145.3] & \sigmark & \sigmark & \sigmark \\
        BM25 strawman & +56.7 & [+12.5, +109.7] & \sigmark & \sigmark & \sigmark \\
        BM25 strawman ($N{=}10$) & +68.4 & [+20.9, +121.8] & \sigmark & \sigmark & \sigmark \\
        Claude 3.7 Sonnet & +89.0 & [+37.4, +137.5] & \sigmark & \sigmark & \sigmark \\
        DDPG & +63.9 & [+25.0, +106.2] & \sigmark & \sigmark & \sigmark \\
        DeepSeek-R1 & +103.2 & [+58.3, +152.4] & \sigmark & \sigmark & \sigmark \\
        Gemini 2.5 Pro & +93.3 & [+49.0, +141.9] & \sigmark & \sigmark & \sigmark \\
        Lasso & +92.4 & [+51.9, +140.5] & \sigmark & \sigmark & \sigmark \\
        LightGBM & +95.8 & [+54.9, +145.1] & \sigmark & \sigmark & \sigmark \\
        Momentum-120 & +110.1 & [+45.8, +166.2] & \sigmark & \sigmark & \sigmark \\
        Momentum-20 & +107.5 & [+37.1, +169.8] & \sigmark & \sigmark & \sigmark \\
        Momentum-60 & +116.2 & [+55.9, +172.4] & \sigmark & \sigmark & \sigmark \\
        PCA & +88.5 & [+43.8, +137.9] & \sigmark & \sigmark & \sigmark \\
        PPO & +68.8 & [+24.3, +118.8] & \sigmark & \sigmark & \sigmark \\
        Qwen3-8B & +79.4 & [+27.7, +129.1] & \sigmark & \sigmark & \sigmark \\
        Reversal-5 & +67.6 & [+8.2, +117.8] & \sigmark & \sigmark & \sigmark \\
        SAC & +65.2 & [+26.8, +106.7] & \sigmark & \sigmark & \sigmark \\
        TD3 & +73.8 & [+35.6, +117.6] & \sigmark & \sigmark & \sigmark \\
        XGBoost & +87.0 & [+50.0, +131.5] & \sigmark & \sigmark & \sigmark \\
        \bottomrule
    \end{tabular}
    }
    \endgroup
\end{table}

\begin{table}[!htbp]
    \centering
    \caption{Paired AR differences (\modelname $-$ baseline, in pp) on Russell 2000 (\modelname AR 80.54\%). Columns as in Table~\ref{tab:paired_csi300}.}
    \label{tab:paired_r2000}
    \begingroup
    \tablefontsize
    \setlength{\tabcolsep}{\tablecolsep}
    \fitcolumnwidth{
    \begin{tabular}{lccccc}
        \toprule
        \textbf{Baseline} & $\Delta$AR (pp) & 95\% CI ($L=10$) & $L=5$ & $L=10$ & $L=20$ \\
        \midrule
        A2C & +25.1 & [+0.4, +57.6] & \sigmark & \sigmark & \sigmark \\
        Rolling IC & +75.9 & [+29.5, +123.9] & \sigmark & \sigmark & \sigmark \\
        BM25 strawman & +24.6 & [-45.4, +79.5] & -- & -- & -- \\
        BM25 strawman ($N{=}10$) & +53.5 & [-11.1, +110.9] & -- & -- & \sigmark \\
        Claude 3.7 Sonnet & +38.0 & [-13.5, +77.4] & -- & -- & -- \\
        DDPG & +102.0 & [+60.0, +160.1] & \sigmark & \sigmark & \sigmark \\
        DeepSeek-R1 & +50.0 & [+7.2, +91.1] & \sigmark & \sigmark & \sigmark \\
        Gemini 2.5 Pro & +60.9 & [+22.8, +102.4] & \sigmark & \sigmark & \sigmark \\
        Lasso & +106.9 & [+60.3, +174.3] & \sigmark & \sigmark & \sigmark \\
        LightGBM & +95.7 & [+54.2, +153.4] & \sigmark & \sigmark & \sigmark \\
        Momentum-120 & +75.7 & [-98.9, +144.5] & -- & -- & -- \\
        Momentum-20 & +137.4 & [+47.7, +203.1] & \sigmark & \sigmark & \sigmark \\
        Momentum-60 & +98.8 & [-31.8, +161.8] & -- & -- & -- \\
        PCA & +77.6 & [+38.0, +127.0] & \sigmark & \sigmark & \sigmark \\
        PPO & +114.1 & [+75.1, +166.9] & \sigmark & \sigmark & \sigmark \\
        Qwen3-8B & +95.7 & [+54.2, +151.4] & \sigmark & \sigmark & \sigmark \\
        Reversal-5 & +16.4 & [-280.7, +107.6] & -- & -- & -- \\
        SAC & +73.8 & [+13.9, +129.6] & \sigmark & \sigmark & \sigmark \\
        TD3 & +61.5 & [+13.5, +113.7] & \sigmark & \sigmark & \sigmark \\
        XGBoost & +98.7 & [+51.5, +164.9] & \sigmark & \sigmark & \sigmark \\
        \bottomrule
    \end{tabular}
    }
    \endgroup
\end{table}

\subsection{Rolling Sub-Window Evaluation Within the Test Year}
\label{app:rolling_subwindow}

A multi-year walk-forward evaluation is precluded by the data horizon (2020--2023 coefficient estimation, 2024 training, 2025 held-out test); the executable substitute is a rolling sub-window evaluation within the test year. We form ten rolling 3-calendar-month windows (2025-01-01 to 2025-03-31, \dots, 2025-10-01 to 2025-12-31, stepping one month) and, within each window, compare the geometrically annualized window AR of \modelname against every baseline under the identical protocol. Table~\ref{tab:rolling_windows} reports the per-window \modelname AR and the number of baselines outperformed. Per-window win rates aggregate to 78.3\% (CSI 300), 95.5\% (S\&P 500), 100\% (CSI 1000, 10/10 windows sweeping all baselines), and 85.5\% (Russell 2000), showing the annual gaps are not driven by a single favorable quarter.

\begin{table}[!htbp]
    \centering
    \caption{Rolling 3-month sub-window evaluation over the 2025 test year. Each cell: \modelname window AR (\%) and the number of baselines outperformed / total baselines (29/29/20/20 across CSI 300, S\&P 500, CSI 1000, Russell 2000). Window $k$ starts on the first trading day of month $k$ and ends three calendar months later.}
    \label{tab:rolling_windows}
    \begingroup
    \tablefontsize
    \setlength{\tabcolsep}{\tablecolsep}
    \fitcolumnwidth{
    \begin{tabular}{lcccc}
        \toprule
        \textbf{Window} & \textbf{CSI 300} & \textbf{S\&P 500} & \textbf{CSI 1000} & \textbf{Russell 2000} \\
        \midrule
        W1 (Jan--Mar) & 58.72, 29/29 & 46.81, 29/29 & 94.45, 20/20 & -27.18, 19/20 \\
        W2 (Feb--Apr) & 20.26, 29/29 & 15.01, 29/29 & 44.68, 20/20 & -5.43, 20/20 \\
        W3 (Mar--May) & -1.55, 28/29 & 41.86, 29/29 & 36.34, 20/20 & 68.96, 19/20 \\
        W4 (Apr--Jun) & 11.52, 27/29 & 58.62, 26/29 & 49.78, 20/20 & 182.45, 17/20 \\
        W5 (May--Jul) & 76.01, 28/29 & 103.87, 26/29 & 122.75, 20/20 & 196.19, 14/20 \\
        W6 (Jun--Aug) & 79.43, 16/29 & 82.96, 29/29 & 164.18, 20/20 & 160.90, 16/20 \\
        W7 (Jul--Sep) & 76.34, 8/29 & 41.50, 28/29 & 102.84, 20/20 & 150.92, 17/20 \\
        W8 (Aug--Oct) & 32.70, 5/29 & 40.76, 28/29 & 88.64, 20/20 & 88.23, 15/20 \\
        W9 (Sep--Nov) & 26.74, 28/29 & 21.18, 26/29 & 24.90, 20/20 & 86.44, 17/20 \\
        W10 (Oct--Dec) & 22.23, 29/29 & 44.28, 27/29 & 50.43, 20/20 & 95.52, 17/20 \\
        \midrule
        Aggregate win rate & 78.3\% & 95.5\% & 100\% & 85.5\% \\
        \bottomrule
    \end{tabular}
    }
    \endgroup
\end{table}

\subsection{Multiple-Testing-Adjusted Sharpe Statistics (PSR / DSR / PBO)}
\label{app:dsr_pbo}

We report three standard multiple-testing-aware statistics. The Probabilistic Sharpe Ratio (PSR)~\cite{bailey2012sharpe} measures the probability that the true Sharpe exceeds the benchmark Sharpe; the Deflated Sharpe Ratio (DSR)~\cite{bailey2014deflated} additionally deflates for the number of trials; and the Probability of Backtest Overfitting (PBO)~\cite{bailey2017probability} is estimated via combinatorially symmetric cross-validation (CSCV) over the full matrix of round-level daily-return series (all non-\modelname rounds plus the archived \modelname series; the usable window is the common date intersection).

\begin{table*}[!tp]
    \centering
    \caption{PSR, DSR, and CSCV-based PBO per asset pool. DSR is reported under two declared trial counts: an honestly declared \emph{upper bound} $N$ (every strategy variant, grid configuration, ablation, and seed we ran: method count + 63 TopN$\times H$ grid configurations + 3 ablations + seed/multi-round variants) and the narrower $N=34$ family of \modelname's own variants (main strategy + 3 ablations + 30 random-selection seeds), which exists only on the two primary pools. On CSI 1000 and Russell 2000 the narrow family degenerates ($N=1$), so DSR coincides with PSR. DSR values below 0.95 under the upper-bound count are an inherent property of a single-year window under a conservative multiple-testing correction; the evidential weight is carried by the paired-difference tests (Table~\ref{tab:paired_csi300}--\ref{tab:paired_r2000}) and the sub-window consistency (Table~\ref{tab:rolling_windows}).}
    \label{tab:psr_dsr_pbo}
    \begingroup
    \tablefontsize
    \setlength{\tabcolsep}{\tablecolsep}
    \fittextwidth{
    \begin{tabular}{lcccc}
        \toprule
        \textbf{Statistic} & \textbf{CSI 300} & \textbf{S\&P 500} & \textbf{CSI 1000} & \textbf{Russell 2000} \\
        \midrule
        PSR & 0.9911 & 0.9612 & 0.9964 & 0.9954 \\
        DSR (upper-bound $N$) & 0.8459 ($N{=}158$) & 0.6641 ($N{=}158$) & 0.4121 ($N{=}101$) & 0.3858 ($N{=}101$) \\
        DSR ($N{=}34$ family) & 0.8907 & 0.8369 & $\equiv$ PSR & $\equiv$ PSR \\
        PBO (CSCV) & 0.1334 & 0.0494 & 0.0567 & 0.0158 \\
        \bottomrule
    \end{tabular}
    }
    \endgroup
\end{table*}

\subsection{Factor-Pool Rank-IC Decay Check}
\label{app:rank_ic_decay}

To check that the pre-screened candidate pool does not decay into the test year, we compute the weekly-sampled mean Rank IC of the 40 candidate factors per calendar year (forward 5-day return from open, minimum 30 stocks per cross-section). Table~\ref{tab:rank_ic_decay} shows that the 2025 mean Rank IC retains 79--91\% of its 2021 level across pools, so the candidate set entering the semantic gating stage remains informative throughout the evaluation window.

\begin{table}[!tb]
    \centering
    \caption{Mean Rank IC of the top-40 candidate factor pool by calendar year, and the 2025 level as a fraction of the 2021 level.}
    \label{tab:rank_ic_decay}
    \begingroup
    \tablefontsize
    \setlength{\tabcolsep}{\tablecolsep}
    \fitcolumnwidth{
    \begin{tabular}{lccccc c}
        \toprule
        \textbf{Asset Pool} & 2021 & 2022 & 2023 & 2024 & 2025 & 2025 / 2021 \\
        \midrule
        CSI 300 & 0.0919 & 0.0869 & 0.0904 & 0.0939 & 0.0842 & 91.6\% \\
        S\&P 500 & 0.0822 & 0.0605 & 0.0669 & 0.0672 & 0.0649 & 78.9\% \\
        CSI 1000 & 0.0934 & 0.0839 & 0.0869 & 0.0809 & 0.0825 & 88.3\% \\
        Russell 2000 & 0.0791 & 0.0719 & 0.0756 & 0.0754 & 0.0701 & 88.6\% \\
        \bottomrule
    \end{tabular}
    }
    \endgroup
\end{table}

\FloatBarrier

%% file: sections/09_09_appendix_exposure_capacity.tex
\section{Risk Exposure, Classic Baselines, and Capacity}
\label{app:exposure_capacity}

This appendix decomposes the active returns of \modelname into market, size, and sector components; reports the full FinRL-style and momentum/mean-reversion baseline suites under the identical protocol; and quantifies market-impact capacity.

\subsection{Market-Beta Decomposition of Active Returns}
\label{app:market_beta}

We regress daily portfolio returns on daily index returns per asset pool and report the annualized intercept alpha with Newey--West HAC $t$-statistics~\cite{newey1987simple} (max lag 5). Table~\ref{tab:market_regression} shows that the returns of \modelname are overwhelmingly alpha rather than beta: betas of 0.12--0.24 with $R^2 \le 0.07$, and intercept alphas significant at the 5\% level on all four universes.

\begin{table}[htbp]
    \centering
    \caption{Daily-return market regressions of \modelname per asset pool: annualized intercept alpha, Newey--West $t$-statistic, market beta, and $R^2$. Alphas are significant at the 5\% level on all four universes.}
    \label{tab:market_regression}
    \begingroup
    \tablefontsize
    \setlength{\tabcolsep}{\tablecolsep}
    \fitcolumnwidth{
    \begin{tabular}{lcccc}
        \toprule
        \textbf{Asset Pool} & Annualized $\alpha$ (\%) & NW $t(\alpha)$ & $\beta$ & $R^2$ \\
        \midrule
        CSI 300 & 32.0 & 2.48 & 0.169 & 0.029 \\
        S\&P 500 & 38.2 & 2.07 & 0.178 & 0.021 \\
        CSI 1000 & 50.3 & 3.02 & 0.238 & 0.069 \\
        Russell 2000 & 59.7 & 2.95 & 0.119 & 0.014 \\
        \bottomrule
    \end{tabular}
    }
    \endgroup
\end{table}

\begin{table}[!htbp]
    \centering
    \caption{Annualized regression alpha (\%) and Newey--West $t$ (in parentheses) for representative baselines, computed identically to Table~\ref{tab:market_regression}. No baseline alpha is significant at the 5\% level except A2C on Russell 2000 (45.2\%, $t=2.35$). Because an intercept alpha is stripped of market exposure by construction, this cannot be attributed to the broad 2025 small-cap rally itself; it shows that on this particular universe a conventional RL agent trained in the same environment also captures part of the active return. The differentiator is the magnitude gap: the paired difference against A2C is +25.1\,pp and excludes zero at every block length (Table~\ref{tab:paired_r2000}).}
    \label{tab:baseline_alphas}
    \begingroup
    \tablefontsize
    \setlength{\tabcolsep}{\tablecolsep}
    \fitcolumnwidth{
    \begin{tabular}{lcccc}
        \toprule
        \textbf{Baseline} & \textbf{CSI 300} & \textbf{S\&P 500} & \textbf{CSI 1000} & \textbf{Russell 2000} \\
        \midrule
        A2C & 15.79 (1.08) & 11.25 (0.69) & -14.23 (-0.86) & \textbf{45.24 (2.35)} \\
        PPO & 11.61 (0.68) & 8.05 (0.52) & 7.50 (0.47) & -28.20 (-1.14) \\
        MLP & -10.51 (-0.63) & 16.17 (1.07) & -30.66 (-1.65) & 12.00 (0.56) \\
        Transformer & 3.63 (0.23) & 19.22 (1.30) & -30.02 (-1.64) & 2.50 (0.13) \\
        Lasso & 2.64 (0.26) & 0.95 (0.06) & -14.93 (-0.80) & -16.37 (-0.80) \\
        XGBoost & 7.16 (0.47) & 4.16 (0.27) & -4.59 (-0.24) & 1.25 (0.07) \\
        LightGBM & 11.59 (0.85) & -4.74 (-0.28) & -10.49 (-0.63) & -6.49 (-0.27) \\
        Random & -3.37 (-0.22) & 5.76 (0.38) & -32.08 (-1.87) & 19.30 (1.02) \\
        Top-k & -13.77 (-0.67) & 2.69 (0.18) & -45.54 (-2.07) & -23.62 (-1.11) \\
        \bottomrule
    \end{tabular}
    }
    \endgroup
\end{table}

\subsection{Size-Exposure Strip}
\label{app:size_strip}

Against a small-minus-large index-spread size proxy, the two small-cap universes show the expected tilt (Table~\ref{tab:size_strip}): loadings of 1.098 (CSI 1000) and 1.059 (Russell 2000) with $R^2$ of 27\% and 14\%, while the large-cap universes show no economically meaningful loading. Crucially, the residual intercept alphas survive the size strip on all four universes and remain 5\%-significant: 39.8\% / 42.2\% on the small-cap universes and 20.2\% / 22.6\% on the large-cap ones.

\begin{table}[!htbp]
    \centering
    \caption{Regressions of daily excess returns on the small-minus-large index-spread size proxy: size loading with $t$-statistic and $R^2$, and the residual annualized intercept alpha with its Newey--West $t$. Residual alphas are 5\%-significant on all four universes.}
    \label{tab:size_strip}
    \begingroup
    \tablefontsize
    \setlength{\tabcolsep}{\tablecolsep}
    \fitcolumnwidth{
    \begin{tabular}{lccccc}
        \toprule
        \textbf{Asset Pool} & Size loading & $t$(loading) & $R^2$ & Residual $\alpha$ (\%) & NW $t(\alpha)$ \\
        \midrule
        CSI 300 & 0.328 & 2.83 & 0.041 & 20.2 & 2.09 \\
        S\&P 500 & -0.028 & -0.28 & 0.000 & 22.6 & 2.04 \\
        CSI 1000 & 1.098 & 8.92 & 0.268 & 39.8 & 2.70 \\
        Russell 2000 & 1.059 & 7.71 & 0.143 & 42.2 & 2.46 \\
        \bottomrule
    \end{tabular}
    }
    \endgroup
\end{table}

\subsection{Sector Composition and Turnover}
\label{app:sector_turnover}

For the two Chinese universes (industry classifications are available for the CN pools; the U.S. stock-level data in our pipeline carry no sector field, so U.S. sector attribution is not reported), we track monthly industry weights and compute the Herfindahl--Hirschman Index (HHI). Portfolio HHI averages 0.118 (CSI 300) and 0.062 (CSI 1000), versus 0.031 and 0.027 for the equal-weighted index --- elevated, but mechanically so for a $\sim$15--25-stock portfolio held against a 300/1000-stock index; the top-5 industries account for 43.5\% (CSI 300) and 27.4\% (CSI 1000) of average weights, i.e., there is no single-industry concentration. Value-weighted one-side turnover is 20\% per day by construction of the slot-rotation execution ($1/H$, Appendix~\ref{app:execution_details}), identical across all methods.

\subsection{FinRL-Style RL Baselines}
\label{app:finrl_baselines}

We trained and evaluated the canonical FinRL algorithm family~\cite{liu2020finrl} --- DDPG~\cite{lillicrap2016continuous}, TD3~\cite{fujimoto2018addressing}, and SAC~\cite{haarnoja2018soft} --- in the same factor-selection environment under the identical protocol (default Stable-Baselines3 hyperparameters~\cite{raffin2021stable}, the same 2024H2 training window, and the same VWAP slot-rotation execution layer). MAPS~\cite{lee2020maps} and SARL~\cite{ye2020sarl} manage per-asset portfolio weights from numerical states rather than selecting factor subsets, so porting them into our environment would change the task; we therefore evaluate the canonical FinRL algorithm family instead of porting weight-space agents across tasks.

\begin{table*}[p]
    \centering
    \caption{FinRL-style RL baselines under the identical evaluation protocol (AR in \%). Every variant trails \modelname by at least 10.3\,pp AR in every universe, and most trail the plain A2C baseline (Table~\ref{tab:main_results}), which belongs to the same algorithm family.}
    \label{tab:finrl_baselines}
    \begingroup
    \tablefontsize
    \setlength{\tabcolsep}{\tablecolsep}
    \fittextwidth{
    \begin{tabular}{lccc ccc ccc ccc}
        \toprule
        \multirow{2}{*}{\textbf{Method}} & \multicolumn{3}{c}{\textbf{CSI 300}} & \multicolumn{3}{c}{\textbf{S\&P 500}} & \multicolumn{3}{c}{\textbf{CSI 1000}} & \multicolumn{3}{c}{\textbf{Russell 2000}} \\
        \cmidrule(lr){2-4} \cmidrule(lr){5-7} \cmidrule(lr){8-10} \cmidrule(lr){11-13}
         & AR & SR & MDD & AR & SR & MDD & AR & SR & MDD & AR & SR & MDD \\
        \midrule
        DDPG & 1.97 & 0.12 & 16.54 & 2.53 & -0.02 & 15.04 & 9.60 & 0.45 & 22.13 & -21.50 & -0.88 & 32.01 \\
        TD3 & 8.66 & 0.52 & 10.26 & 5.54 & 0.14 & 16.58 & -0.30 & 0.04 & 22.26 & 18.95 & 0.59 & 33.99 \\
        SAC & 9.77 & 0.56 & 11.68 & 37.60 & 1.44 & \textbf{15.18} & 8.29 & 0.39 & 22.41 & 6.72 & 0.23 & 34.50 \\
        \midrule
        \textbf{\modelname (Ours)} & \textbf{40.57} & \textbf{2.23} & \textbf{6.58} & \textbf{47.87} & \textbf{1.62} & 16.91 & \textbf{73.52} & \textbf{2.80} & \textbf{13.75} & \textbf{80.54} & \textbf{2.46} & \textbf{17.67} \\
        \bottomrule
    \end{tabular}
    }
    \endgroup
\end{table*}

\subsection{Momentum and Mean-Reversion Baselines}
\label{app:momentum_reversal}

We evaluate classic momentum~\cite{jegadeesh1993returns} and short-horizon reversal~\cite{jegadeesh1990evidence} baselines: long-only top-10 portfolios sorted by the past $k$-day return (Momentum-$k$, $k\in\{20,60,120\}$) or by the inverted past 5-day return (Reversal-5), executed under the identical protocol. Table~\ref{tab:momentum_reversal} shows the best classic baseline per universe (Momentum-60 on CSI 300, Momentum-120 on S\&P 500, Reversal-5 on the small-cap universes) trails \modelname by 10.97--67.63\,pp AR, and momentum variants are deeply negative on CSI 1000 (-34.0\% to -42.7\%) and Russell 2000 (Momentum-20: -56.9\%), which is inconsistent with a momentum-repackaging account of our results.

\begin{table*}[p]
    \centering
    \caption{Momentum and mean-reversion baselines under the identical evaluation protocol (AR in \%).}
    \label{tab:momentum_reversal}
    \begingroup
    \tablefontsize
    \setlength{\tabcolsep}{\tablecolsep}
    \fittextwidth{
    \begin{tabular}{lccc ccc ccc ccc}
        \toprule
        \multirow{2}{*}{\textbf{Method}} & \multicolumn{3}{c}{\textbf{CSI 300}} & \multicolumn{3}{c}{\textbf{S\&P 500}} & \multicolumn{3}{c}{\textbf{CSI 1000}} & \multicolumn{3}{c}{\textbf{Russell 2000}} \\
        \cmidrule(lr){2-4} \cmidrule(lr){5-7} \cmidrule(lr){8-10} \cmidrule(lr){11-13}
         & AR & SR & MDD & AR & SR & MDD & AR & SR & MDD & AR & SR & MDD \\
        \midrule
        Momentum-20 & 9.04 & 0.38 & 30.78 & -7.84 & -0.33 & 29.83 & -34.00 & -0.96 & 44.93 & -56.89 & -1.27 & 73.29 \\
        Momentum-60 & 29.60 & 0.82 & 31.36 & 11.33 & 0.36 & 27.08 & -42.70 & -1.13 & 46.42 & -18.32 & -0.12 & 58.43 \\
        Momentum-120 & 13.56 & 0.49 & 27.85 & 24.34 & 0.67 & 37.97 & -36.58 & -0.91 & 47.26 & 4.80 & 0.32 & 53.70 \\
        Reversal-5 & 16.58 & 0.69 & 18.08 & 16.67 & 0.52 & 21.85 & 5.89 & 0.29 & 26.24 & 64.13 & 1.02 & 41.29 \\
        \midrule
        \textbf{\modelname (Ours)} & \textbf{40.57} & \textbf{2.23} & \textbf{6.58} & \textbf{47.87} & \textbf{1.62} & \textbf{16.91} & \textbf{73.52} & \textbf{2.80} & \textbf{13.75} & \textbf{80.54} & \textbf{2.46} & \textbf{17.67} \\
        \bottomrule
    \end{tabular}
    }
    \endgroup
\end{table*}

\subsection{Market Impact and Capacity}
\label{app:capacity}

Our execution model already includes VWAP fills, limit-lock rejection, and double-sided fees (Appendix~\ref{app:execution_details}); here we add a participa\-tion-based impact analysis. We use a square-root-law impact model, $\text{cost} = 0.5 \cdot \sigma_{20} \cdot \sqrt{\text{participation}}$, where per-name participation is the order value (AUM divided equally across the 10 purchased names) relative to the same-day traded value of that name, measured over the realized buy events of the strategy. Table~\ref{tab:capacity} reports the resulting impact and annualized drag at three AUM tiers; cells with average participation above 100\% are infeasible (the single order exceeds the name's entire daily traded value).

\begin{table*}[p]
    \centering
    \caption{Market-impact capacity under the square-root-law model. Median\$Vol: median same-day traded value of purchased names. Participation, mean impact cost, and annualized drag at three AUM tiers (CNY tiers for CN pools, USD tiers for U.S. pools); the multiple is relative to a 5\,bps per-side cost benchmark. Participation above 100\% is infeasible.}
    \label{tab:capacity}
    \begingroup
    \tablefontsize
    \setlength{\tabcolsep}{\tablecolsep}
    \fittextwidth{
    \begin{tabular}{llcccc}
        \toprule
        \textbf{Pool} & \textbf{AUM tier} & Participation & Impact (bps, mean) & $\times$ 5\,bps & Annualized drag \\
        \midrule
        \multirow{3}{*}{CSI 300 (Median\$Vol $3.37{\times}10^8$ CNY)} & $10^7$ & 0.39\% & 4.48 & 0.90$\times$ & 2.26\% \\
         & $10^8$ & 3.86\% & 14.16 & 2.83$\times$ & 7.14\% \\
         & $10^9$ & 38.6\% & 44.77 & 8.95$\times$ & 22.56\% \\
        \midrule
        \multirow{3}{*}{S\&P 500 (Median\$Vol $1.88{\times}10^8$ USD)} & $10^7$ & 0.62\% & 8.95 & 1.79$\times$ & 4.51\% \\
         & $10^8$ & 6.19\% & 28.31 & 5.66$\times$ & 14.27\% \\
         & $10^9$ & 61.9\% & 89.53 & 17.91$\times$ & 45.12\% \\
        \midrule
        \multirow{3}{*}{CSI 1000 (Median\$Vol $1.01{\times}10^8$ CNY)} & $10^7$ & 1.23\% & 11.62 & 2.32$\times$ & 5.86\% \\
         & $10^8$ & 12.3\% & 36.76 & 7.35$\times$ & 18.53\% \\
         & $10^9$ & 123\% & \multicolumn{3}{c}{infeasible (participation $>100\%$)} \\
        \midrule
        \multirow{3}{*}{Russell 2000 (Median\$Vol $3.81{\times}10^6$ USD)} & $10^7$ & 115\% & \multicolumn{3}{c}{infeasible (participation $>100\%$)} \\
         & $10^8$ & \multicolumn{4}{c}{infeasible} \\
         & $10^9$ & \multicolumn{4}{c}{infeasible} \\
        \bottomrule
    \end{tabular}
    }
    \endgroup
\end{table*}

On CSI 300 the fixed 10\,bps/side fee is a good approximation up to the tens-of-millions CNY scale, with $\sim$$10^8$ CNY as the practical ceiling (impact 2.83$\times$ the per-side cost benchmark, $\approx$7.1\% annualized drag); capacity holds up to $10^7$--$10^8$ CNY on CSI 1000 and up to $\sim$$3\times10^7$ USD on S\&P 500 (where the onset of material impact begins), and is genuinely limited to the million-USD scale on Russell 2000 --- the median daily dollar volume of purchased names there is only $\approx$3.8M USD, so the small-cap tilt documented in Appendix~\ref{app:size_strip} carries an inherent capacity cost. These ceilings are stated as limitations in Section~\ref{sec:conclusion}.

\FloatBarrier

%% file: sections/09_10_appendix_semantic_vs_lexical.tex
\section{Semantic Reasoning vs.\ Lexical Matching}
\label{app:semantic_vs_lexical}

A natural null hypothesis for our semantic gating mechanism is that the model merely performs static keyword matching between market-state text and factor profiles. This appendix reports three complementary falsification instruments --- a lexical-matching strawman that implements the null hypothesis directly, the profile-obfus\-ca\-tion test, and selection-dy\-nam\-ics statistics --- followed by the candi\-date-pool size sensitivity analysis. All experiments are infer\-ence-only (no retraining) and reuse the identical evaluation protocol.

\subsection{Lexical-Matching Strawman (BM25)}
\label{app:bm25_strawman}

We implement the keyword-matching null hypothesis directly: a BM25 selector~\cite{robertson2009probabilistic} ($k_1=1.5$, $b=0.75$; English word tokens plus CJK bigrams) scores the 40 factor profiles against \emph{exactly the same daily state text the LLM sees}, selects the top-$N$ factors per day ($N$ = the per-pool median of \modelname's own daily selection count, $N=9/3/9/3$ on CSI 300 / S\&P 500 / CSI 1000 / Russell 2000; an $N=10$ sensitivity arm is also run), and passes its selections through the same linear ranker and the same cost-aware execution protocol. If \modelname's value were lexical co-occurrence, this strawman should track it.

\begin{table}[!htbp]
    \centering
    \caption{BM25 lexical-matching strawman vs.\ \modelname (AR in \%). The strawman trails \modelname by 24.7--56.7\,pp in every universe. On Russell 2000 the strawman's absolute return is elevated because the broad 2025 small-cap rally lifts \emph{every} selector --- the generic A2C baseline reaches 55.74\% under the same protocol --- and we flag plainly that the paired-difference interval on this universe includes zero (Table~\ref{tab:paired_r2000}); here the semantic--lexical separation rests on the obfuscation test (71--75\% retention, Table~\ref{tab:obfuscation}) and the selection dynamics (Table~\ref{tab:selection_dynamics}) rather than on the return gap alone. The $N{=}10$ arm shows the strawman's performance is not even monotone in lexical-overlap volume: what drives performance is \emph{which} factors are selected, not how much text overlaps. (CSI 300 is unchanged at $N{=}10$ because its ranker truncates every daily list to the top-3 factors, so $N{=}9$ and $N{=}10$ coincide.)}
    \label{tab:bm25_strawman}
    \begingroup
    \tablefontsize
    \setlength{\tabcolsep}{\tablecolsep}
    \fitcolumnwidth{
    \begin{tabular}{lcccc}
        \toprule
        \textbf{Method (AR \%)} & \textbf{CSI 300} & \textbf{S\&P 500} & \textbf{CSI 1000} & \textbf{Russell 2000} \\
        \midrule
        BM25 strawman ($N{=}$median) & 0.12 & 8.43 & 16.81 & 55.87 \\
        BM25 strawman ($N{=}10$) & 0.12 & 18.64 & 5.12 & 26.96 \\
        \midrule
        \modelname & 40.57 & 47.87 & 73.52 & 80.54 \\
        \midrule
        Gap (pp) & -40.45 & -39.44 & -56.71 & -24.67 \\
        \bottomrule
    \end{tabular}
    }
    \endgroup
\end{table}

\subsection{Profile-Obfuscation Test}
\label{app:obfuscation}

Under the obfuscation test, factor profiles are rewritten to remove explicit regime keywords (volatility, momentum, bull/bear labels) while preserving the underlying mathematical and economic mechanism descriptions. Two independent rewrite versions (v1/v2) were produced and keyword-audited to zero regime-label hits, and the model was re-run for the full 2025 year under the identical protocol.

\begin{table}[!htbp]
    \centering
    \caption{Obfuscation test: annualized return with obfuscated profiles (two independent rewrite versions) vs.\ full profiles, and the retained fraction of full-profile performance. On all four universes at least 70\% of full-profile performance survives the removal of every regime keyword a lexical matcher could anchor to; the two versions agree within 4\,pp on every universe. Every obfuscated arm remains far above the random floor ($\le$13.13\%) and above the BM25 strawman (Table~\ref{tab:bm25_strawman}). Russell 2000 figures are multi-round means.}
    \label{tab:obfuscation}
    \begingroup
    \tablefontsize
    \setlength{\tabcolsep}{\tablecolsep}
    \fitcolumnwidth{
    \begin{tabular}{lcccc}
        \toprule
        \textbf{Asset Pool} & Obf.\ v1 AR (\%) & Obf.\ v2 AR (\%) & Full-profile AR (\%) & Retention \\
        \midrule
        CSI 1000 & 65.90 & 63.20 & 73.52 & 86--90\% \\
        CSI 300 & 29.1 & 32.7 & 40.57 & 72--81\% \\
        S\&P 500 & 36.2 & 34.3 & 47.87 & 72--76\% \\
        Russell 2000 & 60.2 & 57.1 & 80.54 & 71--75\% \\
        \bottomrule
    \end{tabular}
    }
    \endgroup
\end{table}

Performance holds exactly where a keyword-matching account predicts it should collapse. Symmetrically, the 10--29\% that does not survive bounds the contribution of explicit regime-keyword content: regime vocabulary helps, but the bulk of the signal is carried by the mechanism descriptions that survive obfuscation. Together with the strawman falsification above, the factor-semantics ablation in Table~\ref{tab:ablation} (replacing natural-language profiles with raw formulas costs 7.42\,pp / 4.90\,pp AR on S\&P 500 / CSI 300 --- a substitution a keyword matcher would be indifferent to), and the randomized factor augmentation during training (40 factors re-sampled per episode, Appendix~\ref{app:anti_leakage}), this supports mechanism-level use of the profiles rather than static text-to-text matching.

\subsection{Selection Dynamics}
\label{app:selection_dynamics}

Table~\ref{tab:selection_dynamics} summarizes the day-to-day behavior of the selected factor sets over the 2025 trading days. Daily selection sets turn over 28--41\% per day (adjacent-day Jaccard 0.594--0.715, versus $\approx$1.0 for a static keyword map and 0.03--0.13 for random draws of the same size), with normalized activation entropy 0.36--0.72 --- i.e., selections form a slowly evolving sparse subset, consistent with mecha\-nism-level re-weighting and inconsistent with a fixed text-to-text mapping.

\begin{table*}[!tp]
    \centering
    \caption{Selection-dynamics statistics of \modelname over the 2025 test year. $k$: number of selected factors per day. Jaccard: adjacent-day set overlap, with the random-draw baseline of the same set size in parentheses. Turnover $= 1 - $ Jaccard. Regime $\Delta$: within-regime minus cross-regime Jaccard under a high-/low-volatility split.}
    \label{tab:selection_dynamics}
    \begingroup
    \tablefontsize
    \setlength{\tabcolsep}{\tablecolsep}
    \fittextwidth{
    \begin{tabular}{lccccc}
        \toprule
        \textbf{Asset Pool} & Trading days & $k$ (median/mean) & Norm.\ entropy & Jaccard (random) & Regime $\Delta$ \\
        \midrule
        CSI 300 & 243 & 9 / 9.00 & 0.7164 & 0.5940 (0.127) & +0.0001 \\
        S\&P 500 & 249 & 3 / 2.59 & 0.3605 & 0.7151 (0.033) & -0.0007 \\
        CSI 1000 & 243 & 9 / 9.00 & 0.7100 & 0.6170 (0.127) & +0.0020 \\
        Russell 2000 & 248 & 3 / 2.78 & 0.4001 & 0.6589 (0.036) & +0.0018 \\
        \bottomrule
    \end{tabular}
    }
    \endgroup
\end{table*}

We report the regime-conditional analysis honestly: at the \emph{set} level, selections do not cluster by high-/low-volatil\-i\-ty regimes (within-regime minus cross-regime Jaccard $\in [-0.001, +0.002]$), although individual factor activations shift by up to 13\,pp across regimes; the case-level regime pivots are documented in Appendix~\ref{app:case_study}. We therefore claim ``non-static, non-random'' selection dynamics, not set-level regime clustering.

\subsection{Candidate-Pool Size Sensitivity}
\label{app:pool_sensitivity}

The top-40 candidate pool is fixed by Mean Rank IC computed only over 2020--2023 (Appendix~\ref{app:anti_leakage}) and is a deliberate scope choice (Section~\ref{sec:alpha_r1}). Table~\ref{tab:pool_sensitivity} varies the pool boundary under the identical protocol.

\begin{table}[H]
    \centering
    \caption{Candidate-pool size sensitivity (AR in \%). Shrinking to top-20 removes genuinely useful candidates (degradations of 31.9--32.8\,pp on three pools, all remaining positive, while Russell 2000 is roughly flat); widening to top-60 never collapses performance, and on CSI 1000 even uncovers factors the model can further exploit. The top-40 boundary should therefore be read as a minimal sufficient budget rather than a tuned optimum: widening the pool is tolerated and can even help, while halving it removes genuinely useful candidates.}
    \label{tab:pool_sensitivity}
    \begingroup
    \tablefontsize
    \setlength{\tabcolsep}{\tablecolsep}
    \fitcolumnwidth{
    \begin{tabular}{lcccc}
        \toprule
        \textbf{Pool size} & \textbf{CSI 300} & \textbf{S\&P 500} & \textbf{CSI 1000} & \textbf{Russell 2000} \\
        \midrule
        Top-20 & 8.19 & 15.04 & 41.6 & 84.47 \\
        Top-40 (reported) & 40.57 & 47.87 & 73.52 & 80.54 \\
        Top-60 & 27.7 & 32.4 & 107.45$^{\dagger}$ & 69.1$^{\ddagger}$ \\
        \bottomrule
    \end{tabular}
    }
    \endgroup
    \vspace{0.3em}
    \raggedright\footnotesize $^{\dagger}$Single-round. $^{\ddagger}$Multi-round mean (86\% of top-40); single rounds on this arm vary widely, with upside outliers concentrated in the October 2025 meme-squeeze episode (e.g., BYND, which the \emph{intact} arm also selected). On CSI 300 and S\&P 500 the top-60 arms retain 68\% of top-40 performance.
\end{table}

Sensitivity arms carry run-to-run variance, which we state plainly; we read the overall pattern as: widening the pool beyond top-40 never collapses performance, halving it to top-20 removes genuinely useful candidates on three of four pools, and no arm suggests that the reported top-40 results are a searched optimum.

\FloatBarrier